\documentclass[12pt,onecolumn]{IEEEtran}
\usepackage[utf8]{inputenc}
\usepackage{float}
\usepackage{amsmath}
\usepackage{amsthm}
\usepackage{amssymb}

\makeatletter

\floatstyle{ruled}
\newfloat{algorithm}{tbp}{loa}
\providecommand{\algorithmname}{Algorithm}
\floatname{algorithm}{\protect\algorithmname}

\theoremstyle{plain}
\newtheorem{thm}{\protect\theoremname}
\theoremstyle{remark}
\newtheorem{claim}[thm]{\protect\claimname}

\usepackage{graphicx}
\usepackage{epic}\usepackage{eepic}\usepackage{epsfig}\usepackage{amsfonts}
\usepackage{latexsym}\usepackage{color}\usepackage{enumerate}\usepackage{bbm}
\allowdisplaybreaks[4]

\usepackage[colorlinks,
linkcolor=blue,
anchorcolor=blue,
citecolor=red,
]{hyperref}

\theoremstyle{plain}
\newtheorem{theorem}{Theorem}\newtheorem{lemma}[theorem]{Lemma}\newtheorem{proposition}[theorem]{Proposition}\theoremstyle{definition}
\theoremstyle{remark}
\newtheorem{remark}{Remark}\newcommand{\supp}{\operatorname{supp}}

\providecommand{\claimname}{Claim}
\providecommand{\theoremname}{Theorem}

\makeatother
\providecommand{\claimname}{Claim}
\providecommand{\theoremname}{Theorem}

\begin{document}
\title{Large Deviation Analysis for the Reverse Shannon Theorem}
\author{Shi-Bing~Li, Ke~Li, and Lei~Yu
\thanks{The work of Shi-Bing Li and Ke Li was supported by the NSFC under
grant 61871156 and grant 12031004. The work of Lei Yu was supported
by the NSFC under grant 62101286 and the Fundamental Research Funds
for the Central Universities of China (Nankai University) under grant
054-63233073. \textit{(Corresponding authors: Ke Li; Lei Yu.)}

Shi-Bing Li is with the Institute for Advanced Study
in Mathematics, School of Mathematics, Harbin Institute of Technology,
Nangang District, Harbin 150001, China (e-mail: shibingli10@gmail.com).
Ke Li is with the Institute for Advanced Study in Mathematics, Harbin
Institute of Technology, Nangang District, Harbin 150001, China (e-mail:
carl.ke.lee@gmail.com). Lei Yu is with the School of Statistics and
Data Science, LPMC, KLMDASR, and LEBPS, Nankai University, Tianjin
300071, China (e-mail: leiyu@nankai.edu.cn).}}
\maketitle
\begin{abstract}
We consider the problem of simulating a noisy channel using noiseless
channels with unlimited shared randomness. This can be interpreted
as the reverse problem of Shannon's noisy channel coding theorem.
In contrast to previous works, we employ R{é}nyi divergence (with
the parameter $\alpha\in[0,\infty]$) to measure the level of approximation,
and we obtain the reverse Shannon theorem under this measure, which
characterizes the R{é}nyi simulation rate, the minimum communication
rate required for the R{é}nyi divergence vanishing asymptotically.
Our derivation is done by a precise large-deviation analysis. When
the communication rate is above the R{é}nyi simulation rate, we
provide a complete characterization of the convergence exponent for
the R{é}nyi divergence, called the reliability function. When the
communication rate is below the R{é}nyi simulation rate, we determine
the linear increasing rate for the R{é}nyi divergence, which implies
the strong converse exponent for the order-$\alpha$ fidelity.
\end{abstract}


\section{Introduction}

The reverse Shannon theorem~\cite{BSST2002entanglement} has established
the optimal rate of simulating a noisy channel by using noiseless
ones. Specifically, it states that, when unlimited
randomness is shared by the sender and the receiver, the minimum
communication rate needed for asymptotically reliable simulation of
a channel $\mathcal{W}:\mathcal{X}\rightarrow\mathcal{Y}$ is given
by its capacity
\[
C(\mathcal{W}):=\max_{P_{X}}I(X:Y),
\]
where $I(X:Y)$ is the mutual information between the input $X$ and
the output $Y$ of the channel, and the maximization ranges over all
the distributions $P_{X}$ on the alphabet $\mathcal{X}$. This result,
together with Shannon's channel coding theorem, implies that the asymptotic
conversion between different channels, with the assistance of common
randomness, is essentially reversible.

Various settings of channel simulation were investigated subsequently;
see e.g.,~\cite{Winter2002compression,HJMR2007communication,CLMW2011zero,Cuff2013distributed,BDHSW2014quantum,YGA2015channel,HYBGA2016simulation,LiEi2018strong,YuTan2020on,YuTan2020exact,CRBT2024channel}.
The work~\cite{HJMR2007communication} employed the method of rejection
sampling for exact simulation with variable-length code and obtained
a tight one-shot achievability bound. Zero-error simulation with fixed-length
code was addressed in~\cite{CLMW2011zero}. The works~\cite{Cuff2013distributed},
\cite{BDHSW2014quantum} and~\cite{YuTan2020exact} are concerned
with the tradeoff between the communication rate and the rate of shared
randomness, assuming that the latter is not provided for free. When
no shared randomness is available, the minimum communication rate
needed for approximately simulating the channel in total variance
or relative entropy, is characterized by Wyner's common information~\cite{Wyner1975common}.
In this case, a difference on the communication rate was found when
one considers exact simulation with variable-length code, or when
one uses R{é}nyi divergence of order infinity to measure the level
of approximation~\cite{YuTan2020on}. More recently, the second-order
asymptotics of the communication rate, with unlimited shared randomness
assistance, was derived in~\cite{CRBT2024channel}. We point out
that, some of the aforementioned works such as~\cite{Cuff2013distributed,YGA2015channel,YuTan2020on,YuTan2020exact}
have assumed a fixed input distribution.

In this paper, we conduct a large-deviation analysis of the reverse
Shannon theorem. To specify the exact setting, we consider the simulation
of a discrete memoryless channel using noiseless communication assisted
by unlimited shared randomness, and we are interested in the best
approximation, for all possible channel input, that can be achieved
by employing fixed-length block codes. The large deviation problem
considered in the present paper is concerned with the optimal exponent
of the discrepancy of the simulation for a given rate of communication
cost. This mirrors the error exponent analysis in the reverse problem
of channel coding, where the low-error regime is characterized by
the reliability function~\cite{Fano1961transmission,Gallager1965simple,SGB1967lower1}
and the hight-error regime by the strong converse exponent~\cite{Arimoto1973converse,DueckKorner1979reliability}.

In contrast to previous works, we use the R{é}nyi divergence~\cite{Renyi1961measures},
which is defined as
\[
D_{\alpha}(P\|Q):=\frac{1}{\alpha-1}\log\sum_{x\in\mathcal{X}}P^{\alpha}(x)Q^{1-\alpha}(x)
\]
for two probability distributions $P$ and $Q$ and R{é}nyi parameter
$\alpha\in[0,\infty]$ (cf. Section~\ref{sec:notation-preliminaries}),
to measure the level of approximation. Our main contributions are
as follows.
\begin{enumerate}
\item We investigate the channel simulation problem under R{é}nyi divergence
measures (with the R{é}nyi parameter $\alpha\in[0,\infty]$). We
obtain achievability and converse results for this setting, which
strengthens the reverse Shannon theorem. In particular, we completely
characterize the R{é}nyi simulation rate, which is defined as the
minimum communication rate such that the R{é}nyi divergence vanishes
asymptotically. It is observed that the R{é}nyi simulation rate
depends on the R{é}nyi parameter $\alpha$: It is strictly larger
than the Shannon capacity for $\alpha>1$ and reduces
to the Shannon capacity for $\alpha\in(0,1]$.
\item We completely characterize the reliability function (i.e., the convergence
exponent of the R{é}nyi divergence), when the communication rate
is larger than the R{é}nyi simulation rate. When the communication
rate is smaller than the R{é}nyi simulation rate, we completely
characterize the strong converse exponent under order-$\alpha$ fidelity
with $\alpha\in[0,\infty]$ (i.e., the linearly increasing rate of
the R{é}nyi divergence).
\item We design two versions of rejection sampling. Both of them constitute
the main technical tool for proving the achievability bounds on the
reliability function and strong converse exponent.
\end{enumerate}

\emph{Note added:} Concurrently and independently, the works~\cite{OCCB2024exponents}
and \cite{OYB2024exponents} have investigated the exponents of channel simulation as well.
The work~\cite{OCCB2024exponents} has obtained the exact error exponents in
total variance. The other work~\cite{OYB2024exponents}, which additionally deals
with the classical-quantum channels, has obtained the exact error exponents
in purified distance, a distance measure closely related to the order-$\frac{1}{2}$
R{é}nyi divergence. Besides the different measure we employed, the techniques that we
used are quite different from those in \cite{OCCB2024exponents}
and \cite{OYB2024exponents}. After the posting of these two related works
and ours on arXiv, we identified a mistake in our initial derivation
of the strong converse exponent for the parameter range $\alpha\in(0,1)$,
and it was fixed in a later revision.

The remainder of this paper is structured as follows. In Section~\ref{sec:notation-preliminaries},
we introduce the basic notation and define the relevant entropic quantities.
In Section~\ref{sec:main-results}, we formulate the problem and
present our main results. In Section~\ref{sec:reliability-function},
we derive the reliability function of channel simulation, which also
ensures the achievability of the R{é}nyi simulation rate. In Section~\ref{sec:strong-con-exponent},
we prove the strong converse exponent of channel simulation, which
implies the optimality of the R{é}nyi simulation rate. Finally,
we conclude the paper with some discussion and open questions in Section~\ref{sec:conclusion-discussion}.

\section{Notation and Preliminaries}

\label{sec:notation-preliminaries}

\subsection{Basic Notation}

Let $P_{X}$ be the probability distribution of a random variable
$X$ on alphabet $\mathcal{X}$. We use $\mathcal{P}(\mathcal{X})$
to denote the set of all probability distributions on $\mathcal{X}$,
and use $\supp(P_{X}):=\{x\in\mathcal{X}:P_{X}(x)\neq0\}$ to denote
the support of $P_{X}\in\mathcal{P}(\mathcal{X})$. For two functions
$A$ and $B$ defined on $\mathcal{X}$, the set $\{x:A(x)>B(x)\}$
is abbreviated as $\{A>B\}$, and
\begin{equation}
P(A>B):=\sum_{x:A(x)>B(x)}P(x)
\end{equation}
is the probability of the set $\{x:A(x)>B(x)\}$ with respect to distribution
$P$. In this paper, we assume that all alphabets are finite.

A discrete memoryless channel $\mathcal{W}:\mathcal{X}\rightarrow\mathcal{Y}$
is a stochastic map from the alphabet $\mathcal{X}$ to $\mathcal{Y}$.
We use the same notation to denote the transition matrix of a channel
$\mathcal{W}$. Namely, $\mathcal{W}_{Y|X}(y|x)$ is the probability
of receiving $Y=y$ given that $X=x$ is the input. Given input distribution
$P_{X}\in\mathcal{P}(\mathcal{X})$, $\mathcal{W}(P):=\sum_{x}\mathcal{W}(\cdot|x)P(x)\in\mathcal{P}(\mathcal{Y})$
represents the output distribution of $\mathcal{W}$, and $P_{X}\cdot\mathcal{W}_{Y|X}\in\mathcal{P}(\mathcal{X}\mathcal{\times Y})$
denotes the joint distribution of the input $X$ and the output $Y$,
i.e., $(P_{X}\cdot\mathcal{W}_{Y|X})(x,y)=P_{X}(x)\mathcal{W}_{Y|X}(y|x)$.
When it is clear from the context, we skip the subscript and abbreviate
$\mathcal{\mathcal{W}}_{Y|X}$ as $\mathcal{W}$, and $P_{X}$ as
$P$.

For a sequence $x^{n}:=(x_{1},x_{2},\ldots,x_{n})\in\mathcal{X}^{\times n}$,
we use $T_{x^{n}}(x):=\frac{1}{n}\sum_{i=1}^{n}1_{\{x_{i}=x\}}$ to
denote the type of $x^{n}$. Let $\mathcal{P}_{n}(\mathcal{X}):=\{T_{x^{n}}:x^{n}\in\mathcal{X}^{\times n}\}$
be the set of all types that the elements of $\mathcal{X}^{\times n}$
can take. It is known that the size of $\mathcal{P}_{n}(\mathcal{X})$
increases polynomially with $n$, and we have $|\mathcal{P}_{n}(\mathcal{X})|\leq(n+1)^{|\mathcal{X}|}$~\cite{Csiszar1998method}.
For a type $S\in\mathcal{P}_{n}(\mathcal{X})$, the set of sequences
of length $n$ and type $S$ is called the type class of $S$, denoted
by
\begin{equation}
\mathcal{T}_{S}:=\{x^{n}\in\mathcal{X}^{\times n}:~T_{x^{n}}=S\}.
\end{equation}

Let $G_{n}$ be the symmetric group over the set $\{1,2,\ldots,n\}$.
For any $\pi\in G_{n}$, we use $\mathcal{G}_{\mathcal{X}^{\times n}}^{\pi}$
to denote the permutation channel on $\mathcal{X}^{\times n}$ that
maps $(x_{1},\ldots,x_{n})$ deterministically to $(x_{\pi^{-1}(1)},\ldots,x_{\pi^{-1}(n)})$.
In other words, the transition matrix of this channel is given by
$\mathcal{G^{\pi}}(\bar{x}^{n}|x^{n})=1$ if $(\bar{x}_{1},\ldots,\bar{x}_{n})=(x_{\pi^{-1}(1)},\ldots,x_{\pi^{-1}(n)})$
and $\mathcal{G^{\pi}}(\bar{x}^{n}|x^{n})=0$ otherwise. We say that
a probability distribution $P\in\mathcal{P}(\mathcal{X}^{\times n})$
is symmetric if $P=\mathcal{G}^{\pi}(P)$ for all $\pi\in G_{n}$.

For positive functions $f$ and $g$, we write $f(n)\dot{\leq}g(n)$ if $\limsup_{n\rightarrow\infty}\frac{1}{n}\log\frac{f(n)}{g(n)}\leq0$,
and $f(n)\dot{=}g(n)$ if both $f(n)\dot{\leq}g(n)$ and $g(n)\dot{\leq}f(n)$.
we let $|x|^{+}$ denote $\max\{x,0\}$. Throughout this paper, the
functions $\log$ and $\exp$ are with base $2$, and $\ln$ is with
base $e$.

\subsection{R{é}nyi Divergence and Information Measures}

For $\alpha\in(0,1)\cup(1,\infty)$, the order-$\alpha$ fidelity
between two probability distributions $P,Q\in\mathcal{P}(\mathcal{X})$
is defined as
\begin{equation}
F_{\alpha}(P,Q):=\bigg(\sum_{x\in\mathcal{X}}P^{\alpha}(x)Q^{1-\alpha}(x)\bigg)^{\frac{1}{1-\alpha}}.
\end{equation}
Note that when $\alpha>1$, we adopt the conventions $\frac{0}{0}=0$
and $\frac{a}{0}=\infty$ for $a>0$. The fidelity of order $0$,
$1$ and $\infty$ are defined by taking the limits. That is,
\begin{align}
F_{0}(P,Q) & :=\lim_{\alpha\to0}F_{\alpha}(P,Q)=\sum_{x:P(x)>0}Q(x),\\
F_{1}(P,Q) & :=\lim_{\alpha\to1}F_{\alpha}(P,Q)=\prod_{x\in\mathcal{X}}\left(\frac{Q(x)}{P(x)}\right)^{P(x)},\\
F_{\infty}(P,Q) & :=\lim_{\alpha\to\infty}F_{\alpha}(P,Q)=\min_{x\in\mathcal{X}}\frac{Q(x)}{P(x)}.
\end{align}
With the fidelity defined as above, we can now introduce the order-$\alpha$
R{é}nyi divergence for $P,Q\in\mathcal{P}(\mathcal{X})$ and $\alpha\geq0$:
\begin{equation}
D_{\alpha}(P\|Q):=-\log F_{\alpha}(P,Q).
\end{equation}
When $\alpha=1$, the R{é}nyi divergence $D_{1}(P\|Q)$ is equal
to the relative entropy
\begin{align}
D(P\|Q):=\sum_{x\in\mathcal{X}}P(x)\log\dfrac{P(x)}{Q(x)},
\end{align}
which is also known as the Kullback--Leibler divergence. For later
use, we further extend the above definitions, without changing the
formulas, to the case where $Q$ is a general nonnegative vector on
$\mathcal{X}$.

For a joint probability distribution $P_{XY}\in\mathcal{P}(\mathcal{X}\times\mathcal{Y})$
and $\alpha\in[0,+\infty]$, the R{é}nyi mutual information of order
$\alpha$ is defined as
\begin{equation}
I_{\alpha}(X:Y)_{P_{XY}}:=\min_{Q_{Y}\in\mathcal{P}(\mathcal{Y})}D_{\alpha}(P_{XY}\|P_{X}\times Q_{Y}).
\end{equation}
The minimization above can be solved, and it can
be expressed as~\cite{Verdu2015alpha,HoVerdu2015convexity}
\begin{equation}
I_{\alpha}(X:Y)_{P_{XY}}=\begin{cases}
\frac{\alpha}{\alpha-1}\log\sum_{y}\left(\sum_{x}P_{X}(x)P_{Y|X}^{\alpha}(y|x)\right)^{\frac{1}{\alpha}}, & \alpha\in(0,+\infty)\\
-\max_{y}\log\sum_{x:P_{Y|X}(y|x)>0}P_{X}(x), & \alpha=0\\
\log\sum_{y}\max_{x:P_{X}(x)>0}P_{Y|X}(y|x), & \alpha=\infty.
\end{cases}\label{eq:Iinfty}
\end{equation}

Let $\mathcal{W}:\mathcal{X}\rightarrow\mathcal{Y}$ be a discrete
memoryless channel. For $\alpha\in[0,+\infty]$, the order-$\alpha$
R{é}nyi capacity~\cite{Arimoto1977information} of $\mathcal{W}$
is defined as
\begin{equation}
I_{\alpha}(\mathcal{W}):=\max_{P_{X}\in\mathcal{P}(\mathcal{X})}I_{\alpha}(X:Y)_{P_{X}\cdot\mathcal{W}_{Y|X}}.\label{eq:-6}
\end{equation}
This quantity has two alternative expressions given by~\cite{Csiszar1995generalized}
\begin{align}
I_{\alpha}(\mathcal{W}) & =\max_{P_{X}\in\mathcal{P}(\mathcal{X})}\min_{Q_{Y}\in\mathcal{P}(\mathcal{Y})}\sum_{x\in\mathcal{X}}P_{X}(x)D_{\alpha}(\mathcal{W}_{Y|X}(\cdot|x)\|Q_{Y})\nonumber \\
 & =\min_{Q_{Y}\in\mathcal{P}(\mathcal{Y})}\max_{x}D_{\alpha}(\mathcal{W}_{Y|X}(\cdot|x)\|Q_{Y}).\label{eq:Ichannel}
\end{align}

The function $t\mapsto tI_{1+t}(\mathcal{W})$ appears in the expression
of the reliability function. On the interval $[0,\infty)$, this function
is continuous, but not necessarily differentiable~\cite{Gallager1968information}.
As stated in Lemma~\ref{lem:RenyiC-properties}, on this interval
it is convex, which implies that its right derivative always exists.
We denote its right derivative by
\begin{align}
\hat{R}(t)\equiv\partial_{t}^{+}tI_{1+t}(\mathcal{W}) & :=\lim_{t'\searrow t}\frac{t'I_{1+t'}(\mathcal{W})-tI_{1+t}(\mathcal{W})}{t'-t}.
\end{align}
By Danskin's theorem (e.g., \cite[Theorem 4.16]{bonnans2013perturbation}),
it holds that for $t\in(-1,+\infty)$,
\[
\hat{R}(t)=\max_{P_{X}\in\mathcal{P}_{t}}\frac{\partial}{\partial t}\,(1+t)\log\sum_{y}\left(\sum_{x}P_{X}(x)\mathcal{W}_{Y|X}^{1+t}(y|x)\right)^{\frac{1}{1+t}},
\]
where $\mathcal{P}_{t}$ is the set of optimal solutions (capacity-achieving
distributions) in \eqref{eq:-6} with $\alpha=1+t$.  Similarly,
the left derivative of $t\mapsto tI_{1+t}(\mathcal{W})$ is equal
to
\[
\min_{P_{X}\in\mathcal{P}_{t}}\frac{\partial}{\partial t}\,(1+t)\log\sum_{y}\left(\sum_{x}P_{X}(x)\mathcal{W}_{Y|X}^{1+t}(y|x)\right)^{\frac{1}{1+t}}.
\]
If the capacity-achieving distribution is unique, then $t\mapsto tI_{1+t}(\mathcal{W})$
is differentiable.

Properties of the above entropic quantities are summarized in Lemma~\ref{lem:RenyiD-properties}
and Lemma~\ref{lem:RenyiC-properties} in Appendix~\ref{app:Miscellaneous-Lemmas}.

\section{Problem Statement and Main Results}

\label{sec:main-results}

\subsection{Channel Simulation}

The simulation of a discrete memoryless channel $\mathcal{W}:\mathcal{X}\rightarrow\mathcal{Y}$
with no more than $c$ bits of communication consists of the following
steps. Let $N\in\mathbb{N}$ satisfy $\log N\leq c$. At first, the
sender (Alice) and the receiver (Bob) share an arbitrary random variable
$K$, with distribution $P_{K}$ on alphabet $\mathcal{K}$ of arbitrary
size. Then, for the channel input $X$, Alice applies a local stochastic
map $\mathcal{E}:\mathcal{X}\times\mathcal{K}\rightarrow\{1,\ldots,N\}$
to generate a random variable $J$, and sends $J$ to Bob. At last,
upon receiving $J$, Bob applies a local stochastic map $\mathcal{D}:\{1,\ldots,N\}\times\mathcal{K}\rightarrow\mathcal{Y}$
to transform $(J,K)$ into the channel output $Y$. Denote the above
simulation procedure by $\mathcal{N}:\mathcal{X}\rightarrow\mathcal{Y}$.
Then its transition matrix is
\begin{equation}
\mathcal{N}_{Y|X}(y|x)=\sum_{k\in\mathcal{K}}\sum_{j\in\{1,\ldots,N\}}P_{K}(k)\mathcal{E}_{J|XK}(j|x,k)\mathcal{D}_{Y|JK}(y|j,k).\label{eq:simchannel}
\end{equation}

We employ the order-$\alpha$ R{é}nyi divergence with $\alpha\in[0,\infty]$
to measure the level of approximation of the simulation, defined as
\begin{align}
D_{\alpha}(\mathcal{W},\mathcal{N}):= & \max_{P_{X}\in\mathcal{P}(\mathcal{X})}D_{\alpha}(P_{X}\cdot\mathcal{W}_{Y|X}\|P_{X}\cdot\mathcal{N}_{Y|X})\nonumber \\
= & \max_{x\in\mathcal{X}}D_{\alpha}(\mathcal{W}_{Y|X}(\cdot|x)\|\mathcal{N}_{Y|X}(\cdot|x)).
\end{align}
Let $\mathcal{A}(\mathcal{W},c)$ be the set of all the simulations
for $\mathcal{W}$ whose communication cost is upper bounded by $c$
bits. Then the optimal performance of the simulation for $\mathcal{W}$,
that can be achieved by consuming at most $c$ bits of communication,
is characterized by
\begin{equation}
\mathsf{D}_{\alpha}(\mathcal{W},c):=\min_{\mathcal{N}\in\mathcal{A}(\mathcal{W},c)}D_{\alpha}(\mathcal{W},\mathcal{N}).\label{eq:def-simperform}
\end{equation}

We are interested in the asymptotic performance of simulating $n$
uses of the channel $\mathcal{W}$. Let $\{c_{n}\}_{n=1}^{\infty}$
be a sequence of communication cost. Then the corresponding communication
rate is given by $\limsup_{n\rightarrow\infty}\frac{c_{n}}{n}$. The
minimum communication rate required for asymptotically perfect simulation
of $\mathcal{W}$ is defined as the $\alpha$-R{ényi} simulation
rate and we denote it by $R_{{\rm {RST}}}^{(\alpha)}(\mathcal{W})$.
That is,
\begin{equation}
R_{{\rm {RST}}}^{(\alpha)}(\mathcal{W}):=\inf\left\{ \limsup_{n\rightarrow\infty}\frac{c_{n}}{n}\Big|\limsup_{n\rightarrow\infty}\mathsf{D}_{\alpha}\left(\mathcal{W}^{\times n},c_{n}\right)=0\right\} .
\end{equation}

When the communication rate is strictly larger than $R_{{\rm {RST}}}^{(\alpha)}(\mathcal{W})$,
we expect that the simulation performance $\mathsf{D}_{\alpha}(\mathcal{W}^{\times n},c_{n})$
decay exponentially to $0$ as $n$ increases. The
reliability function characterizes the best rate of such exponential
decay, defined as
\begin{equation}
E_{{\rm rf}}^{(\alpha)}\!(\mathcal{W}\!,r)\!:=\!\sup\!\left\{ \!\liminf_{n\rightarrow\infty}\!\frac{-1}{n}\!\log\mathsf{D}_{\alpha}\!\left(\mathcal{W}^{\times n}\!,\!c_{n}\right)\!\Big|\!\limsup_{n\rightarrow\infty}\!\frac{c_{n}}{n}\!\leq\!r\!\right\} \!.\label{eq:def-rf}
\end{equation}

On the other hand, when the communication rate is strictly smaller
than $R_{{\rm {RST}}}^{(\alpha)}(\mathcal{W})$, the simulation performance
$\mathsf{D}_{\alpha}(\mathcal{W}^{\times n},c_{n})$ actually increases
at least linearly in $n$. The strong converse exponent
characterizes the smallest rate of such linear increase,
defined as
\begin{equation}
E_{{\rm sc}}^{(\alpha)}(\mathcal{W},r):=\inf\left\{ \limsup_{n\to\infty}\frac{1}{n}\mathsf{D}_{\alpha}\left(\mathcal{W}^{\times n},c_{n}\right)\Big|\limsup_{n\rightarrow\infty}\frac{c_{n}}{n}\leq r\right\} .\label{eq:def-sce}
\end{equation}

\begin{remark} The above definition of the strong converse exponent
based on the order-$\alpha$ R{é}nyi divergence is equivalent to
a definition based on the order-$\alpha$ fidelity. Specifically,
in Eq.~\eqref{eq:def-sce}, $\mathsf{D}_{\alpha}(\mathcal{W}^{\times n},c_{n})$
can be replaced by $-\log\mathsf{F}_{\alpha}(\mathcal{W}^{\times n},c_{n})$
without changing the strong converse exponent, where
\begin{equation}
\mathsf{F}_{\alpha}\left(\mathcal{W}^{\times n},c_{n}\right):=\max_{\mathcal{N}^{(n)}\in\mathcal{A}(\mathcal{W}^{\times n},c_{n})}F_{\alpha}\left(\mathcal{W}^{\times n},\mathcal{N}^{(n)}\right)
\end{equation}
with
\begin{equation}
F_{\alpha}(\mathcal{W},\mathcal{N}):=\min_{P_{X}\in\mathcal{P}(\mathcal{X})}F_{\alpha}(P_{X}\cdot\mathcal{W},P_{X}\cdot\mathcal{N}).
\end{equation}
In this way it is easy to see that $E_{{\rm sc}}^{(\alpha)}(\mathcal{W},r)$
characterizes the slowest rate of exponential convergence of the simulation
performance towards the useless, measured by the order-$\alpha$ fidelity.
\end{remark}

\subsection{Main Results}

\label{subsec:Main-Results}

The first main result is the reverse Shannon theorem under the R{é}nyi
divergence with parameter $\alpha\in[0,\infty]$.
Interestingly, when $\alpha\in(0,1]$, the $\alpha$-R{é}nyi simulation
rate $R_{{\rm {RST}}}^{(\alpha)}(\mathcal{W})$ is shown to equal
the channel capacity, while when $\alpha\in\{0\}\cup(1,\infty]$,
the $\alpha$-R{é}nyi simulation rate $R_{{\rm {RST}}}^{(\alpha)}(\mathcal{W})$
changes from the channel capacity to the order-$\alpha$ R{é}nyi
capacity.

\begin{theorem}\label{thm:RST} For any discrete
memoryless channel $\mathcal{W}:\mathcal{X}\rightarrow\mathcal{Y}$,
we have
\begin{equation}
R_{{\rm {RST}}}^{(\alpha)}(\mathcal{W})=\left\{ \begin{array}{ll}
I(\mathcal{W}), & \alpha\in(0,1]\\
I_{\alpha}(\mathcal{W}), & \alpha\in\{0\}\cup(1,\infty].
\end{array}\right.
\end{equation}
\end{theorem}

Theorem~\ref{thm:RST} follows from a more refined large-deviation
analysis of the problem. In the following Theorem~\ref{thm:reliability-function}
and Theorem~\ref{thm:strong-con-exponent}, we completely characterize
the reliability function and the strong converse exponent under the
R{é}nyi divergence with parameter $\alpha\in[0,\infty]$.

\begin{theorem} \label{thm:reliability-function}
For any discrete memoryless channel $\mathcal{W}:\mathcal{X}\rightarrow\mathcal{Y}$,
we have that for $0\le r<I_{\infty}(\mathcal{W})$,
\begin{equation}
E_{{\rm rf}}^{(\alpha)}(\mathcal{W},r)=\left\{ \begin{array}{ll}
\sup\limits_{\beta\geq1}(\beta-1)\left(r-I_{\beta}(\mathcal{W})\right), & \alpha\in(0,1]\\
\left|\sup\limits_{\beta\geq\alpha}(\beta-1)\left(r-I_{\beta}(\mathcal{W})\right)\right|^{+}, & \alpha\in(1,\infty],
\end{array}\right.\label{eq:rel-func}
\end{equation}
and for $r\ge I_{\infty}(\mathcal{W})$, $E_{{\rm rf}}^{(\alpha)}(\mathcal{W},r)=\infty$.
Moreover, $E_{{\rm rf}}^{(0)}(\mathcal{W},r)=\infty$ for $r>I_{0}(\mathcal{W})$ and $E_{{\rm rf}}^{(0)}(\mathcal{W},r)=0$ for $r<I_{0}(\mathcal{W})$.
\end{theorem}

In fact, the result that for $r\ge I_{\infty}(\mathcal{W})$,
$E_{{\rm rf}}^{(\alpha)}(\mathcal{W},r)=\infty$ was immediately implied
by \cite{CLMW2011zero}, since in \cite{CLMW2011zero}, Cubitt et
al. showed that exact simulation of the channel $\mathcal{W}$
can be realized as long as $r\ge I_{\infty}(\mathcal{W})$.

\begin{theorem}\label{thm:strong-con-exponent}
For any discrete memoryless channel $\mathcal{W}:\mathcal{X}\rightarrow\mathcal{Y}$
and $r\geq0$, we have
\begin{equation}
E_{{\rm sc}}^{(\alpha)}(\mathcal{W},r)=\left\{ \begin{array}{ll}
\max\limits_{\alpha\leq\beta\leq1}\frac{\alpha(1-\beta)}{\beta(1-\alpha)}\left(I_{\beta}(\mathcal{W})-r\right), & \alpha\in(0,1)\\
|I_{\alpha}(\mathcal{W})-r|^{+}, & \alpha\in\{0\}\cup[1,\infty].
\end{array}\right.\label{eq:str-con}
\end{equation}
\end{theorem}

In the next two sections, we prove Theorem~\ref{thm:reliability-function} and Theorem~\ref{thm:strong-con-exponent} for $\alpha>0$. The results for the special
case $\alpha=0$ will be derived in Appendix~\ref{app:proof-order-zero}. It can be easily seen from Theorems~\ref{thm:reliability-function}
and \ref{thm:strong-con-exponent} that when the communication rate is larger
than the Rényi simulation rate, the reliability function is strictly
positive, and when the communication rate is smaller than the Rényi
simulation rate, the strong converse exponent is strictly positive.
This proves Theorem~\ref{thm:RST}; see Appendix~\ref{app:proof-RST}
for details.

The limiting case $\alpha=\infty$ is worth further discussion.
Theorem~\ref{thm:RST} gives that
\begin{equation}
R_{{\rm{RST}}}^{(\infty)}(\mathcal{W})
=I_{\infty}(\mathcal{W})=\log\sum_{y}\max_{x}\mathcal{W}_{Y|X}(y|x),
\end{equation}
where the second equality can be easily verified by using Eq.~\eqref{eq:Iinfty}.
An interesting observation is that this $\infty$-R{é}nyi simulation
rate coincides with the minimum communication rate for exact channel
simulation; see \cite{CLMW2011zero}. Similar phenomena were observed
in \cite{YuTan2020on,YuTan2020exact} for the common information problem
and the channel synthesis problem (with limited shared randomness).
Moreover, an explicit equivalence between the $\infty$-R{é}nyi
common information and the exact common information and an explicit
equivalence between the $\infty$-R{é}nyi channel synthesis and
the exact channel synthesis were respectively provided in \cite{YuTan2020on}
and \cite{YuTan2020exact}. As for the channel simulation problem
here, the equivalent relation between the $\infty$-R{é}nyi simulation
and the exact simulation is still unclear. Specifically, given an
$\infty$-Rényi simulation code, is there a simple way to modify it
to an exact simulation code? It is interesting to investigate this
in the future.

\section{Reliability Function}

\label{sec:reliability-function}

This section is devoted to the proof of the reliability function of
Theorem~\ref{thm:reliability-function}. In Section~\ref{subsec:rf-achi},
we design a simulation scheme and analyze its performance, resulting
in an achievability bound for the reliability function. An alternative
simulation scheme is given in Appendix~\ref{app:altsimulation}.
In Section~\ref{subsec:rf-conv}, we prove that the obtained bound
is indeed optimal.

\subsection{Achievability Bound}

\label{subsec:rf-achi} We consider use the rejection sampling (see
e.g., \cite{HJMR2007communication,CRBT2024channel}) to design a simulation
scheme. We first introduce the rejection sampling.

\subsubsection{Rejection Sampling Procedure}

Let $P,Q$ be two probability distributions satisfying $\supp(P)\subseteq\supp(Q)$,
and $N,\tilde{N}$ be positive integers. We design a rejection sampling
procedure as follows.
\begin{algorithm}[H]
\textbf{Computing Sub-distribution:} We introduce the sub-distribution
$\bar{P}$ with $\bar{P}(x)=\min\{P(x),NQ(x)\}$.

\textbf{Rejection Sampling Procedure:} Let $X_{j}\sim Q$, $0\leq j\leq\widetilde{N}$
be sampled independently.
\begin{enumerate}
\item For $j\leftarrow1$ to $\widetilde{N}$ do \\
 With probability $\frac{\bar{P}(X_{j})}{NQ(X_{j})}$, halt and output
$j$; with probability $1-\frac{\bar{P}(X_{j})}{NQ(X_{j})}$,
continue.
\item Abort and output $j=0$ (this happens if the procedure does not yield
an output in the $\widetilde{N}$ iterations).
\end{enumerate}
\caption{Rejection sampling with parameters $(P,Q,N,\widetilde{N})$.}
\label{alg:rejsamp}
\end{algorithm}

The final random variable generated by this procedure
is $X_{J}$, where $J$ is the output of the procedure. Conditioning
on that the iteration goes to the $j$-th, the probability that the
procedure halts and $X_{j}=x$ is $\frac{\bar{P}(x)}{N}$ (which is
a sub-distribution if $x$ is considered as variable), and the probability
that the procedure continues to go into the next iteration is $1-\frac{1}{N}\sum_{x}\bar{P}(x)$.
So, the probability that the output $J$ is in $\{1,\ldots,\widetilde{N}\}$
and $X_{J}=x$ is  proportional to $\bar{P}(x)$. The situation
that is not this case is when the output $j=0$, which occurs with
probability $p:=(1-\frac{1}{N}\sum_{x}\bar{P}(x))^{\widetilde{N}}$.
Hence, the total procedure samples a distribution $S$, which is given
by
\begin{equation}
S(x)=\left(1-p\right)\frac{\bar{P}(x)}{\sum_{x'}\bar{P}(x')}+pQ(x).\label{eq:rej-sam}
\end{equation}
For $s\in(0,\infty]$, the order-$(1+s)$ R{é}nyi divergence between
$P$ and the sampling distribution $S$ is
\begin{align}
D_{1+s}(P\|S) & =D_{1+s}\left(P\big\|(1-p)\frac{\bar{P}}{\sum_{x}\bar{P}(x)}+pQ\right)\nonumber \\
 & \leq D_{1+s}\left(P\|(1-p)\bar{P}\right)\nonumber \\
 & =D_{1+s}\left(P\|\bar{P}\right)+\log\left\{ 1+\frac{p}{1-p}\right\} .\label{eq:key-eq-2}
\end{align}
When $s=\infty$, we have
\begin{align}
D_{\infty}(P\|\bar{P})= & \log\max_{x}\frac{P(x)}{\min\{P(x),NQ(x)\}}\nonumber \\
= & \Big|\log\max_{x}\frac{P(x)}{NQ(x)}\Big|^{+}\nonumber \\
= & |D_{\infty}(P\|Q)-\log N|^{+}.\label{eq:key-eq-infty}
\end{align}
For the other case $s\in(0,\infty)$, it holds that
\begin{align}
 & D_{1+s}\left(P\|\bar{P}\right)\nonumber \\
= & \frac{1}{s}\log\left\{ P(\mathcal{O}^{c})+N^{-s}\sum_{x\in\mathcal{O}}P(x)\left(\frac{P(x)}{Q(x)}\right)^{s}\right\} \nonumber \\
\leq & \frac{1}{s}\log\left\{ 1+N^{-s}\sum_{x}P(x)\left(\frac{P(x)}{Q(x)}\right)^{s}\right\} \nonumber \\
= & \frac{1}{s}\log\big\{1+\exp\{-s(\log N-D_{1+s}(P\|Q))\}\big\}\label{eq:re-sim}\\
\leq & \frac{1}{s\ln2}\exp\{-s(\log N-D_{1+s}(P\|Q))\},\label{eq:re-sim1}
\end{align}
where the last line is by the inequality $\ln(1+x)\leq x$ for $x\geq0$.
Using this inequality again, from Eqs.~\eqref{eq:key-eq-2}, \eqref{eq:key-eq-infty}
and \eqref{eq:re-sim1}, we obtain that for $s\in(0,\infty)$
\begin{align}
D_{1+s}(P\|S)\leq\frac{1}{s\ln2}\exp\{-s(\log N-D_{1+s}(P\|Q))\}+\frac{p}{(1-p)\ln2},\label{eq:eq-re-1}
\end{align}
and
\begin{align}
D_{\infty}(P\|S)\leq|D_{\infty}(P\|Q)-\log N|^{+}+\frac{p}{(1-p)\ln2}.\label{eq:eq-re-infty}
\end{align}
To bound the sampling error, we need to estimate $p$. By the definition
of $\bar{P}(x)$, we have
\begin{align}
\sum_{x}\bar{P}(x)\geq & 1-P\left(P>NQ\right)\nonumber \\
\geq & 1-\sum_{x}P(x)\left(\frac{P(x)}{NQ(x)}\right)^{s}\nonumber \\
= & 1-\exp\{-s(\log N-D_{1+s}(P\|Q))\}.\label{eq:for-strong-converse}
\end{align}
Therefore,
\begin{align}
p\leq\left(1-\frac{1}{N}(1-\exp\{-s(\log N-D_{1+s}(P\|Q))\})\right)^{\widetilde{N}}.\label{eq:re-one-shot-p}
\end{align}

Using the above rejection sampling procedure, we construct a simulation
scheme for $\mathcal{W}^{\times n}$.

\subsubsection{Simulation Scheme}\label{subsec:Simulation-Scheme}
Let $s>0$ and $r\geq0$. We set $N:=2^{nr}$,  $\widetilde{N}:=(\ln2)Nn\log n$ and let $Q_{Y}$ be the minimizer of the optimization $\min_{Q_{Y}}\max_{x}D_{1+s}(\mathcal{W}_{Y|X}(\cdot|x)\|Q_{Y})$.
The simulation scheme with parameters $(\mathcal{W}_{Y|X}^{\times n},Q_{Y}^{\times n},N,\widetilde{N})$ consists of the following steps.
\begin{enumerate}
\item Alice and Bob share $Y_{j}^{(n)}\sim Q_{Y}^{\times n}$ for $j=0,1,2,\cdots,\widetilde{N}$.
\item For any input $x^{n}$, Alice uses the rejection sampling procedure
of Algorithm~\ref{alg:rejsamp}, with parameter $(\mathcal{W}_{Y|X}^{\times n}(\cdot|x^{n}),Q_{Y}^{\times n},N,\widetilde{N})$,
to generate a random index $J\in\{0,1,\cdots,\widetilde{N}\}$.
\item Alice sends the index $J$ to Bob by using $\log(\widetilde{N}+1)$
bits.
\item Bob picks $Y_{J}^{(n)}$ as the channel output.
\end{enumerate}
We denote the above simulation by $\mathcal{N}^{(n)}$. Then
\begin{equation}
\mathcal{N}_{Y^{n}|X^{n}}^{(n)}(\cdot|x^{n})=(1-p_{n})\frac{\overline{\mathcal{W}}_{Y^{n}|X^{n}}(\cdot|x^{n})}{\sum_{y^{n}}\overline{\mathcal{W}}_{Y^{n}|X^{n}}(y^{n}|x^{n})}+p_{n}Q_{Y}^{\times n},
\end{equation}
where
\begin{equation}
\overline{\mathcal{W}}_{Y^{n}|X^{n}}(y^{n}|x^{n})=\min\{\mathcal{W}_{Y|X}^{\times n}(y^{n}|x^{n}),NQ_{Y}^{\times n}(y^{n})\}
\end{equation}
and
\begin{equation}
p_{n}=\big(1-\frac{1}{N}\sum_{y^{n}}\overline{\mathcal{W}}_{Y^{n}|X^{n}}(y^{n}|x^{n})\big)^{\widetilde{N}}.
\end{equation}
The communication rate is
\begin{equation}
\lim\limits_{n\to\infty}\frac{1}{n}\log(\widetilde{N}+1)=\lim\limits_{n\to\infty}\frac{1}{n}\log\big\{(\ln2)2^{nr}n\log n+1\big\}=r.
\end{equation}
To estimate the performance of simulation, we consider the following three cases.

\textit{Case 1:} $s\in(0,\infty)$ and $r\in(I_{1+s}(\mathcal{W}),I_{\infty}(\mathcal{W}))$.
By Eq.~\eqref{eq:eq-re-1}, we have
\begin{align}
 & D_{1+s}\left(\mathcal{W}_{Y|X}^{\times n}(\cdot|x^{n})\big{\|}\mathcal{N}_{Y^{n}|X^{n}}^{(n)}(\cdot|x^{n})\right)\nonumber \\
\leq & \frac{1}{s\ln2}\exp\left\{ -ns\left(r-\frac{1}{n}D_{1+s}(\mathcal{W}_{Y|X}^{\times n}(\cdot|x^{n})\big{\|}Q_{Y}^{\times n})\right)\right\} +\frac{p_{n}}{(1-p_{n})\ln2}\nonumber \\
\leq & \frac{1}{s\ln2}\exp\left\{ -ns\left(r-I_{1+s}(\mathcal{W})\right)\right\} +\frac{p_{n}}{(1-p_{n})\ln2}.\label{eq:next-for}
\end{align}
In the above derivation, we have used the relation
\begin{align}
D_{1+s}(\mathcal{W}_{Y|X}^{\times n}(\cdot|x^{n})\big{\|}Q_{Y}^{\times n})
\leq nI_{1+s}(\mathcal{W}),\label{eq:K-2}
\end{align}
which follows from the additivity of R{é}nyi divergence
(cf. Lemma~\ref{lem:RenyiD-properties}~(\romannumeral2)), Eq.~\eqref{eq:Ichannel}
and the choice of $Q_{Y}$.
From Eqs.~\eqref{eq:re-one-shot-p} and \eqref{eq:K-2}, we deduce that
\begin{align}
p_{n}\leq\left(1-\frac{1}{N}(1-\exp\{-ns(r-I_{1+s}(\mathcal{W}))\})\right)^{\widetilde{N}}.\label{eq:for-next-2}
\end{align}
Noticing that the bound of Eq.~\eqref{eq:for-next-2} is independent
of $x^{n}$, we obtain the following uniform estimation:
\begin{align}
 \frac{p_{n}}{(1-p_{n})\ln2}\dot{=} p_n
\dot{\leq} \exp\{-n\log n\}\dot{\leq}\exp\{-ns(r-I_{1+s}(\mathcal{W}))\}.\label{eq:bound-pn}
\end{align}
The combination of Eqs.~\eqref{eq:next-for} and \eqref{eq:bound-pn} gives
\begin{align}
 & \max_{x^{n}}D_{1+s}\left(\mathcal{W}_{Y|X}^{\times n}(\cdot|x^{n})\big{\|}\mathcal{N}_{Y^{n}|X^{n}}^{(n)}(\cdot|x^{n})\right)\nonumber \\
\dot{\leq} &\exp\left\{ -ns\left(r-I_{1+s}(\mathcal{W})\right)\right\}. \label{eq:for-sc}
\end{align}
By the definition of $E_{{\rm rf}}^{(1+s)}(\mathcal{W},r)$
(cf. Eq.~\eqref{eq:def-rf}), we get from Eq.~\eqref{eq:for-sc} that
\begin{align}
E_{{\rm rf}}^{(1+s)}(\mathcal{W},r)
\geq s\left(r-I_{1+s}(\mathcal{W})\right).\label{eq:final-1}
\end{align}

\textit{Case 2:} $s=\infty$ and $r\geq I_{\infty}(\mathcal{W})$. Since Eq.~\eqref{eq:K-2}
holds for $s=\infty$ as well, for all $x^{n}\in\mathcal{X}^{\times n}$ we have
\begin{equation}
	|D_{\infty}(\mathcal{W}_{Y|X}^{\times n}(\cdot|x^{n})\|Q_{Y}^{\times n})-nr|^{+}=0.
\end{equation}
Using Eq.~\eqref{eq:eq-re-infty}, we deduce that
\begin{equation}
	D_{\infty}\left(\mathcal{W}_{Y|X}^{\times n}(\cdot|x^{n})\big{\|}\mathcal{N}_{Y^{n}|X^{n}}^{(n)}(\cdot|x^{n})\right)
	\leq\frac{p_{n}}{(1-p_{n})\ln2}.\label{eq:re-D-infty}
\end{equation}
Because $r\geq I_{\infty}(\mathcal{W})$, by Eq.~\eqref{eq:Ichannel}
and the choice of $Q_{Y}$, we have $\overline{\mathcal{W}}_{Y^{n}|X^{n}}(\cdot|x^{n})=\mathcal{W}_{Y|X}^{\times n}(\cdot|x^{n})$. Therefore,
\begin{equation}
	p_{n}=\left(1-\frac{1}{N}\right)^{\widetilde{N}}\dot{=}\exp\{-n\log n\}.\label{eq:re-pn-infty}
\end{equation}
By the definition of $E_{{\rm rf}}^{(\infty)}(\mathcal{W},r)$
(cf. Eq.~\eqref{eq:def-rf}), Eqs.~\eqref{eq:re-D-infty} and \eqref{eq:re-pn-infty}
imply that for $r\geq I_{\infty}(\mathcal{W})$,
\begin{equation}
	E_{{\rm rf}}^{(\infty)}(\mathcal{W},r)=\infty.\label{eq:final-0}
\end{equation}

\textit{Case 3:} $s\in(0,\infty]$ and $r\in[0,I_{1+s}(\mathcal{W})]$.
Because the function $Q\mapsto D_{1+s}(P\|Q)$ is convex, we have
\begin{align}
 & D_{1+s}\left(\mathcal{W}_{Y|X}^{\times n}(\cdot|x^{n})\big{\|}\mathcal{N}_{Y^{n}|X^{n}}^{(n)}(\cdot|x^{n})\right)\nonumber \\
\leq & (1-p_{n})D_{1+s}\left(\mathcal{W}_{Y|X}^{\times n}(\cdot|x^{n})\Big{\|}\frac{\overline{\mathcal{W}}_{Y^{n}|X^{n}}(\cdot|x^{n})}{\sum_{y^{n}}\overline{\mathcal{W}}_{Y^{n}|X^{n}}(y^{n}|x^{n})}\right)+p_{n}D_{1+s}\left(\mathcal{W}_{Y|X}^{\times n}(\cdot|x^{n})\big{\|}Q_{Y}^{\times n}\right)\nonumber \\
\leq & (1-p_{n})D_{1+s}\left(\mathcal{W}_{Y|X}^{\times n}(\cdot|x^{n})\big{\|}\overline{\mathcal{W}}_{Y^{n}|X^{n}}(\cdot|x^{n})\right)+p_{n}nI_{1+s}\left(\mathcal{W}\right).\label{eq:next0-1}
\end{align}
For $s=\infty$, by Eqs.~\eqref{eq:key-eq-infty} and \eqref{eq:K-2},
\begin{align}
 & D_{\infty}\left(\mathcal{W}_{Y|X}^{\times n}(\cdot|x^{n})\big{\|}\overline{\mathcal{W}}_{Y^{n}|X^{n}}(\cdot|x^{n})\right)\nonumber \\
\leq & D_{\infty}(\mathcal{W}_{Y|X}^{\times n}(\cdot|x^{n})\big{\|}Q_{Y}^{\times n})-nr\nonumber \\
\leq & n(I_{\infty}(\mathcal{W})-r).\label{eq:re-next-infty}
\end{align}
For $s\in(0,\infty)$, by Eqs.~\eqref{eq:re-sim} and \eqref{eq:K-2},
\begin{align}
 & D_{1+s}\left(\mathcal{W}_{Y|X}^{\times n}(\cdot|x^{n})\big{\|}\overline{\mathcal{W}}_{Y^{n}|X^{n}}(\cdot|x^{n})\right)\nonumber \\
\leq &\frac{1}{s}\log\left\{ 1+\exp\left\{ ns(I_{1+s}\left(\mathcal{W}\right)-r)\right\} \right\} \nonumber \\
\leq & n(I_{1+s}\left(\mathcal{W}\right)-r)+\frac{1}{s}.\label{eq:next10}
\end{align}
So, Eq.~\eqref{eq:next0-1} provides a bound which is independent of $x^n$ and increases at most linearly with $n$.
This yields
\begin{equation}
E_{{\rm rf}}^{(1+s)}(\mathcal{W},r)\geq0.\label{eq:final-2}
\end{equation}

Combining Eqs.~\eqref{eq:final-1} and \eqref{eq:final-2}, we obtain that
for any $s\in(0,\infty]$ and $r\in[0,I_{\infty}(\mathcal{W}))$,
\begin{align}
E_{{\rm rf}}^{(1+s)}(\mathcal{W},r)\geq
\left|s\left(r-I_{1+s}(\mathcal{W})\right)\right|^{+}.\label{eq:key-1}
\end{align}
By the monotonicity of R{é}nyi divergence with respect to its order,
we have $E_{{\rm rf}}^{(1+s_{1})}(\mathcal{W},r)\geq E_{{\rm rf}}^{(1+s_{2})}(\mathcal{W},r)$
when $s_{1}\leq s_{2}$. Then, Eqs.~\eqref{eq:final-0} and \eqref{eq:key-1} directly imply
the follow theorem.
\begin{theorem}\label{thm:reliability-achi}
For any discrete memoryless channel $\mathcal{W}:\mathcal{X}\rightarrow\mathcal{Y}$,
we have that for $0\le r<I_{\infty}(\mathcal{W})$,
\begin{align}
E_{{\rm rf}}^{(1+s)}(\mathcal{W},r)\geq\left\{ \begin{array}{ll}
\sup\limits_{t\geq0}t\left(r-I_{1+t}(\mathcal{W})\right), & s\in(-1,0]\\
\left|\sup\limits_{t\geq s}t\left(r-I_{1+t}(\mathcal{W})\right)\right|^{+}, & s\in(0,\infty],
\end{array}\right.\label{eq:final-result}
\end{align}
and for $r\geq I_{\infty}(\mathcal{W})$, $E_{{\rm rf}}^{(1+s)}(\mathcal{W},r)=\infty$.
\end{theorem}

\subsection{Converse Bound}

\label{subsec:rf-conv}

The following lemma is crucial in the proof of the converse bound
of the reliability function. It will also be used in the next section
for proving the converse bound of the strong converse exponent.

\begin{lemma}\label{lem:u-bound} Let $\mathcal{N}:\mathcal{X}\rightarrow\mathcal{Y}$
be a channel representing a simulation procedure with communication
cost no more than $c$ bits. There exists probability distribution
$Q_{Y}\in\mathcal{P}(\mathcal{Y})$, such that for all $P_{X}\in\mathcal{P}(\mathcal{X})$,
\begin{equation}
P_{X}\cdot\mathcal{N}_{Y|X}\leq2^{c}P_{X}\times Q_{Y}.
\end{equation}
\end{lemma}
\begin{IEEEproof}
Let $K$ with distribution $P_{K}$ be the shared randomness in the
simulation. Assume communication cost is $\log N$
bits such that $\log N\leq c$. Let $\mathcal{E}_{J|XK}$ and $\mathcal{D}_{Y|JK}$
be the transition matrices of the local stochastic maps.
According to Eq.~\eqref{eq:simchannel}, we have
\begin{align}
 & (P_{X}\cdot\mathcal{N}_{Y|X})(x,y)\nonumber \\
= & P_{X}(x)\sum_{k\in\mathcal{K}}P_{K}(k)\sum_{j\in\{1,\ldots,N\}}\mathcal{E}(j|x,k)\mathcal{D}(y|j,k)\nonumber \\
\leq & P_{X}(x)\sum_{k\in\mathcal{K}}P_{K}(k)\sum_{j\in\{1,\ldots,N\}}\frac{2^{c}}{N}\mathcal{D}(y|j,k),
\end{align}
Let
\begin{align}
Q_{Y}(y)=\sum_{k\in\mathcal{K}}P_{K}(k)\sum_{j\in\{1,\ldots,N\}}\frac{1}{N}\mathcal{D}(y|j,k).
\end{align}
Then $\sum_{y\in\mathcal{Y}}Q_{Y}(y)=1$, and so $Q_{Y}\in\mathcal{P}(\mathcal{Y})$.
Therefore, we have $P_{X}\cdot\mathcal{N}_{Y|X}\leq2^{c}P_{X}\times Q_{Y}$.
\end{IEEEproof}
The symmetric probability distribution on $\mathcal{Y}^{\times n}$
with equal weight for each type will play a key role in our proof.
We denote it by
\begin{equation}
\Phi_{Y^{n}}=\frac{1}{|\mathcal{P}_{n}(\mathcal{Y})|}\sum_{S\in\mathcal{P}_{n}(\mathcal{Y})}\Psi_{Y^{n}}^{S},\label{eq:unidistribution}
\end{equation}
where $\Psi_{Y^{n}}^{S}$ is the uniform distribution on the type
class $\mathcal{T}_{S}$. For any symmetric probability distribution
$P_{Y^{n}}\in\mathcal{P}(\mathcal{Y}^{\times n})$, since it can be
written as a convex combination of the distributions $\Psi_{Y^{n}}^{S}$,
we have
\begin{equation}
P_{Y^{n}}\leq|\mathcal{P}_{n}(\mathcal{Y})|\Phi_{Y^{n}}\leq(n+1)^{|\mathcal{Y}|}\Phi_{Y^{n}}.\label{eq:symbound}
\end{equation}

\begin{proposition}\label{prop:exponent-u-symm} Let $P_{XY}\in\mathcal{P}(\mathcal{X}\times\mathcal{Y})$,
$\Phi_{Y^{n}}$ be the symmetric probability distribution given in
Eq.~\eqref{eq:unidistribution} and $g(n)$ be a sub-exponential
function of $n\in\mathbb{N}$ satisfying $g(n)>(n+1)^{|\mathcal{Y}|}$.
For fixed $r\in\mathbb{R}$, consider the sequence
\begin{equation}
p_{n}:=P_{XY}^{\times n}\left(P_{XY}^{\times n}\geq g(n)2^{nr}P_{X}^{\times n}\times\Phi_{Y^{n}}\right).
\end{equation}
It holds that
\begin{equation}
\lim_{n\rightarrow\infty}\frac{-1}{n}\log p_{n}=\sup_{t\geq0}t\left(r-I_{1+t}(X:Y)_{P_{XY}}\right).\label{eq:exp-u-s}
\end{equation}
\end{proposition}

Proposition~\ref{prop:exponent-u-symm}
can be regarded as the classical case of the quantum version in~\cite[Proposition 16]{LiYao2024reliability}.
We give a proof in Appendix~\ref{app:proof-of-propexu} for completeness.

We will apply Lemma~\ref{lem:u-bound}, Proposition~\ref{prop:exponent-u-symm}
as well as Eq.~\eqref{eq:symbound} to show the following converse
bound. Note that the converse for $r\geq I_{\infty}(\mathcal{W})$
is trivial, and hence, only the case $r\in[0,I_{\infty}(\mathcal{W}))$
is considered.

\begin{theorem}\label{thm:rel-converse-bound} For any discrete memoryless
channel $\mathcal{W}:\mathcal{X}\rightarrow\mathcal{Y}$, we have that for
$0\le r<I_{\infty}(\mathcal{W})$,
\begin{align}
E_{{\rm rf}}^{(1+s)}(\mathcal{W},r)\leq\left\{ \begin{array}{ll}
\sup\limits_{t\geq0}t(r-I_{1+t}(\mathcal{W})), & \!\!s\in(-1,0]\\
\left|\sup\limits_{t\geq s}t(r-I_{1+t}(\mathcal{W}))\right|^{+}, & \!\!s\in(0,\infty].
\end{array}\right.\label{eq:con-final-result}
\end{align}
\end{theorem}
\begin{IEEEproof}
Let $\{c_{n}\}_{n=1}^{\infty}$ be an arbitrary sequence of nonnegative
numbers satisfying
\begin{equation}
\limsup_{n\rightarrow\infty}\frac{c_{n}}{n}\leq r,\label{eq:eq-rate}
\end{equation}
which is equivalent to the existence of a sub-exponential function
$g(n)\geq1$ such that $2^{c_{n}}\leq g(n)2^{nr}$. By the definition
of the reliability function of Eq.~\eqref{eq:def-rf}, we see that
to prove the theorem, it suffices to show that the same upper bound
applies to
\begin{equation}
\liminf_{n\rightarrow\infty}\frac{-1}{n}\log\mathsf{D}_{1+s}\left(\mathcal{W}^{\times n},c_{n}\right).
\end{equation}
In the following, we will prove an even stronger result, with the
limit inferior replaced by the limit superior.

Fix an input distribution $P_{X}$, and write $P_{XY}:=P_{X}\cdot\mathcal{W}_{Y|X}$.
Let $\mathcal{N}^{(n)}:\mathcal{X}^{\times n}\rightarrow\mathcal{Y}^{\times n}$
be a simulation for $\mathcal{W}^{\times n}:\mathcal{X}^{\times n}\rightarrow\mathcal{Y}^{\times n}$
with communication cost no more than $c_{n}$ bits. Then we have
\begin{align}
 & D_{1+s}\left(\mathcal{W}^{\times n},\mathcal{N}^{(n)}\right)\nonumber \\
\geq & D_{1+s}\left(P_{XY}^{\times n}\big{\|}P_{X}^{\times n}\cdot\mathcal{N}_{Y^{n}|X^{n}}^{(n)}\right)\nonumber \\
\geq & D_{1+s}\left(\mathcal{G}_{(\mathcal{X}\times\mathcal{Y})^{\times n}}\big(P_{XY}^{\times n}\big)\big{\|}\mathcal{G}_{(\mathcal{X}\times\mathcal{Y})^{\times n}}\big(P_{X}^{\times n}\cdot\mathcal{N}_{Y^{n}|X^{n}}^{(n)}\big)\right)\nonumber \\
= & D_{1+s}\left(P_{XY}^{\times n}\big{\|}\mathcal{G}_{(\mathcal{X}\times\mathcal{Y})^{\times n}}\big(P_{X}^{\times n}\cdot\mathcal{N}_{Y^{n}|X^{n}}^{(n)}\big)\right),\label{eq:rel-conv-a}
\end{align}
where $\mathcal{G}_{(\mathcal{X}\times\mathcal{Y})^{\times n}}:=\sum_{\pi\in G_{n}}\frac{1}{|G_{n}|}\mathcal{G}_{(\mathcal{X}\times\mathcal{Y})^{\times n}}^{\pi}$
is the random permutation map. In Eq.~\eqref{eq:rel-conv-a}, the
first inequality is by definition, the second inequality is due to
the data processing inequality of Lemma~\ref{lem:RenyiD-properties}~(\romannumeral3).
Lemma~\ref{lem:u-bound} implies that there is a probability distribution
$S_{Y^{n}}$, such that
\begin{align}
 & \mathcal{G}_{(\mathcal{X}\times\mathcal{Y})^{\times n}}\left(P_{X}^{\times n}\cdot\mathcal{N}_{Y^{n}|X^{n}}^{(n)}\right)\nonumber \\
\leq & \mathcal{G}_{(\mathcal{X}\times\mathcal{Y})^{\times n}}\left(2^{c_{n}}P_{X}^{\times n}\times S_{Y^{n}}\right)\nonumber \\
= & 2^{c_{n}}P_{X}^{\times n}\times\mathcal{G}_{\mathcal{Y}^{\times n}}\left(S_{Y^{n}}\right)\nonumber \\
\leq & (n+1)^{|\mathcal{Y}|}g(n)2^{nr}P_{X}^{\times n}\times\Phi_{Y^{n}},\label{eq:rel-conv-b}
\end{align}
where for the last inequality note that $\mathcal{G}_{\mathcal{Y}^{\times n}}(S_{Y^{n}})$
is symmetric and we use the inequality of Eq.~\eqref{eq:symbound}.
Now, combining Eqs.~\eqref{eq:rel-conv-a} and \eqref{eq:rel-conv-b},
we get from the definition of Eq.~\eqref{eq:def-simperform} that
\begin{equation}
\mathsf{D}_{1+s}\left(\mathcal{W}^{\times n},c_{n}\right)\geq\min_{Q_{X^{n}Y^{n}}^{(n)}}D_{1+s}\left(P_{XY}^{\times n}\big{\|}Q_{X^{n}Y^{n}}^{(n)}\right),\label{eq:rel-conv-c}
\end{equation}
where the minimization is over all probability distributions $Q_{X^{n}Y^{n}}^{(n)}$
on $\mathcal{X}^{\times n}\times\mathcal{Y}^{\times n}$ such that
\begin{equation}
Q_{X^{n}Y^{n}}^{(n)}\leq(n+1)^{|\mathcal{Y}|}g(n)2^{nr}P_{X}^{\times n}\times\Phi_{Y^{n}}.
\end{equation}
For brevity, in the following we will use $Q_{X^{n}Y^{n}}^{(n)}$
to denote the minimizer in Eq.~\eqref{eq:rel-conv-c}. The rest of
the proof is divided into four cases.

\textit{Case 1:} $s\in(0,\infty)$ and $r\geq\hat{R}(s)$. We choose
a positive constant $a$ such that $a^{s}-(1+s)\geq1$, i.e., $a\geq(2+s)^{\frac{1}{s}}$.
Denote
\begin{align}
p_{n}= & P_{XY}^{\times n}\left(P_{XY}^{\times n}\geq a(n+1)^{|\mathcal{Y}|}g(n)2^{nr}P_{X}^{\times n}\times\Phi_{Y^{n}}\right),\\
q_{n}= & Q_{X^{n}Y^{n}}^{(n)}\left(P_{XY}^{\times n}\geq a(n+1)^{|\mathcal{Y}|}g(n)2^{nr}P_{X}^{\times n}\times\Phi_{Y^{n}}\right).
\end{align}
Then
\begin{equation}
p_{n}\geq aq_{n}.
\end{equation}
Due to the data processing inequality~(Lemma~\ref{lem:RenyiD-properties}~(\romannumeral3)),
it holds that
\begin{align}
\mathsf{D}_{1+s}\left(\mathcal{W}^{\times n},c_{n}\right) & \geq D_{1+s}\left(P_{XY}^{\times n}\big{\|}Q_{X^{n}Y^{n}}^{(n)}\right)\nonumber \\
 & \geq\dfrac{1}{s}\log\left\{ p_{n}^{1+s}q_{n}^{-s}+(1-p_{n})^{1+s}(1-q_{n})^{-s}\right\} \nonumber \\
 & \geq\dfrac{1}{s}\log\left\{ a^{s}p_{n}+(1-p_{n})^{1+s}\right\} \nonumber \\
 & \geq\dfrac{1}{s}\log\left\{ 1+(a^{s}-(1+s))p_{n}\right\} \nonumber \\
 & \geq\dfrac{1}{s}\log\left\{ 1+p_{n}\right\} \nonumber \\
 & \dot{=}p_{n},\label{eq:rel-conv-d}
\end{align}
where for the fourth inequality we have used $(1-x)^{y}\geq1-xy$
for $0\leq x\leq1\leq y$. Applying Proposition~\ref{prop:exponent-u-symm},
we get from Eq.~\eqref{eq:rel-conv-d} that
\begin{align}
\limsup_{n\rightarrow\infty}\frac{-1}{n}\log\mathsf{D}_{1+s}\left(\mathcal{W}^{\times n},c_{n}\right)\leq\sup_{t\geq0}t\left(r-I_{1+t}(X:Y)_{P_{XY}}\right).
\end{align}
Since $P_{X}$ is picked arbitrarily, we can minimize the right hand
side over all probability distributions $P_{X}$. Therefore,
\begin{align}
 & \limsup_{n\rightarrow\infty}\frac{-1}{n}\log\mathsf{D}_{1+s}\left(\mathcal{W}^{\times n},c_{n}\right)\nonumber \\
\leq & \min_{P_{X}}\sup_{t\geq0}t\left(r-I_{1+t}(X:Y)_{P_{X}\cdot\mathcal{W}_{Y|X}}\right)\nonumber \\
= & \sup_{t\geq0}\min_{P_{X}}t\left(r-I_{1+t}(X:Y)_{P_{X}\cdot\mathcal{W}_{Y|X}}\right)\nonumber \\
= & \sup_{t\geq0}t\left(r-I_{1+t}(\mathcal{W})\right)\nonumber \\
= & \left|\sup_{t\geq0}t\left(r-I_{1+t}(\mathcal{W})\right)\right|^{+}\nonumber \\
= & \left|\sup\limits_{t\geq s}t(r-I_{1+t}(\mathcal{W}))\right|^{+}.
\end{align}
In the above equation, the first equality is by Sion's minimax theorem~(Lemma~\ref{lem:minimax-Thm}
in Appendix~\ref{app:Miscellaneous-Lemmas}), and the last equality
comes from the condition $r\geq\hat{R}(s)$ and Lemma~\ref{lem:Renyi-maximization}~(\romannumeral1).
To see that Sion's minimax theorem applies here, we have (a) the function
$t\mapsto t(r-I_{1+t}(X:Y)_{P_{X}\cdot\mathcal{W}_{Y|X}})$ is continuous
and concave by Lemma~\ref{lem:RenyiC-properties}~(\romannumeral4),
and (b) the function $P_{X}\mapsto t(r-I_{1+t}(X:Y)_{P_{X}\cdot\mathcal{W}_{Y|X}})$
is continuous and convex by Lemma~\ref{lem:RenyiC-properties}~(\romannumeral2).

\textit{Case 2:} $s\in(0,\infty)$ and $0\leq r<\hat{R}(s)$. Denote
\begin{align}
p_{n}= & P_{XY}^{\times n}\left(P_{XY}^{\times n}\geq(n+1)^{|\mathcal{Y}|}g(n)2^{n\hat{R}(s)}P_{X}^{\times n}\times\Phi_{Y^{n}}\right),\\
q_{n}= & Q_{X^{n}Y^{n}}^{(n)}\left(P_{XY}^{\times n}\geq(n+1)^{|\mathcal{Y}|}g(n)2^{n\hat{R}(s)}P_{X}^{\times n}\times\Phi_{Y^{n}}\right).
\end{align}
Then
\begin{equation}
p_{n}\geq2^{n(\hat{R}(s)-r)}q_{n}.
\end{equation}
By the data processing inequality, it holds that
\begin{align}
 & \mathsf{D}_{1+s}\left(\mathcal{W}^{\times n},c_{n}\right)\nonumber \\
\geq & D_{1+s}\left(P_{XY}^{\times n}\big{\|}Q_{X^{n}Y^{n}}^{(n)}\right)\nonumber \\
\geq & \dfrac{1}{s}\log\left\{ p_{n}^{1+s}q_{n}^{-s}+(1-p_{n})^{1+s}(1-q_{n})^{-s}\right\} \nonumber \\
\geq & \dfrac{1}{s}\log\left\{ 2^{ns(\hat{R}(s)-r)}p_{n}+(1-p_{n})^{1+s}\right\} \nonumber \\
\geq & \dfrac{1}{s}\log\left\{ 1+\left(2^{ns(\hat{R}(s)-r)}-(1+s)\right)p_{n}\right\} .
\end{align}
This lets us derive that
\begin{align}
 & \limsup_{n\rightarrow\infty}\frac{-1}{n}\log\mathsf{D}_{1+s}\left(\mathcal{W}^{\times n},c_{n}\right)\nonumber \\
\leq & \left|\lim_{n\rightarrow\infty}\frac{-1}{n}\log\left\{ \left(2^{ns(\hat{R}(s)-r)}-(1+s)\right)p_{n}\right\} \right|^{+}\nonumber \\
= & \left|\sup_{t\geq0}t\left(\hat{R}(s)-I_{1+t}(X:Y)_{P_{XY}}\right)-s(\hat{R}(s)-r)\right|^{+},
\end{align}
where the equality is by Proposition~\ref{prop:exponent-u-symm}.
Similar to the first case, we can minimize the right hand side over
all probability distributions $P_{X}$ to obtain
\begin{align}
 & \limsup_{n\rightarrow\infty}\frac{-1}{n}\log\mathsf{D}_{1+s}\left(\mathcal{W}^{\times n},c_{n}\right)\nonumber \\
\leq & \min_{P_{X}}\left|\sup_{t\geq0}t\left(\hat{R}(s)-I_{1+t}(X:Y)_{P_{X}\cdot\mathcal{W}_{Y|X}}\right)-s(\hat{R}(s)-r)\right|^{+}\nonumber \\
= & \left|\min_{P_{X}}\sup_{t\geq0}t\left(\hat{R}(s)-I_{1+t}(X:Y)_{P_{X}\cdot\mathcal{W}_{Y|X}}\right)-s(\hat{R}(s)-r)\right|^{+}\nonumber \\
\stackrel{(a)}{=} & \left|\sup_{t\geq0}\min_{P_{X}}t\left(\hat{R}(s)-I_{1+t}(X:Y)_{P_{X}\cdot\mathcal{W}_{Y|X}}\right)-s(\hat{R}(s)-r)\right|^{+}\nonumber \\
= & \left|\sup_{t\geq0}t\left(\hat{R}(s)-I_{1+t}(\mathcal{W})\right)-s(\hat{R}(s)-r)\right|^{+}\nonumber \\
\stackrel{(b)}{=} & \left|s\left(r-I_{1+s}(\mathcal{W})\right)\right|^{+}\nonumber \\
\stackrel{(c)}{=} & \left|\sup_{t\geq s}t\left(r-I_{1+t}(\mathcal{W})\right)\right|^{+}.
\end{align}
In the above equation, $(a)$ is by Sion's minimax theorem, $(b)$
is because the maximum of the function $t\mapsto t(\hat{R}(s)-I_{1+t}(\mathcal{W}))$
is attained at $t=s$ (cf. Lemma~\ref{lem:Renyi-maximization}~(\romannumeral3)),
and $(c)$ comes from the condition $0\leq r<\hat{R}(s)$ and Lemma~\ref{lem:Renyi-maximization}~(\romannumeral2).

\textit{Case 3:} $s\in(-1,-\frac{1}{2})$. We pick a positive constant
$a=(s+1)^{\frac{-s}{2s+1}}$ which is larger than $1$. Denote
\begin{align}
p_{n}= & P_{XY}^{\times n}\left(P_{XY}^{\times n}\geq a(n+1)^{|\mathcal{Y}|}g(n)2^{nr}P_{X}^{\times n}\times\Phi_{Y^{n}}\right),\\
q_{n}= & Q_{X^{n}Y^{n}}^{(n)}\left(P_{XY}^{\times n}\geq a(n+1)^{|\mathcal{Y}|}g(n)2^{nr}P_{X}^{\times n}\times\Phi_{Y^{n}}\right).
\end{align}
Then
\begin{equation}
p_{n}\geq aq_{n}.
\end{equation}
We have
\begin{align}
 & \mathsf{D}_{1+s}\left(\mathcal{W}^{\times n},c_{n}\right)\nonumber \\
\geq & D_{1+s}\left(P_{XY}^{\times n}\big{\|}Q_{X^{n}Y^{n}}^{(n)}\right)\nonumber \\
\stackrel{(a)}{\geq} & \dfrac{1}{s}\log\left\{ p_{n}^{1+s}q_{n}^{-s}+(1-p_{n})^{1+s}(1-q_{n})^{-s}\right\} \nonumber \\
= & -\log\left\{ p_{n}^{1+s}q_{n}^{-1-s}q_{n}+(1-p_{n})^{1+s}(1-q_{n})^{-1-s}(1-q_{n})\right\} ^{-\frac{1}{s}}\nonumber \\
\stackrel{(b)}{\geq} & -\log\left\{ q_{n}p_{n}^{\frac{1+s}{-s}}q_{n}^{\frac{1+s}{s}}+(1-q_{n})(1-p_{n})^{\frac{1+s}{-s}}(1-q_{n})^{\frac{1+s}{s}}\right\} \nonumber \\
= & -\log\left\{ p_{n}^{\frac{1+s}{-s}}q_{n}^{\frac{1+2s}{s}}+(1-q_{n})^{\frac{1+2s}{s}}(1-p_{n})^{\frac{1+s}{-s}}\right\} \nonumber \\
\stackrel{(c)}{\geq} & -\log\left\{ p_{n}a^{\frac{2s+1}{-s}}+1-\frac{1+s}{-s}p_{n}\right\} \nonumber \\
= & -\log\left\{ 1-\left(\frac{1}{-s}-s-2\right)p_{n}\right\} \nonumber \\
\dot{=} & p_{n},
\end{align}
where $(a)$ is by the data processing inequality, $(b)$ comes from
Jensen's inequality, $(c)$ is by $p_{n}\geq aq_{n}$ and the inequality
$(1-x)^{y}\leq1-xy$ for $0<x,y<1$, and the last line is because
of $\frac{1}{-s}-s-2>0$. Similar to the first two cases, this lets
us obtain
\begin{align}
 & \limsup_{n\rightarrow\infty}\frac{-1}{n}\log\mathsf{D}_{1+s}\left(\mathcal{W}^{\times n},c_{n}\right)\nonumber \\
\leq & \min_{P_{X}}\sup_{t\geq0}t\left(r-I_{1+t}(X:Y)_{P_{X}\cdot\mathcal{W}_{Y|X}}\right)\nonumber \\
= & \sup_{t\geq0}\min_{P_{X}}t\left(r-I_{1+t}(X:Y)_{P_{X}\cdot\mathcal{W}_{Y|X}}\right)\nonumber \\
= & \sup_{t\geq0}t\left(r-I_{1+t}(\mathcal{W})\right).\label{eq:rel-conv-e}
\end{align}

\textit{Case 4:} $s\in[-\frac{1}{2},0]\cup\{\infty\}$. Due to the
monotonicity of R{é}nyi divergence of Lemma~\ref{lem:RenyiD-properties}~(\romannumeral1),
we have that Eq.~\eqref{eq:rel-conv-e} still holds for $s\in[-\frac{1}{2},0]$.
Using the monotonicity of R{é}nyi divergence again and the results
of Cases 1 and 2, the desired result follows for $s=\infty$.
\end{IEEEproof}

\section{Strong Converse Exponent}

\label{sec:strong-con-exponent}

This section is devoted to the proof of the strong converse exponent
of Theorem~\ref{thm:strong-con-exponent}. In Section~\ref{subsec:sce-variantionE},
we present a variational expression. In Section~\ref{subsec:sce-achi},
we prove the achievability part. In Section~\ref{subsec:sce-conv},
we prove the converse part.

\subsection{Variational Expression}

\label{subsec:sce-variantionE}

\begin{proposition}\label{prop:variational-expression} Let $\mathcal{W}:\mathcal{X}\rightarrow\mathcal{Y}$
be a discrete memoryless channel. Let $\alpha\in(0,1)$ and $r\geq0$.
We have
\begin{align}
 & \max_{\alpha\leq\beta\leq1}\frac{\alpha(1-\beta)}{\beta(1-\alpha)}(I_{\beta}(\mathcal{W})-r)\nonumber \\
= & \max_{P_{X}\in\mathcal{P}(\mathcal{X})}\min_{\widehat{\mathcal{W}}_{Y|X}}\left\{ \frac{\alpha}{1-\alpha}D\left(P_{X}\cdot\widehat{\mathcal{W}}_{Y|X}\big\| P_{X}\cdot\mathcal{W}_{Y|X}\right)+\big|I(X:Y)_{P_{X}\cdot\widehat{\mathcal{W}}_{Y|X}}-r\big|^{+}\right\} ,
\end{align}
where the minimization is over the set of all channels from $\mathcal{X}$
to $\mathcal{Y}$. \end{proposition}
\begin{IEEEproof}
Equation~\eqref{eq:Ichannel} provides that
\begin{equation}
I_{\beta}(\mathcal{W})=\max_{P_{X}\in\mathcal{P}(\mathcal{X})}\min_{Q_{Y}\in\mathcal{P}(\mathcal{Y})}\sum_{x\in\mathcal{X}}P_{X}(x)D_{\beta}(\mathcal{W}_{Y|X}(\cdot|x)\|Q_{Y}).\label{eq:sce-var-1}
\end{equation}
Using the variational expression of Lemma~\ref{lem:RenyiD-properties}~(\romannumeral4),
we can write
\begin{align}
 & \min_{Q_{Y}\in\mathcal{P}(\mathcal{Y})}\sum_{x\in\mathcal{X}}P_{X}(x)D_{\beta}\left(\mathcal{W}_{Y|X}(\cdot|x)\|Q_{Y}\right)\nonumber \\
= & \min_{Q_{Y}\in\mathcal{P}(\mathcal{Y})}\sum_{x\in\mathcal{X}}P_{X}(x)\min_{U_{Y}^{(x)}\in\mathcal{P}(\mathcal{Y})}\left\{ \frac{\beta}{1-\beta}D\left(U_{Y}^{(x)}\big\|\mathcal{W}_{Y|X}(\cdot|x)\right)+D\left(U_{Y}^{(x)}\big\| Q_{Y}\right)\right\} \nonumber \\
= & \min_{Q_{Y}\in\mathcal{P}(\mathcal{Y})}\min_{\widehat{\mathcal{W}}_{Y|X}}\left\{ \frac{\beta}{1-\beta}D\left(P_{X}\cdot\widehat{\mathcal{W}}_{Y|X}\big\| P_{X}\cdot\mathcal{W}_{Y|X}\right)+D\left(P_{X}\cdot\widehat{\mathcal{W}}_{Y|X}\big\| P_{X}\times Q_{Y}\right)\right\} \nonumber \\
= & \min_{\widehat{\mathcal{W}}_{Y|X}}\left\{ \frac{\beta}{1-\beta}D\left(P_{X}\cdot\widehat{\mathcal{W}}_{Y|X}\big\| P_{X}\cdot\mathcal{W}_{Y|X}\right)+\min_{Q_{Y}\in\mathcal{P}(\mathcal{Y})}D\left(P_{X}\cdot\widehat{\mathcal{W}}_{Y|X}\big\| P_{X}\times Q_{Y}\right)\right\} \nonumber \\
= & \min_{\widehat{\mathcal{W}}_{Y|X}}\left\{ \frac{\beta}{1-\beta}D\left(P_{X}\cdot\widehat{\mathcal{W}}_{Y|X}\big\| P_{X}\cdot\mathcal{W}_{Y|X}\right)+I(X:Y)_{P_{X}\cdot\widehat{\mathcal{W}}_{Y|X}}\right\} ,\label{eq:sce-var-2}
\end{align}
where in the second equality we have identified $\widehat{\mathcal{W}}_{Y|X}(\cdot|x)$
with $U_{Y}^{(x)}$. By Eqs.~\eqref{eq:sce-var-1} and \eqref{eq:sce-var-2},
we have
\begin{align}
 & \max_{\alpha\leq\beta\leq1}\frac{\alpha(1-\beta)}{\beta(1-\alpha)}\left(I_{\beta}(\mathcal{W})-r\right)\nonumber \\
= & \max_{\alpha\leq\beta\leq1}\max_{P_{X}\in\mathcal{P}(\mathcal{X})}\min_{\widehat{\mathcal{W}}_{Y|X}}\left\{ \frac{\alpha}{1-\alpha}D\left(P_{X}\cdot\widehat{\mathcal{W}}_{Y|X}\big\| P_{X}\cdot\mathcal{W}_{Y|X}\right)+\frac{\alpha(1-\beta)}{\beta(1-\alpha)}\left(I(X:Y)_{P_{X}\cdot\widehat{\mathcal{W}}_{Y|X}}-r\right)\right\} \nonumber \\
= & \max_{P_{X}\in\mathcal{P}(\mathcal{X})}\max_{0\leq\lambda\leq1}\min_{\widehat{\mathcal{W}}_{Y|X}}\left\{ \frac{\alpha}{1-\alpha}D\left(P_{X}\cdot\widehat{\mathcal{W}}_{Y|X}\big\| P_{X}\cdot\mathcal{W}_{Y|X}\right)+\lambda\left(I(X:Y)_{P_{X}\cdot\widehat{\mathcal{W}}_{Y|X}}-r\right)\right\} .
\end{align}
The function $\widehat{\mathcal{W}}_{Y|X}\mapsto D(P_{X}\cdot\widehat{\mathcal{W}}_{Y|X}\|P_{X}\cdot\mathcal{W}_{Y|X})$
is convex and lower semi-continuous. The function $\widehat{\mathcal{W}}_{Y|X}\mapsto I(X:Y)_{P_{X}\cdot\widehat{\mathcal{W}}_{Y|X}}$
is convex and continuous by Lemma~\ref{lem:RenyiC-properties}~(\romannumeral3).
Also, the set of all channels from $\mathcal{X}$ to $\mathcal{Y}$
is tight. So, we can employ Sion's minimax theorem to exchange the
minimization and the second maximization. This leads to
\begin{align}
 & \max_{\alpha\leq\beta\leq1}\frac{\alpha(1-\beta)}{\beta(1-\alpha)}\left(I_{\beta}(\mathcal{W})-r\right)\nonumber \\
= & \max_{P_{X}\in\mathcal{P}(\mathcal{X})}\min_{\widehat{\mathcal{W}}_{Y|X}}\max_{0\leq\lambda\leq1}\left\{ \frac{\alpha}{1-\alpha}D\left(P_{X}\cdot\widehat{\mathcal{W}}_{Y|X}\big\| P_{X}\cdot\mathcal{W}_{Y|X}\right)+\lambda\left(I(X:Y)_{P_{X}\cdot\widehat{\mathcal{W}}_{Y|X}}-r\right)\right\} \nonumber \\
= & \max_{P_{X}\in\mathcal{P}(\mathcal{X})}\min_{\widehat{\mathcal{W}}_{Y|X}}\left\{ \frac{\alpha}{1-\alpha}D\left(P_{X}\cdot\widehat{\mathcal{W}}_{Y|X}\big\| P_{X}\cdot\mathcal{W}_{Y|X}\right)+\big|I(X:Y)_{P_{X}\cdot\widehat{\mathcal{W}}_{Y|X}}-r\big|^{+}\right\} .
\end{align}
\end{IEEEproof}

\subsection{Achievability Bound}

\label{subsec:sce-achi}

In this section, we first prove the achievability part for any $\alpha\in(0,1)$
using the variational expression given in Section~\ref{subsec:sce-variantionE}.
Then, based on the result of the reliability function, we complete
the proof for any $\alpha\geq1$.

\begin{theorem}\label{thm:str-alpha-1} Let $\alpha\in(0,1)$. For
any discrete memoryless channel $\mathcal{W}:\mathcal{X}\rightarrow\mathcal{Y}$
and $r\geq0$, we have
\begin{equation}
E_{{\rm {sc}}}^{(\alpha)}(\mathcal{W},r)\leq\max_{\alpha\leq\beta\leq1}\frac{\alpha(1-\beta)}{\beta(1-\alpha)}(I_{\beta}(\mathcal{W})-r).
\end{equation}
\end{theorem}
\begin{IEEEproof}
By the definition of $E_{{\rm sc}}^{(\alpha)}(\mathcal{W},r)$ (cf.
Eq.~\eqref{eq:def-sce}), we see that it suffices to show that for
any $\delta>0$, there exists simulation $\mathcal{N}^{(n)}$ for
$\mathcal{W}^{\times n}$ with communication cost $c_{n}$ bits for
all $n\in\mathbb{N}$, such that
\begin{equation}
\limsup_{n\rightarrow\infty}\frac{1}{n}\max_{x^{n}}D_{\alpha}\left(\mathcal{W}_{Y|X}^{\times n}(\cdot|x^{n})\big{\|}\mathcal{N}_{Y^{n}|X^{n}}^{(n)}(\cdot|x^{n})\right)\leq\max_{\alpha\leq\beta\leq1}\frac{\alpha(1-\beta)}{\beta(1-\alpha)}(I_{\beta}(\mathcal{W})-r)+\delta\label{eq:fin-eq}
\end{equation}
and
\begin{equation}
\limsup_{n\to\infty}\frac{c_{n}}{n}\leq r.\label{eq:comrate}
\end{equation}

Now, we design a simulation channel for $\mathcal{W}^{\times n}$
using the rejection sampling procedure. Before the simulation starts,
let Alice and Bob share enough uniform randomness, say, $\{U_{j}\}_{j}$
that are independent and each uniformly distributed on the interval
$(0,1)$. In the simulation, Alice and Bob do the following for any
channel input $x^{n}\in\mathcal{X}^{\times n}$:
\begin{enumerate}
\item Upon receiving the input $x^{n}$, Alice tells Bob its type $T_{x^{n}}$
by communicating $\log|\mathcal{P}_{n}(\mathcal{X})|$ bits, enabling
Bob to determine the distribution of random variables that need to
be shared.
\item Alice and Bob compute
\begin{equation}
\widetilde{\mathcal{W}}=\underset{\widehat{\mathcal{W}}}{\mathrm{arg\,min}}\left\{ \!\frac{\alpha}{1-\alpha}D\left(T_{x^{n}}\cdot\widehat{\mathcal{W}}\big{\|}T_{x^{n}}\cdot\mathcal{W}\right)\!+\!\left|I(X:Y)_{T_{x^{n}}\cdot\widehat{\mathcal{W}}}-r\right|^{+}\!\right\} \label{eq:sc-optch}
\end{equation}
and $Q_{Y}=\widetilde{\mathcal{W}}(T_{x^{n}})=\sum_{x}T_{x^{n}}(x)\widetilde{\mathcal{W}}_{Y|X}(\cdot|x)$.
Alice sets $N=\exp\{n(I(X:Y)_{T_{x^{n}}\cdot\widetilde{\mathcal{W}}_{Y|X}}+\delta)\}$.
\item Alice and Bob, using the pre-shared uniform randomness $\{U_{j}\}_{j}$,
generate shared independent random variables $Y_{j}^{(n)}\sim Q_{Y}^{\times n}$,
$j=0,1,\cdots,\widetilde{N}$ with $\widetilde{N}=2^{nr+1}$. (This
can be done by employing the inverse cumulative distribution function
sampling.)
\item Alice uses the rejection sampling procedure of Algorithm~\ref{alg:rejsamp}
with parameters $(\widetilde{\mathcal{W}}_{Y|X}^{\times n}(\cdot|x^{n}),Q_{Y}^{\times n},N,\widetilde{N})$
to generate a random index $J\in\{0,1,\cdots,\widetilde{N}\}$
and sends $J$ to Bob with $\log(\widetilde{N}+1)$ bits.
\item Bob picks $Y_{J}^{(n)}$ as the channel output.
\end{enumerate}
We point out that the minimizer in Eq.~\eqref{eq:sc-optch} is unique.
This is because $\widehat{\mathcal{W}}\mapsto D(T_{x^{n}}\cdot\widehat{\mathcal{W}}\|T_{x^{n}}\cdot\mathcal{W})$
is strictly convex and $\widehat{\mathcal{W}}\mapsto|I(X:Y)_{T_{x^{n}}\cdot\widehat{\mathcal{W}}}-r|^{+}$
is convex. Denote the above simulation by $\mathcal{N}^{(n)}$. Then
according to Eq.~\eqref{eq:rej-sam} we have
\begin{equation}
\mathcal{N}_{Y^{n}|X^{n}}^{(n)}(\cdot|x^{n})=(1-p_{n})\frac{\overline{\mathcal{W}}_{Y^{n}|X^{n}}(\cdot|x^{n})}{\sum_{y^{n}}\overline{\mathcal{W}}_{Y^{n}|X^{n}}(y^{n}|x^{n})}+p_{n}Q_{Y}^{\times n},\label{eq:simch-sce}
\end{equation}
where
\begin{equation}
\overline{\mathcal{W}}_{Y^{n}|X^{n}}(\cdot|x^{n}):=\min\{\widetilde{\mathcal{W}}_{Y|X}^{\times n}(\cdot|x^{n}),NQ_{Y}^{\times n}\}\label{eq:conch-1}
\end{equation}
and
\begin{equation}
p_{n}:=(1-\frac{1}{N}\sum_{y^{n}}\overline{\mathcal{W}}_{Y^{n}|X^{n}}(y^{n}|x^{n}))^{\widetilde{N}}.\label{eq:conch-2}
\end{equation}

The communication rate is
\begin{equation}
\lim_{n\to\infty}\frac{1}{n}\left(\log|\mathcal{P}_{n}(\mathcal{X})|+\log(1+\widetilde{N})\right)=r,
\end{equation}
which confirms Eq.~\eqref{eq:comrate}. To estimate the performance
of the simulation, we need upper bound
\begin{equation}
\limsup_{n\to\infty}\max_{x^{n}}\Gamma_{n}(x^{n}),
\end{equation}
where
\begin{equation}
\Gamma_{n}(x^{n}):=\frac{1}{n}D_{\alpha}\left(\mathcal{W}_{Y|X}^{\times n}(\cdot|x^{n})\big{\|}\mathcal{N}_{Y^{n}|X^{n}}^{(n)}(\cdot|x^{n})\right).
\end{equation}
For each $n$, let $\tilde{x}^{n}$ be a maximizer for the maximization
above. Then,
\begin{equation}
\limsup_{n\to\infty}\max_{x^{n}}\Gamma_{n}(x^{n})=\limsup_{n\to\infty}\Gamma_{n}(\tilde{x}^{n}).\label{eq:}
\end{equation}
By the definition of limsup, there is an increasing sequence $\{n_{k}\}_{k\in\mathbb{N}}$
such that
\begin{equation}
\limsup_{n\to\infty}\Gamma_{n}(\tilde{x}^{n})=\lim_{k\to\infty}\Gamma_{n_{k}}(\tilde{x}^{n_{k}}).\label{eq:-2}
\end{equation}
Since the probability simplex is compact, there is a subsequence $\{m_{k}\}_{k\in\mathbb{N}}$
of $\{n_{k}\}_{k\in\mathbb{N}}$ such that $T_{\tilde{x}^{m_{k}}}\cdot\widetilde{\mathcal{W}}_{Y|X}$
converges to some distribution $S_{X}\cdot\widetilde{\mathcal{W}}_{Y|X}^{*}$,
where $\widetilde{\mathcal{W}}_{Y|X}$ denotes the associated channel
for $\tilde{x}^{m_{k}}$ defined in Eq.~\eqref{eq:sc-optch}. For
such a subsequence, it obviously holds that
\begin{equation}
\lim_{k\to\infty}\Gamma_{n_{k}}(\tilde{x}^{n_{k}})=\lim_{k\to\infty}\Gamma_{m_{k}}(\tilde{x}^{m_{k}}).\label{eq:-3}
\end{equation}
Combining Eqs.~\eqref{eq:}-\eqref{eq:-3}, we only need to bound
$\lim_{k\to\infty}\Gamma_{m_{k}}(\tilde{x}^{m_{k}})$.

Let
\begin{equation}
\theta(Q_{X},\widehat{\mathcal{W}}):=\frac{\alpha}{1-\alpha}D\left(Q_{X}\cdot\widehat{\mathcal{W}}\big{\|}Q_{X}\cdot\mathcal{W}\right)\!+\!\left|I(X:Y)_{Q_{X}\cdot\widehat{\mathcal{W}}}-r\right|^{+}.
\end{equation}
Observe that
\begin{align}
\Gamma_{m_{k}}(\tilde{x}^{m_{k}})\stackrel{(a)}{\leq} & \frac{\alpha}{m_{k}(1-\alpha)}D\left(\widetilde{\mathcal{W}}_{Y|X}^{\times m_{k}}(\cdot|\tilde{x}^{m_{k}})\big{\|}\mathcal{W}_{Y|X}^{\times m_{k}}(\cdot|\tilde{x}^{m_{k}})\right)+\frac{1}{m_{k}}D\left(\widetilde{\mathcal{W}}_{Y|X}^{\times m_{k}}(\cdot|\tilde{x}^{m_{k}})\big{\|}\mathcal{N}_{Y^{m_{k}}|X^{m_{k}}}^{(m_{k})}(\cdot|\tilde{x}^{m_{k}})\right)\nonumber \\
= & \frac{\alpha}{1-\alpha}D\left(T_{\tilde{x}^{m_{k}}}\cdot\widetilde{\mathcal{W}}_{Y|X}\big{\|}T_{\tilde{x}^{m_{k}}}\cdot\mathcal{W}_{Y|X}\right)+\frac{1}{m_{k}}D\left(\widetilde{\mathcal{W}}_{Y|X}^{\times m_{k}}(\cdot|\tilde{x}^{m_{k}})\big{\|}\mathcal{N}_{Y^{m_{k}}|X^{m_{k}}}^{(m_{k})}(\cdot|\tilde{x}^{m_{k}})\right)\nonumber \\
\stackrel{(b)}{\leq} & \theta(S_{X},\widetilde{\mathcal{W}}_{Y|X}^{*})+\delta+o(1),\label{eq:-4}
\end{align}
where $(a)$ is by Lemma~\ref{lem:RenyiD-properties}~(\romannumeral4),
and $(b)$ is shown in Lemma~\ref{lem:infty-upper}
given below. To proceed, we now show that
\begin{equation}
\widetilde{\mathcal{W}}_{Y|X}^{*}=\underset{\widehat{\mathcal{W}}}{\mathrm{arg\,min}}\,\theta(S_{X},\widehat{\mathcal{W}}).\label{eq:-5}
\end{equation}
It is easy to check that the function $(Q_{X},\widehat{\mathcal{W}})\mapsto\theta(Q_{X},\widehat{\mathcal{W}})$
is continuous on $\mathcal{P}(\mathcal{X})\times\Omega$, where $\Omega$
is the set of all channels $\widehat{\mathcal{W}}_{Y|X}$ such that
$\supp(\widehat{\mathcal{W}}(\cdot|x))\subseteq\supp(\mathcal{W}(\cdot|x))$
for all $x$. Fix an arbitrary channel $\mathcal{V}_{Y|X}\in\Omega$.
we have
\begin{equation}
\theta(S_{X},\widetilde{\mathcal{W}}_{Y|X}^{*})=\lim_{k\to\infty}\theta(T_{\tilde{x}^{m_{k}}},\widetilde{\mathcal{W}}_{Y|X})\le\lim_{k\to\infty}\theta(T_{\tilde{x}^{m_{k}}},\mathcal{V}_{Y|X})=\theta(S_{X},\mathcal{V}_{Y|X}),
\end{equation}
where the inequality follows since $\widetilde{\mathcal{W}}_{Y|X}$
is a minimizer of $\widehat{\mathcal{W}}\mapsto\theta(T_{\tilde{x}^{m_{k}}},\widehat{\mathcal{W}})$,
and the equalities follow by continuity. This confirms Eq.~\eqref{eq:-5},
which lets us get from Eq.~\eqref{eq:-4} that
\begin{align}
\lim_{k\to\infty}\Gamma_{m_{k}}(\tilde{x}^{m_{k}})\le & \theta(S_{X},\widetilde{\mathcal{W}}_{Y|X}^{*})+\delta\nonumber \\
\leq & \max_{\alpha\leq\beta\leq1}\frac{\alpha(1-\beta)}{\beta(1-\alpha)}(I_{\beta}(\mathcal{W})-r)+\delta,\label{eq:conkey}
\end{align}
where the second inequality follows from Proposition~\ref{prop:variational-expression}.
This leads to Eq.~\eqref{eq:fin-eq} and we are done.
\end{IEEEproof}
\begin{lemma}\label{lem:infty-upper} Let $\{m_{k}\}_{k\in\mathbb{N}}$
be a sequence of increasing natural numbers and $\{\tilde{x}^{m_{k}}\}_{k\in\mathbb{N}}$
be a sequence such that $T_{\tilde{x}^{m_{k}}}\cdot\widetilde{\mathcal{W}}_{Y|X}$
converges to some distribution $S_{X}\cdot\widetilde{\mathcal{W}}_{Y|X}^{*}$,
where $\widetilde{\mathcal{W}}_{Y|X}$ denotes the associated channel
for $\tilde{x}^{m_{k}}$ defined in Eq.~\eqref{eq:sc-optch}. Let
$\mathcal{N}_{Y^{m_{k}}|X^{m_{k}}}^{(m_{k})}$ be the simulation channel
given in Eq.~\eqref{eq:simch-sce} (with $n$ set to $m_{k}$). Then,
we have
\begin{equation}
\lim_{k\to\infty}\frac{1}{m_{k}}D\left(\widetilde{\mathcal{W}}_{Y|X}^{\times m_{k}}(\cdot|\tilde{x}^{m_{k}})\big{\|}\mathcal{N}_{Y^{m_{k}}|X^{m_{k}}}^{(m_{k})}(\cdot|\tilde{x}^{m_{k}})\right)\leq\big|I(X:Y)_{S_{X}\cdot\widetilde{\mathcal{W}}_{Y|X}^{*}}-r\big|^{+}+\delta.\label{eq:desire-result}
\end{equation}
\end{lemma}
\begin{IEEEproof}
We have
\begin{align}
 & \lim_{k\to\infty}\frac{1}{m_{k}}D\left(\widetilde{\mathcal{W}}_{Y|X}^{\times m_{k}}(\cdot|\tilde{x}^{m_{k}})\big{\|}\mathcal{N}_{Y^{m_{k}}|X^{m_{k}}}^{(m_{k})}(\cdot|\tilde{x}^{m_{k}})\right)\nonumber \\
\leq & \lim_{k\to\infty}\frac{1}{m_{k}}D\left(\widetilde{\mathcal{W}}_{Y|X}^{\times m_{k}}(\cdot|\tilde{x}^{m_{k}})\big{\|}\overline{\mathcal{W}}_{Y^{m_{k}}|X^{m_{k}}}(\cdot|\tilde{x}^{m_{k}})\right)-\lim_{k\to\infty}\frac{1}{m_{k}}\log(1-p_{m_{k}}),\label{Eq:sc-eq1}
\end{align}
where $\overline{\mathcal{W}}_{Y^{m_{k}}|X^{m_{k}}}(\cdot|\tilde{x}^{m_{k}})$
and $p_{m_{k}}$ are given in Eqs.~\eqref{eq:conch-1} and \eqref{eq:conch-2},
respectively. First, we bound the first term in Eq.~\eqref{Eq:sc-eq1}.
Let $s_{k}=m_{k}^{-\frac{1}{2}}$. From the monotonicity of R{é}nyi
divergence and Eq.~\eqref{eq:re-sim1}, it follows that
\begin{align}
 & D\left(\widetilde{\mathcal{W}}_{Y|X}^{\times m_{k}}(\cdot|\tilde{x}^{m_{k}})\big{\|}\overline{\mathcal{W}}_{Y^{m_{k}}|X^{m_{k}}}(\cdot|\tilde{x}^{m_{k}})\right)\nonumber \\
\leq & D_{1+s_{k}}\left(\widetilde{\mathcal{W}}_{Y|X}^{\times m_{k}}(\cdot|\tilde{x}^{m_{k}})\big{\|}\overline{\mathcal{W}}_{Y^{m_{k}}|X^{m_{k}}}(\cdot|\tilde{x}^{m_{k}})\right)\nonumber \\
\leq & \frac{1}{s_{k}\ln2}\exp\Big\{-s_{k}m_{k}\Big(I(X:Y)_{T_{\tilde{x}^{m_{k}}}\cdot\widetilde{\mathcal{W}}}+\delta-\sum_{x}T_{\tilde{x}^{m_{k}}}(x)D_{1+s_{k}}\left(\widetilde{\mathcal{W}}_{Y|X}(\cdot|x)\big{\|}\widetilde{\mathcal{W}}(T_{\tilde{x}^{m_{k}}})\right)\Big)\Big\}.
\end{align}
Since $s_{k}\searrow0$ and $(T_{\tilde{x}^{m_{k}}},\widetilde{\mathcal{W}}_{Y|X})\to(S_{X},\widetilde{\mathcal{W}}_{Y|X}^{*})$,
by continuity we have
\begin{align}
 & \lim_{k\to\infty}\sum_{x}T_{\tilde{x}^{m_{k}}}(x)D_{1+s_{k}}\left(\widetilde{\mathcal{W}}_{Y|X}(\cdot|x)\big{\|}\widetilde{\mathcal{W}}(T_{\tilde{x}^{m_{k}}})\right)\nonumber \\
= & \sum_{x}S_{X}(x)D(\widetilde{\mathcal{W}}_{Y|X}^{*}(\cdot|x)\|\widetilde{\mathcal{W}}_{Y|X}^{*}(S_{X}))=I(X:Y)_{S_{X}\cdot\widetilde{\mathcal{W}}^{*}}\label{eq:limita}
\end{align}
and $I(X:Y)_{T_{\tilde{x}^{m_{k}}}\cdot\widetilde{\mathcal{W}}}\to I(X:Y)_{S_{X}\cdot\widetilde{\mathcal{W}}^{*}}$.
The limit in Eq.~\eqref{eq:limita} is not obvious and is confirmed
in Lemma~\ref{lem:continuity}. Hence, for sufficiently large $k$,
we have
\begin{equation}
\sum_{x}T_{\tilde{x}^{m_{k}}}(x)D_{1+s_{k}}\left(\widetilde{\mathcal{W}}_{Y|X}(\cdot|x)\big{\|}\widetilde{\mathcal{W}}(T_{\tilde{x}^{m_{k}}})\right)-I(X:Y)_{T_{\tilde{x}^{m_{k}}}\cdot\widetilde{\mathcal{W}}}\leq\frac{\delta}{2}.\label{eq:sc-for-next}
\end{equation}
Therefore, it holds that
\begin{equation}
\lim_{k\to\infty}\frac{1}{m_{k}}D\left(\widetilde{\mathcal{W}}_{Y|X}^{\times m_{k}}(\cdot|\tilde{x}^{m_{k}})\big{\|}\overline{\mathcal{W}}_{Y^{m_{k}}|X^{m_{k}}}(\cdot|\tilde{x}^{m_{k}})\right)\leq\lim_{k\to\infty}\frac{m_{k}^{-\frac{1}{2}}}{\ln2}\exp\Big(-\frac{\delta}{2}m_{k}^{\frac{1}{2}}\Big)=0.\label{Eq:sc-eq2}
\end{equation}
Next, we estimate the second term in Eq.~\eqref{Eq:sc-eq1}. For
sufficiently large $k$, from Eqs.~\eqref{eq:for-strong-converse}
and \eqref{eq:sc-for-next}, we get
\begin{align}
 & \sum\limits_{y^{m_{k}}}\overline{\mathcal{W}}_{Y^{m_{k}}|X^{m_{k}}}(y^{m_{k}}|\tilde{x}^{m_{k}})\nonumber \\
\geq & 1-\exp\Big\{-s_{k}m_{k}\Big(I(X:Y)_{T_{\tilde{x}^{m_{k}}}\cdot\widetilde{\mathcal{W}}}+\delta-\sum_{x}T_{\tilde{x}^{m_{k}}}(x)D_{1+s_{k}}\left(\widetilde{\mathcal{W}}_{Y|X}(\cdot|x)\big{\|}\widetilde{\mathcal{W}}(T_{\tilde{x}^{m_{k}}})\right)\Big)\Big\}\nonumber \\
\geq & 1-\exp\Big(-\frac{\delta}{2}m_{k}^{\frac{1}{2}}\Big)\geq\frac{1}{2}.
\end{align}
Because $(1-\frac{1}{a})^{a}\nearrow\frac{1}{e}$ as $a\to\infty$,
for sufficiently large $k$, we obtain
\begin{align}
p_{m_{k}} & =\Big(1-\frac{1}{N}\sum\limits_{y^{m_{k}}}\overline{\mathcal{W}}_{Y^{m_{k}}|X^{m_{k}}}(y^{m_{k}}|\tilde{x}^{m_{k}})\Big)^{\widetilde{N}}\nonumber \\
 & \leq\Big(1-\frac{1}{2N}\Big)^{2N\frac{\widetilde{N}}{2N}}\nonumber \\
 & \leq e^{-\exp\left\{ -m_{k}\left(I(X:Y)_{T_{\tilde{x}^{m_{k}}}\cdot\widetilde{\mathcal{W}}}-r+\delta\right)\right\} }\nonumber \\
 & \leq e^{-\exp\left\{ -m_{k}\left(|I(X:Y)_{T_{\tilde{x}^{m_{k}}}\cdot\widetilde{\mathcal{W}}}-r|^{+}+\delta\right)\right\} }\nonumber \\
 & \leq1-\frac{1}{2}\exp\left\{ -m_{k}\left(|I(X:Y)_{T_{\tilde{x}^{m_{k}}}\cdot\widetilde{\mathcal{W}}}-r|^{+}+\delta\right)\right\} ,
\end{align}
where the last inequality follows from the relation $e^{-a}\leq1-\frac{a}{2}$
for $a\in[0,1]$. By the continuity of the function $(T_{\tilde{x}^{m_{k}}},\widetilde{\mathcal{W}})\mapsto|I(X:Y)_{T_{\tilde{x}^{m_{k}}}\cdot\widetilde{\mathcal{W}}}-r|^{+}$,
it follows that
\begin{align}
\lim_{k\to\infty}\frac{-1}{m_{k}}\log(1-p_{m_{k}})\leq|I(X:Y)_{S_{X}\cdot\widetilde{\mathcal{W}}^{*}}-r|^{+}+\delta.\label{eq:sc-final-2}
\end{align}
Combining Eqs.~\eqref{Eq:sc-eq1}, \eqref{Eq:sc-eq2} and \eqref{eq:sc-final-2},
we obtain the desired result.
\end{IEEEproof}

Next, we establish the achievability bound of the strong converse
exponent for $\alpha\geq1$.

\begin{theorem} Let $\alpha\in[1,\infty]$.
	For any discrete memoryless channel $\mathcal{W}:\mathcal{X}\rightarrow\mathcal{Y}$
	and $r\geq0$, we have
	\begin{equation}
		E_{{\rm {sc}}}^{(\alpha)}(\mathcal{W},r)\leq|I_{\alpha}(\mathcal{W})-r|^{+}.\label{eq:infty-upper}
	\end{equation}
\end{theorem}
\begin{IEEEproof}
	Let $\delta>0$ and $r':=I_{\alpha}(\mathcal{W})+\delta$. According
	to the result of the reliability function (cf. Theorem~\ref{thm:reliability-function}),
	we have
	\begin{equation}
		E_{{\rm rf}}^{(\alpha)}(\mathcal{W},r')>0.
	\end{equation}
	Therefore, there exists simulation $\widetilde{\mathcal{N}}^{(n)}$
	for $\mathcal{W}^{\times n}$ with communication cost $c_{n}$ bits
	for all $n\in\mathbb{N}$, such that
	\begin{equation}
		\limsup_{n\rightarrow\infty}\frac{1}{n}\max_{x^{n}}D_{\alpha}\left(\mathcal{W}_{Y|X}^{\times n}(\cdot|x^{n})\big{\|}\widetilde{\mathcal{N}}_{Y^{n}|X^{n}}^{(n)}(\cdot|x^{n})\right)=0\label{eq:sc-geq1}
	\end{equation}
	and
	\begin{equation}
		\limsup_{n\to\infty}\frac{c_{n}}{n}\leq r'.\label{eq:communication-cost}
	\end{equation}
	This immediately implies the desired result for the case $r>I_{\alpha}(\mathcal{W})$.
	So we only need to prove the other case $r\in[0,I_{\alpha}(\mathcal{W})]$. Let $n'=\lfloor\frac{nr}{r'}\rfloor$, and let $Q_{Y}$ be the minimizer in the optimization \begin{equation}
		\min_{Q_{Y}}\max_{x}D_{\alpha}(\mathcal{W}_{Y|X}(\cdot|x)\|Q_{Y}).
	\end{equation}
	We consider a simulation scheme for $\mathcal{W}^{\times n}$ given by
	\begin{equation}
		\mathcal{N}_{Y^{n}|X^{n}}^{(n)}(\cdot|x^{n})
		=\widetilde{\mathcal{N}}_{Y^{n'}|X^{n'}}^{(n')}(\cdot|x^{n'})\times Q_{Y}^{\times(n-n')}.
	\end{equation}
	That is, We employ $\widetilde{\mathcal{N}}^{(n')}$ to simulate the first $n'$ copies of
	the channel $\mathcal{W}$, and always set the output distribution to be $Q_{Y}^{\times(n-n')}$ for the remaining $n-n'$ copies. Because communication only occurs when we simulate the first $n'$ copies, the total communication cost is $c_{n'}$
	bits. From Eq.~\eqref{eq:communication-cost}, it follows that the communication rate is bounded by
	\begin{equation}
		\limsup _{n\to\infty}\frac{c_{n'}}{n}
		=\limsup _{n\to\infty}\frac{c_{n'}}{n'}\lim_{n\to\infty}\frac{n'}{n}\leq r.
	\end{equation}
	Moreover,
	\begin{align}
		& \frac{1}{n}\max_{x^{n}}D_{\alpha}\left(\mathcal{W}_{Y|X}^{\times n}(\cdot|x^{n})\big{\|}\mathcal{N}_{Y^{n}|X^{n}}^{(n)}(\cdot|x^{n})\right)\nonumber \\
		= & \frac{1}{n}\max_{x^{n}}\left\{D_{\alpha}\left(\mathcal{W}_{Y|X}^{\times n'}(\cdot|x^{n'})\big{\|}\widetilde{\mathcal{N}}_{Y^{n'}|X^{n'}}^{(n')}(\cdot|x^{n'})\right)+\sum_{k=n'+1}^{n}D_{\alpha}\left(\mathcal{W}_{Y|X}(\cdot|x_{k})\big{\|}Q_{Y}\right)\right\}\nonumber \\
		\leq & \frac{1}{n}\max_{x^{n'}}D_{\alpha}\left(\mathcal{W}_{Y|X}^{\times n'}(\cdot|x^{n'})\big{\|}\widetilde{\mathcal{N}}_{Y^{n'}|X^{n'}}^{(n')}(\cdot|x^{n'})\right)+\frac{n-n'}{n}I_{\alpha}(\mathcal{W}),\label{eq:sc-next}
	\end{align}
	where the equality come from the additivity of R{é}nyi divergence
	(cf. Lemma~\ref{lem:RenyiD-properties}~(\romannumeral2)) and the
	inequality follows from the definition of order-$\alpha$ R{é}nyi
	capacity (cf. Eq.~\eqref{eq:Ichannel}). By Eq.~\eqref{eq:sc-geq1},
	we obtain
	\begin{equation}
		\limsup_{n\rightarrow\infty}\frac{1}{n}\max_{x^{n}}D_{\alpha}\left(\mathcal{W}_{Y|X}^{\times n}(\cdot|x^{n})\big{\|}\mathcal{N}_{Y^{n}|X^{n}}^{(n)}(\cdot|x^{n})\right)\leq I_{\alpha}(\mathcal{W})-\frac{I_{\alpha}(\mathcal{W})}{I_{\alpha}(\mathcal{W})+\delta}r.
	\end{equation}
	Since $\delta>0$ is arbitrary, the desired result holds for $r\in[0,I_{\alpha}(\mathcal{W})]$. This concludes the proof of the theorem.
\end{IEEEproof}

\subsection{Converse Bound}
\label{subsec:sce-conv}

At first, we prove a one-shot version of the converse bound. Then,
we apply it directly to deal with the asymptotic situation.

\begin{theorem}\label{them:one-shot-con-bound}
For any discrete memoryless channel $\mathcal{W}:\mathcal{X}\rightarrow\mathcal{Y}$
and $c\geq0$, we have
\begin{equation}
\mathsf{D}_{\alpha}(\mathcal{W},c)\geq\left\{ \begin{array}{ll}
\max\limits_{\alpha\leq\beta\leq1}\frac{\alpha(1-\beta)}{\beta(1-\alpha)}\left(I_{\beta}(\mathcal{W})-c\right), & \!\!\alpha\in(0,1)\\
|I_{\alpha}(\mathcal{W})-c|^{+}, & \!\!\alpha\in[1,\infty].
\end{array}\right.
\end{equation}
\end{theorem}
\begin{IEEEproof}
Let $\mathcal{N}:\mathcal{X}\rightarrow\mathcal{Y}$ be a simulation
for $\mathcal{W}:\mathcal{X}\rightarrow\mathcal{Y}$ with communication
cost no more than $c$ bits. For any input distribution $P_{X}\in\mathcal{P}(\mathcal{X})$,
we write $P_{XY}:=P_{X}\cdot\mathcal{W}_{Y|X}$ and $\widetilde{P}_{XY}:=P_{X}\cdot\mathcal{N}_{Y|X}$.
By Lemma~\ref{lem:u-bound}, there exists $Q_{Y}\in\mathcal{P}(\mathcal{Y})$
such that $\widetilde{P}_{XY}\leq2^{c}P_{X}\times Q_{Y}$.

For $\alpha\in(0,1)$, we have
\begin{align}
 & (\alpha-1)D_{\alpha}\left(P_{X}\cdot\mathcal{W}_{Y|X}\|P_{X}\cdot\mathcal{N}_{Y|X}\right)\nonumber \\
= & \log\sum_{x,y}P_{XY}^{\alpha}(x,y)\widetilde{P}_{XY}^{1-\alpha}(x,y)\nonumber \\
\stackrel{(a)}{=} & \log\sum_{x,y}P_{XY}^{\alpha}(x,y)(P_{X}(x)Q_{Y}(y))^{\lambda}(P_{X}(x)Q_{Y}(y))^{-\lambda}\widetilde{P}_{XY}^{1-\alpha}(x,y)\nonumber \\
\stackrel{(b)}{\leq} & \log\big(\sum_{x,y}P_{XY}^{\beta}(x,y)(P_{X}(x)Q_{Y}(y))^{\frac{\beta}{\alpha}\lambda}\big)^{\frac{\alpha}{\beta}}+\log\big(\sum_{x,y}\widetilde{P}_{XY}^{(1-\alpha)\frac{\gamma}{\alpha}}(x,y)(P_{X}(x)Q_{Y}(y))^{-\frac{\gamma}{\alpha}\lambda}\big)^{\frac{\alpha}{\gamma}}\nonumber \\
\stackrel{(c)}{=} & \frac{\alpha}{\beta}\log\sum_{x,y}P_{XY}^{\beta}(x,y)(P_{X}(x)Q_{Y}(y))^{1-\beta}+\frac{\beta-\alpha}{\beta}\log\sum_{x,y}\widetilde{P}_{XY}^{\frac{(1-\alpha)\beta}{\beta-\alpha}}(x,y)(P_{X}(x)Q_{Y}(y))^{\frac{-\alpha(1-\beta)}{\beta-\alpha}}\nonumber \\
\stackrel{(d)}{\leq} & \frac{\alpha(\beta-1)}{\beta}D_{\beta}(P_{XY}\|P_{X}\times Q_{Y})+\frac{\beta-\alpha}{\beta}\log\sum_{x,y}\widetilde{P}_{XY}^{\frac{(1-\alpha)\beta}{\beta-\alpha}}(x,y)(2^{-c}\widetilde{P}_{XY}(x,y))^{\frac{-\alpha(1-\beta)}{\beta-\alpha}}\nonumber \\
\leq & \frac{\alpha(\beta-1)}{\beta}\min_{Q_{Y}\in\mathcal{P}(\mathcal{Y})}D_{\beta}(P_{XY}\|P_{X}\times Q_{Y})-\frac{\alpha(\beta-1)}{\beta}c\nonumber \\
= & \frac{\alpha(\beta-1)}{\beta}\left(I_{\beta}(X:Y)_{P_{XY}}-c\right).
\end{align}
In the above derivation, $(a)$ holds for any $\lambda\in\mathbb{R}$,
$(b)$ is by the H{ö}lder inequality with $\frac{\alpha}{\beta}+\frac{\alpha}{\gamma}=1$
and $\beta\in[\alpha,1]$, $(c)$ is by setting $\lambda=\frac{\alpha(1-\beta)}{\beta}$,
and $(d)$ follows from the inequality $\widetilde{P}_{XY}\leq2^{c}P_{X}\times Q_{Y}$.
Therefore, for any $\beta\in[\alpha,1]$, we obtain
\begin{align}
D_{\alpha}(\mathcal{W},\mathcal{N}) & =\max_{P_{X}\in\mathcal{P}(\mathcal{X})}D_{\alpha}\left(P_{X}\cdot\mathcal{W}_{Y|X}\|P_{X}\cdot\mathcal{N}_{Y|X}\right)\nonumber \\
 & \geq\frac{\alpha(1-\beta)}{\beta(1-\alpha)}\left(I_{\beta}(\mathcal{W})-c\right).\label{eq:con-con-ltone}
\end{align}
Minimizing the left hand side over all $\mathcal{N}\in\mathcal{A}(\mathcal{W},c)$
and maximizing the right hand side over all $\beta\in[\alpha,1]$,
we get from Eq.~\eqref{eq:con-con-ltone} that
\begin{align}
\mathsf{D}_{\alpha}(\mathcal{W},c)\geq\max_{\alpha\leq\beta\leq1}\frac{\alpha(1-\beta)}{\beta(1-\alpha)}\left(I_{\beta}(\mathcal{W})-c\right).
\end{align}

Now, we consider the case $\alpha\in[1,\infty]$. At first, for $\alpha\in(1,\infty)$,
we have
\begin{align}
D_{\alpha}(P_{XY}\|\widetilde{P}_{XY}) & =\frac{1}{\alpha-1}\log\sum_{x,y}P_{XY}^{\alpha}(x,y)\widetilde{P}_{XY}^{1-\alpha}(x,y)\nonumber \\
 & \geq\frac{1}{\alpha-1}\log\sum_{x,y}P_{XY}^{\alpha}(x,y)(2^{c}P_{X}(x)Q_{Y}(y))^{1-\alpha}\nonumber \\
 & =D_{\alpha}(P_{XY}\|P_{X}\times Q_{Y})-c\nonumber \\
 & \geq I_{\alpha}(X:Y)_{P_{XY}}-c,
\end{align}
where the first inequality follows from the relation $\widetilde{P}_{XY}\leq2^{c}P_{X}\times Q_{Y}$.
Similar to the situation $\alpha\in(0,1)$ discussed above, this leads
to
\begin{equation}
\mathsf{D}_{\alpha}(\mathcal{W},c)\geq I_{\alpha}(\mathcal{W})-c.
\end{equation}
By definition, the quantity $\mathsf{D}_{\alpha}(\mathcal{W},c)$
is nonnegative. So, we actually have
\begin{equation}
\mathsf{D}_{\alpha}(\mathcal{W},c)\geq|I_{\alpha}(\mathcal{W})-c|^{+}.\label{eq:con-con-gtone}
\end{equation}
For the cases $\alpha=1$ and $\alpha=\infty$,
Eq.~\eqref{eq:con-con-gtone} can be proved in a similar way. This
finishes the proof.
\end{IEEEproof}
With Theorem~\ref{them:one-shot-con-bound} established, the converse
bound for the strong converse exponent follows directly.

\begin{theorem} For any discrete memoryless channel
$\mathcal{W}:\mathcal{X}\rightarrow\mathcal{Y}$ and $r\geq0$, we
have
\begin{align}
E_{{\rm sc}}^{(\alpha)}(\mathcal{W},r)\geq\left\{ \begin{array}{ll}
\max\limits_{\alpha\leq\beta\leq1}\frac{\alpha(1-\beta)}{\beta(1-\alpha)}\left(I_{\beta}(\mathcal{W})-r\right), & \!\!\alpha\in(0,1)\\
|I_{\alpha}(\mathcal{W})-r|^{+}, & \!\!\alpha\in[1,\infty].
\end{array}\right.
\end{align}
\end{theorem}
\begin{IEEEproof}
Let $\{c_{n}\}_{n=1}^{\infty}$ be an arbitrary nonnegative sequence
of communication cost that satisfies
\begin{equation}
\limsup_{n\rightarrow\infty}\frac{c_{n}}{n}\leq r.
\end{equation}
Applying Theorem~\ref{them:one-shot-con-bound} with the substitutions
$\mathcal{W}\leftarrow\mathcal{W}^{\times n}$ and $c\leftarrow c_{n}$,
and making use of the additivity of R{é}nyi capacity~(Lemma~\ref{lem:RenyiC-properties}~(\romannumeral1)),
we get
\begin{equation}
\limsup_{n\rightarrow\infty}\frac{1}{n}\mathsf{D}_{\alpha}(\mathcal{W}^{\times n},c_{n})\geq\left\{ \begin{array}{ll}
\max\limits_{\alpha\leq\beta\leq1}\frac{\alpha(1-\beta)}{\beta(1-\alpha)}\left(I_{\beta}(\mathcal{W})-r\right), & \!\!\alpha\in(0,1)\\
|I_{\alpha}(\mathcal{W})-r|^{+}, & \!\!\alpha\in[1,\infty].
\end{array}\right.
\end{equation}
This, by the definition of strong converse exponent (cf. Eq.~\eqref{eq:def-sce}),
confirms the statement.
\end{IEEEproof}

\section{Conclusion and Discussion}

\label{sec:conclusion-discussion} In this paper, we study the large-deviation
behavior of channel simulation under the measure of R{é}nyi divergence
with parameter $\alpha\in[0,\infty]$. We have completely characterized
the R{é}nyi simulation rate, the reliability function and the strong
converse exponent. Notably, while for channel coding the reliability
function is only known when the communication rate is larger than
a critical value~\cite{Gallager1968information}, for the reverse
problem of channel simulation we have obtained the
complete solution.

As mentioned in Section \ref{subsec:Main-Results}, the $\infty$-R{é}nyi
simulation rate coincides with the minimum communication rate for
exact channel simulation~\cite{CLMW2011zero}. However, it remains
unclear why these two quantities coincide. It is interesting to investigate
this in the future.

Another natural question is how the present study can be extended
to the quantum setting. The quantum reverse Shannon theorem under
the quantum version of total variance, namely, the diamond norm, has
been established in the work~\cite{BDHSW2014quantum} (see also \cite{BCR2011the}).
As for the reliability function, it has been solved under the purified
distance, in the case where the communication rate is below a critical
value~\cite{LiYao2025reliable}. The present work, as well as the
independent works~\cite{OYB2024exponents} and \cite{OCCB2024exponents},
suggests that a complete characterization may be possible
to obtain.

\appendices{}

\section{Proofs of Theorem~\ref{thm:reliability-function} and Theorem~\ref{thm:strong-con-exponent} for $\alpha=0$}
\label{app:proof-order-zero}

At first, we prove Theorem~\ref{thm:reliability-function}
for the case $\alpha=0$. That is, $E_{{\rm rf}}^{(0)}(\mathcal{W},r)=\infty$
for all $r>I_{0}(\mathcal{W})$, and $E_{{\rm rf}}^{(0)}(\mathcal{W},r)=0$
for all $r<I_{0}(\mathcal{W})$.
\begin{IEEEproof}[Proof of Theorem~\ref{thm:reliability-function} for $\alpha=0$]
A key observation here is that $D_{0}(\mathcal{W}_{Y|X}^{\times n}(\cdot|x^{n})\big{\|}\mathcal{N}_{Y^{n}|X^{n}}^{(n)}(\cdot|x^{n}))$
does not decrease if we replace $\mathcal{W}_{Y|X}$ with any channel
$\mathcal{V}_{Y|X}$ such that $\mathrm{supp}(\mathcal{V}_{Y|X=x})\subseteq\mathrm{supp}(\mathcal{W}_{Y|X=x})$
for any $x$. Let $Q_{Y}$ be the minimizer of
\begin{align}
I_{0}(\mathcal{W}) & =\min_{Q_{Y}\in\mathcal{P}(\mathcal{Y})}\max_{x}D_{0}(\mathcal{W}_{Y|X}(\cdot|x)\|Q_{Y}).\label{eq:Ichannel-2}
\end{align}
Here we choose $\mathcal{V}_{Y|X}$ such that $\mathcal{V}_{Y|X=x}=Q_{Y}(\cdot|\mathrm{supp}(\mathcal{W}_{Y|X=x}))$,
where $Q_{Y}(\cdot|\mathrm{supp}(\mathcal{W}_{Y|X=x}))$ is the conditional
distribution of $Q_{Y}$ on the event $\mathrm{supp}(\mathcal{W}_{Y|X=x})$,
i.e., $Q_{Y}(y|\mathrm{supp}(\mathcal{W}_{Y|X=x}))=\frac{Q_{Y}(y)}{Q_{Y}(\mathrm{supp}(\mathcal{W}_{Y|X=x}))}$
for $y\in\mathrm{supp}(\mathcal{W}_{Y|X=x})$. So, for $s>0$,
\begin{align}
 & D_{0}\left(\mathcal{W}_{Y|X}^{\times n}(\cdot|x^{n})\big{\|}\mathcal{N}_{Y^{n}|X^{n}}^{(n)}(\cdot|x^{n})\right)\nonumber \\
 & \le D_{0}\left(\mathcal{V}_{Y|X}^{\times n}(\cdot|x^{n})\big{\|}\mathcal{N}_{Y^{n}|X^{n}}^{(n)}(\cdot|x^{n})\right)\nonumber \\
 & \le D_{1+s}\left(\mathcal{V}_{Y|X}^{\times n}(\cdot|x^{n})\big{\|}\mathcal{N}_{Y^{n}|X^{n}}^{(n)}(\cdot|x^{n})\right).
\end{align}

Set $N:=2^{nr}$, $\widetilde{N}:=(\ln2)Nn\log n$.
To further upper bound the last line above, we adopt the simulation
scheme with parameters $(\mathcal{V}_{Y|X}^{\times n},Q_{Y}^{\times n},N,\widetilde{N})$
in \ref{subsec:Simulation-Scheme} to generate a channel $\mathcal{N}_{Y^{n}|X^{n}}^{(n)}$
that approximates the target channel $\mathcal{V}_{Y|X}^{\times n}$.
For such a channel $\mathcal{N}_{Y^{n}|X^{n}}^{(n)}$, by Eq.~\eqref{eq:eq-re-1},
\begin{align}
 & D_{1+s}\left(\mathcal{V}_{Y|X}^{\times n}(\cdot|x^{n})\big{\|}\mathcal{N}_{Y^{n}|X^{n}}^{(n)}(\cdot|x^{n})\right)\nonumber \\
 & \leq\frac{1}{s\ln2}\exp\left\{ -ns\left(r-\frac{1}{n}D_{1+s}(\mathcal{V}_{Y|X}^{\times n}(\cdot|x^{n})\big{\|}Q_{Y}^{\times n})\right)\right\} +\frac{p_{n}}{(1-p_{n})\ln2},\label{eq:-10}
\end{align}
Here,
\begin{align}
\frac{1}{n}D_{1+s}(\mathcal{V}_{Y|X}^{\times n}(\cdot|x^{n})\big{\|}Q_{Y}^{\times n}) & =\frac{1}{ns}\log\sum_{y^{n}}\frac{\mathcal{V}_{Y|X}^{\times n}(y^{n}|x^{n})^{1+s}}{Q_{Y}^{\times n}(y^{n})^{s}}\nonumber \\
 & =\frac{1}{ns}\log\prod_{i=1}^{n}\sum_{y_{i}}\frac{\mathcal{V}_{Y|X}(y_{i}|x_{i})^{1+s}}{Q_{Y}(y_{i})^{s}}\nonumber \\
 & =-\frac{1}{ns}\log\prod_{i=1}^{n}Q_{Y}(\mathrm{supp}(\mathcal{W}_{Y|X=x_{i}}))^{s}\nonumber \\
 & =-\frac{1}{n}\sum_{i=1}^{n}\log Q_{Y}(\mathrm{supp}(\mathcal{W}_{Y|X=x_{i}}))\nonumber \\
 & =\frac{1}{n}\sum_{i=1}^{n}D_{0}(\mathcal{W}_{Y|X=x_{i}}\|Q_{Y})\nonumber \\
 & \le\max_{x}D_{0}(\mathcal{W}_{Y|X=x}\|Q_{Y})\nonumber \\
 & =I_{0}(\mathcal{W}).\label{eq:-9}
\end{align}
Combining Eqs.~\eqref{eq:re-one-shot-p} and \eqref{eq:-9},
we have
\begin{align}
p_{n}\leq\left(1-\frac{1}{N}(1-\exp\{-ns(r-I_{0}(\mathcal{W}))\})\right)^{\widetilde{N}},\label{eq:for-next-2-1}
\end{align}
where the right hand side of Eq.~\eqref{eq:for-next-2-1} is independent
of $x^{n}$. Moreover,
\begin{align}
 & \left(1-\frac{1}{N}(1-\exp\{-ns(r-I_{0}(\mathcal{W}))\})\right)^{\widetilde{N}}\nonumber \\
\dot{=} & \exp\{-n\log n\}.\label{eq:bound-pn-1}
\end{align}
Then, continuing Eq.~\eqref{eq:-10}, we have for sufficiently large
$n$,
\begin{align}
 & D_{0}\left(\mathcal{W}_{Y|X}^{\times n}(\cdot|x^{n})\big{\|}\mathcal{N}_{Y^{n}|X^{n}}^{(n)}(\cdot|x^{n})\right)\nonumber \\
\leq & \max_{x^{n}}\left\{ \frac{1}{s\ln2}\exp\big\{-ns\big(r-I_{0}(\mathcal{W})\big)\big\}+2p_{n}\right\} \nonumber \\
\dot{\leq} & \frac{1}{s\ln2}\exp\left\{ -ns\left(r-I_{0}(\mathcal{W})\right)\right\} ,\label{eq:for-sc-1}
\end{align}
where the first inequality follows from Eq.~\eqref{eq:-9} and the
fact that $p_{n}$ vanishes as $n\to\infty$, and the second inequality
comes from Eqs.~\eqref{eq:for-next-2-1}, \eqref{eq:bound-pn-1}
and
\begin{equation}
\exp\{-ns(r-I_{0}(\mathcal{W}))\}\dot{\geq}\exp\{-n\log n\}.
\end{equation}
Then, by the definition of $E_{{\rm rf}}^{(0)}(\mathcal{W},r)$ (cf.
Eq.~\eqref{eq:def-rf}) and Eq.~\eqref{eq:for-sc-1}, we obtain
\begin{align}
E_{{\rm rf}}^{(0)}(\mathcal{W},r)\geq s\left(r-I_{0}(\mathcal{W})\right).\label{eq:final-1-1}
\end{align}
Here, $s>0$ is arbitrary. So, $E_{{\rm rf}}^{(0)}(\mathcal{W},r)=\infty$
for all $r>I_{0}(\mathcal{W})$. On the other hand, by definition, Theorem~\ref{thm:strong-con-exponent} with $\alpha=0$ proved below implies
immediately that $E_{{\rm rf}}^{(0)}(\mathcal{W},r)=0$
for all $r<I_{0}(\mathcal{W})$.
\end{IEEEproof}

\smallskip
We now prove Theorem~\ref{thm:strong-con-exponent}
for the case $\alpha=0$. That is, $E_{{\rm sc}}^{(0)}(\mathcal{W},r)=|I_{0}(\mathcal{W})-r|^{+}$
for all $r\ge0$.
\begin{IEEEproof}[Proof of Theorem \ref{thm:strong-con-exponent} for $\alpha=0$]
By the monotonicity of the Rényi divergence in
its order, it holds that for $\delta>0$ and $0<\epsilon<1$,
\begin{align}
 & E_{{\rm sc}}^{(0)}(\mathcal{W},r)\nonumber \\
 & \geq\limsup_{n\to\infty}\frac{1}{n}\mathsf{D}_{0}\left(\mathcal{W}^{\times n},n(r+\delta)\right)\nonumber \\
 & =\limsup_{n\to\infty}\min_{\mathcal{N}^{(n)}\in\mathcal{A}(\mathcal{W}^{\times n},n(r+\delta))}\frac{1}{n}D_{0}(\mathcal{W}^{\times n},\mathcal{N}^{(n)})\nonumber \\
 & \stackrel{(a)}{=}\inf_{n\ge1}\min_{\mathcal{N}^{(n)}\in\mathcal{A}(\mathcal{W}^{\times n},n(r+\delta))}\frac{1}{n}D_{0}(\mathcal{W}^{\times n},\mathcal{N}^{(n)})\nonumber\\
 & =\inf_{n\ge1}\min_{\mathcal{N}^{(n)}\in\mathcal{A}(\mathcal{W}^{\times n},n(r+\delta))}\inf_{\alpha\in(0,\epsilon)}\frac{1}{n}D_{\alpha}(\mathcal{W}^{\times n},\mathcal{N}^{(n)})\nonumber \\
 & =\inf_{\alpha\in(0,\epsilon)}\inf_{n\ge1}\min_{\mathcal{N}^{(n)}\in\mathcal{A}(\mathcal{W}^{\times n},n(r+\delta))}\frac{1}{n}D_{\alpha}(\mathcal{W}^{\times n},\mathcal{N}^{(n)})\nonumber \\
 & \stackrel{(b)}{=}\inf_{\alpha\in(0,\epsilon)}\limsup_{n\to\infty}\frac{1}{n}\mathsf{D}_{\alpha}\left(\mathcal{W}^{\times n},n(r+\delta)\right)\nonumber\\
 & \geq\inf_{\alpha\in(0,\epsilon)}E_{{\rm sc}}^{(\alpha)}(\mathcal{W},r+\delta)\nonumber \\
 & =\inf_{\alpha\in(0,\epsilon)}\max\limits_{\alpha\leq\beta\leq1}\frac{\alpha(1-\beta)}{\beta(1-\alpha)}\left(I_{\beta}(\mathcal{W})-(r+\delta)\right),
\end{align}
where (a) follows by Fekete's lemma since $n\mapsto\min_{\mathcal{N}^{(n)}\in\mathcal{A}(\mathcal{W}^{\times n},n(r+\delta))}D_{0}(\mathcal{W}^{\times n},\mathcal{N}^{(n)})$
is subadditive, and in (b) we change the expression back
to $\mathsf{D}_{\alpha}\left(\mathcal{W}^{\times n},n(r+\delta)\right)$
which follows by arguments similar to the derivations in (a).
To see the subadditivity, using the additivity
of the R{é}nyi divergence, for any $\alpha\geq0$, it holds that
\begin{align}
 & \min_{\mathcal{N}^{(n+m)}\in\mathcal{A}(\mathcal{W}^{\times(n+m)},(n+m)(r+\delta))}D_{\alpha}(\mathcal{W}^{\times(n+m)},\mathcal{N}^{(n+m)})\nonumber \\
\leq & D_{\alpha}(\mathcal{W}^{\times(n+m)},\mathcal{\widehat{N}}^{(n)}\times\mathcal{\widehat{N}}^{(m)})\nonumber \\
= & D_{\alpha}(\mathcal{W}^{\times n},\mathcal{\widehat{N}}^{(n)})+D_{\alpha}(\mathcal{W}^{\times m},\mathcal{\widehat{N}}^{(m)})\nonumber \\
= & \min_{\mathcal{N}^{(n)}\in\mathcal{A}(\mathcal{W}^{\times n},n(r+\delta))}D_{\alpha}(\mathcal{W}^{\times n},\mathcal{N}^{(n)})+\min_{\mathcal{N}^{(m)}\in\mathcal{A}(\mathcal{W}^{\times m},m(r+\delta))}D_{\alpha}(\mathcal{W}^{\times m},\mathcal{N}^{(m)}),
\end{align}
where $\mathcal{\widehat{N}}^{(n)}$ is the minimizer of $\min_{\mathcal{N}^{(n)}\in\mathcal{A}(\mathcal{W}^{\times n},n(r+\delta))}D_{\alpha}(\mathcal{W}^{\times n},\mathcal{N}^{(n)})$.
Using the monotonicity of the Rényi divergence in its order again, it holds
that
\begin{align}
E_{{\rm sc}}^{(0)}(\mathcal{W},r) & \le\inf_{\alpha\in(0,\epsilon)}E_{{\rm sc}}^{(\alpha)}(\mathcal{W},r)\nonumber \\
 & =\inf_{\alpha\in(0,\epsilon)}\max\limits_{\alpha\leq\beta\leq1}\frac{\alpha(1-\beta)}{\beta(1-\alpha)}\left(I_{\beta}(\mathcal{W})-r\right).
\end{align}
Letting $\epsilon\to0$ and $\delta\to0$ in the
upper and lower bounds, and using Lemma~\ref{lem:sc-ahpha-0} below,
we obtain the desired result.
\end{IEEEproof}

\begin{lemma}\label{lem:sc-ahpha-0}
	For any $r\ge0$ and discrete memoryless channel $\mathcal{W}$, we have
	\begin{equation}
		\lim_{\alpha\searrow0}\max\limits_{\alpha\leq\beta\leq1}\frac{\alpha(1-\beta)}{\beta(1-\alpha)}\left(I_{\beta}(\mathcal{W})-r\right)=|I_{0}(\mathcal{W})-r|^{+}.\label{eq:sc-alpha-0}
	\end{equation}
\end{lemma}
\begin{IEEEproof}
	When $r\geq I(\mathcal{W})$, both sides of Eq.$\,$\eqref{eq:sc-alpha-0}
	are zero. Thus it suffices to consider the case $r\in[0,I(\mathcal{W})]$.
	For any $\beta\in[\alpha,1]$, define
	\begin{equation}
		f_{\alpha}(\beta):=\frac{\alpha(1-\beta)}{\beta(1-\alpha)}\left(I_{\beta}(\mathcal{W})-r\right).
	\end{equation}
	It suffices to show that for any $r\in[0,I(\mathcal{W})]$, we have
	\begin{align}
		\lim_{\alpha\searrow0}\max_{\beta\in[\alpha,1]}f_{\alpha}(\beta) & \le|I_{0}(\mathcal{W})-r|^{+},\label{eq:alpha-0-result-1}\\
		\lim_{\alpha\searrow0}\max_{\beta\in[\alpha,1]}f_{\alpha}(\beta) & \geq|I_{0}(\mathcal{W})-r|^{+}.\label{eq:alpha-0-result-2}
	\end{align}
	
	First, we prove the lower bound. Since
	\[
	\max_{\beta\in[\alpha,1]}f_{\alpha}(\beta)\geq\max\{f_{\alpha}(1),f_{\alpha}(\alpha)\}=|I_{\alpha}(\mathcal{W})-r|^{+}.
	\]
	Letting $\alpha\searrow0$ and using the continuity of the function
	$\alpha\mapsto|I_{\alpha}(\mathcal{W})-r|^{+}$, equation \eqref{eq:alpha-0-result-2}
	follows.
	
	Next, we prove the upper bound. Let $\epsilon>0$
	be arbitrary. Since $I_{\alpha}(\mathcal{W})\searrow I_{0}(\mathcal{W})$
	as $\alpha\searrow0$, there exists $\delta\in(0,1)$ such that $0\le I_{\delta}(\mathcal{W})-I_{0}(\mathcal{W})<\varepsilon.$
	Let $\alpha\in(0,\delta]$. We split the interval $[\alpha,1]$ into
	two parts: $[\alpha,1]=[\alpha,\delta]\cup[\delta,1].$ For the case
	$r\in[0,I(\mathcal{W})]$ and $\beta\in[\delta,1]$, we have
	\begin{equation}
		f_{\alpha}(\beta)\le\frac{\alpha}{\delta(1-\alpha)}I(\mathcal{W}),
	\end{equation}
	which tend to zero as $\alpha\searrow0$. So, there exists $\alpha'>0$,
	independent of $\beta$, such that for all $\alpha\in(0,\alpha')$
	we have $f_{\alpha}(\beta)<\epsilon$. For the case $r\in[I_{\delta}(\mathcal{W}),I(\mathcal{W})]$
	and $\beta\in[\alpha,\delta]$, we have $f_{\alpha}(\beta)\leq0$.
	For the case $r\in[0,I_{\delta}(\mathcal{W})]$ and $\beta\in[\alpha,\delta]$,
	since $I_{\beta}(\mathcal{W})\le I_{\delta}(\mathcal{W})$ and $\frac{\alpha(1-\beta)}{\beta(1-\alpha)}\in(0,1]$,
	we have
	\begin{equation}
		f_{\alpha}(\beta)\le\frac{\alpha(1-\beta)}{\beta(1-\alpha)}(I_{\delta}(\mathcal{W})-r)\le I_{\delta}(\mathcal{W})-r\leq I_{0}(\mathcal{W})-r+\epsilon.
	\end{equation}
	Let $\alpha^{*}=\min\{\alpha',\delta\}$. When $\alpha\in(0,\alpha^{*})$,
	we have
	\begin{align}
		\max_{\beta\in[\alpha,1]}f_{\alpha}(\beta) & =\max\left\{ \max_{\beta\in[\alpha,\delta]}f_{\alpha}(\beta),\max_{\beta\in[\delta,1]}f_{\alpha}(\beta)\right\} \nonumber \\
		& \leq\left\{
		\begin{array}{ll}
			\epsilon, &r\in[I_{\delta}(\mathcal{W}),I(\mathcal{W})]\\
			\max\{I_{0}(\mathcal{W})-r+\epsilon,\epsilon\}, &r\in[0,I_{\delta}(\mathcal{W})].
		\end{array}\right.
	\end{align}
	Taking the limit $\alpha\searrow0$ and then letting $\epsilon\to0$,
	we confirm Eq.$\,$\eqref{eq:alpha-0-result-1}. This completes the
	proof.
\end{IEEEproof}

\section{Proof of Theorem~\ref{thm:RST} }
\label{app:proof-RST}
By using Theorems~\ref{thm:reliability-function}
and \ref{thm:strong-con-exponent}, we derive the reverse Shannon
theorem under the R{é}nyi divergence with parameter $\alpha\in[0,\infty]$.
\begin{IEEEproof}[Proof of Theorem~\ref{thm:RST}]
	As stated in Lemma~\ref{lem:RenyiC-properties}~(\romannumeral5),
	$I_{\alpha}(\mathcal{W})$ is an increasing and continuous function
	of $\alpha$ on $(0,\infty)$, converging to $I_{0}(\mathcal{W})$
	as $\alpha\to0$, $I(\mathcal{W})$ as $\alpha\to1$ and to $I_{\infty}(\mathcal{W})$
	as $\alpha\to\infty$.
	
	From Theorem \ref{thm:reliability-function},
	we have that $E_{{\rm rf}}^{(\alpha)}(\mathcal{W},r)>0$ for $r>I(\mathcal{W})$
	and $\alpha\in(0,1]$, and $E_{{\rm rf}}^{(\alpha)}(\mathcal{W},r)>0$
	for $r>I_{\alpha}(\mathcal{W})$ and $\alpha\in\{0\}\cup(1,\infty]$.
	In fact, $E_{{\rm rf}}^{(\infty)}(\mathcal{W},r)=\infty$ for $r\geq I_{\infty}(\mathcal{W})$
	and $E_{{\rm rf}}^{(0)}(\mathcal{W},r)=\infty$ for $r>I_{0}(\mathcal{W})$.
	From Theorem~\ref{thm:strong-con-exponent}, we obtain that $E_{{\rm sc}}^{(\alpha)}(\mathcal{W},r)>0$
	for $r<I(\mathcal{W})$ and $\alpha\in(0,1]$, and $E_{{\rm sc}}^{(\alpha)}(\mathcal{W},r)>0$
	for $r<I_{\alpha}(\mathcal{W})$ and $\alpha\in\{0\}\cup(1,\infty]$.
	Thus, for asymptotically reliable simulation, the minimum communication
	rates are given by $I(\mathcal{W})$ for $\alpha\in(0,1]$ and $I_{\alpha}(\mathcal{W})$
	for $\alpha\in\{0\}\cup(1,\infty]$, respectively. This completes
	the proof.
\end{IEEEproof}

\section{Proof of Proposition~\ref{prop:exponent-u-symm}}
\label{app:proof-of-propexu} \setcounter{equation}{0}
\global\long\def\theequation{B.\arabic{equation}}%
The proof of Proposition~\ref{prop:exponent-u-symm} relies on the
G{ä}rtner-Ellis theorem of large deviation theory. We will use a
variant version proved in~\cite{Chen2000generalization}, and more
specifically in~\cite[Lemma 37]{MosonyiOgawa2015two}.

\begin{lemma} \label{lem:Gartner-Ellis} Given a sequence of random
variables $\mathcal{Z}\equiv\{Z_{n}\}_{n\in\mathbb{N}}$, we introduce
its asymptotic cumulant generating function as
\begin{equation}
\Lambda_{\mathcal{Z}}(t):=\lim_{n\rightarrow\infty}\frac{1}{n}\log\mathbb{E}\left[e^{ntZ_{n}}\right],
\end{equation}
if the limit exists and the function $t\mapsto\Lambda_{\mathcal{Z}}(t)$
is differentiable in some interval $(a,b)$. Then for any $z\in(\lim\limits_{t\searrow a}\Lambda_{\mathcal{Z}}'(t),\lim\limits_{t\nearrow b}\Lambda_{\mathcal{Z}}'(t))$,
we have
\begin{equation}
\limsup_{n\rightarrow\infty}\frac{-1}{n}\log\operatorname{Pr}\left[Z_{n}\geq z\right]\leq\sup_{t\in(a,b)}\left\{ zt-\Lambda_{\mathcal{Z}}(t)\right\} .
\end{equation}
\end{lemma}

We also need the following asymptotic expression for the Rényi mutual
information. It was proved in~\cite{TomamichelHayashi2017operational}
in a more general setting. For reader's convenience, we include the
proof here.

\begin{lemma}\label{lem:D-u-symm} Let $P_{XY}\in\mathcal{P}(\mathcal{X}\times\mathcal{Y})$
and $\Phi_{Y^{n}}$ be the symmetric probability distribution given
in Eq.~\eqref{eq:unidistribution}. For any $\alpha>0$, we have
\begin{equation}
\frac{1}{n}D_{\alpha}\left(P_{XY}^{\times n}\|P_{X}^{\times n}\times\Phi_{Y^{n}}\right)=I_{\alpha}(X:Y)_{P_{XY}}+O\left(\dfrac{\log n}{n}\right).
\end{equation}
\end{lemma}
\begin{IEEEproof}
For any $Q_{Y}\in\mathcal{P}(\mathcal{Y})$, as shown in Eq.~\eqref{eq:symbound}
we have $Q_{Y}^{\times n}\leq(n+1)^{|\mathcal{Y}|}\Phi_{Y^{n}}$.
Then,
\begin{equation}
D_{\alpha}\left(P_{XY}^{\times n}\|P_{X}^{\times n}\times\Phi_{Y^{n}}\right)\leq D_{\alpha}\left(P_{XY}^{\times n}\|P_{X}^{\times n}\times Q_{Y}^{\times n}\right)+\log(n+1)^{|\mathcal{Y}|}.
\end{equation}
By the additivity of R{é}nyi divergence~(Lemma~\ref{lem:RenyiD-properties}~(\romannumeral2)),
we obtain
\begin{align}
\frac{1}{n}D_{\alpha}\left(P_{XY}^{\times n}\|P_{X}^{\times n}\times\Phi_{Y^{n}}\right) & \leq\min_{Q_{Y}\in\mathcal{P}(\mathcal{Y})}D_{\alpha}\left(P_{XY}\|P_{X}\times Q_{Y}\right)+\frac{\log(n+1)^{|\mathcal{Y}|}}{n}\nonumber \\
 & =I_{\alpha}(X:Y)_{P_{XY}}+O\left(\dfrac{\log n}{n}\right).
\end{align}
On the other hand, by definition and the additivity of Rényi mutual
information for product distributions~\cite{Verdu2015alpha}, we
get
\begin{equation}
\frac{1}{n}D_{\alpha}\left(P_{XY}^{\times n}\|P_{X}^{\times n}\times\Phi_{Y^{n}}\right)\geq\frac{1}{n}I_{\alpha}(X^{n}:Y^{n})_{P_{XY}^{\times n}}=I_{\alpha}(X:Y)_{P_{XY}}.
\end{equation}
Combining the upper and lower bounds leads to the statement.
\end{IEEEproof}
\medskip{}

\begin{IEEEproof}[Proof of Proposition~\ref{prop:exponent-u-symm}]
For convenience, we use the following abbreviations
\begin{equation}
P_{n}:=P_{XY}^{\times n},\quad Q_{n}:=P_{X}^{\times n}\times\Phi_{Y^{n}}.
\end{equation}
By Lemma~\ref{lem:D-u-symm}, we have
\begin{equation}
\lim_{n\rightarrow\infty}\frac{1}{n}D_{1+t}(P_{n}\|Q_{n})=I_{1+t}(X:Y)_{P_{XY}}.
\end{equation}

We first prove the ``$\geq$'' part. For the case $r\geq I_{\infty}(X:Y)_{P_{XY}}$,
let $Q_{Y}$ be the minimizer in the optimization $\min_{Q_{Y}}D_{\infty}(P_{XY}\|P_{X}\times Q_{Y})$.
We have
\begin{align}
\max_{x^{n},y^{n}}\frac{P_{n}(x^{n},y^{n})}{Q_{n}(x^{n},y^{n})}\leq & (n+1)^{|\mathcal{Y}|}\max_{x^{n},y^{n}}\frac{P_{Y|X}^{\times n}(y^{n}|x^{n})}{Q_{Y}^{\times n}(y^{n})}\nonumber \\
= & (n+1)^{|\mathcal{Y}|}2^{nI_{\infty}(X:Y)_{P_{XY}}}.
\end{align}
This implies $P_{n}\left(P_{n}\geq g(n)2^{nr}Q_{n}\right)=0$ because
of $g(n)>(n+1)^{|\mathcal{Y}|}$. Therefore,
\begin{equation}
\liminf_{n\rightarrow\infty}\frac{-1}{n}\log p_{n}=\infty.
\end{equation}
Consider the other case $r\in[0,I_{\infty}(X:Y)_{P_{XY}})$. For any
$t>0$,
\begin{align}
\frac{-1}{n}\log p_{n} & =\frac{-1}{n}\log P_{n}\left(P_{n}\geq g(n)2^{nr}Q_{n}\right)\nonumber \\
 & \geq\frac{-1}{n}\log\sum_{x^{n},y^{n}}P_{n}(x^{n},y^{n})\left(\frac{P_{n}(x^{n},y^{n})}{g(n)2^{nr}Q_{n}(x^{n},y^{n})}\right)^{t}\nonumber \\
 & =tr-\frac{t}{n}D_{1+t}(P_{n}\|Q_{n})+\frac{t\log g(n)}{n}\nonumber \\
 & \to t\left(r-I_{1+t}(X:Y)_{P_{XY}}\right),\quad\text{as }n\rightarrow\infty.
\end{align}
Since $t>0$ is arbitrary, we have
\begin{equation}
\liminf_{n\rightarrow\infty}\frac{-1}{n}\log p_{n}\geq\sup_{t\geq0}t\left(r-I_{1+t}(X:Y)_{P_{XY}}\right).\label{eq:pn-lower-bound}
\end{equation}

For the ``$\leq$'' part, the Gärther-Ellis theorem plays a key
role. Let
\begin{equation}
Z_{n}(x^{n},y^{n}):=\frac{1}{n}\log\frac{P_{n}(x^{n},y^{n})}{g(n)2^{nr}Q_{n}(x^{n},y^{n})}.
\end{equation}
Then
\begin{align}
p_{n} & =P_{n}\left(P_{n}\geq g(n)2^{nr}Q_{n}\right)\nonumber \\
 & =P_{n}\left(Z_{n}\geq0\right).
\end{align}
We calculate the asymptotic cumulant generating function of the sequence
$\mathcal{Z}:=\{Z_{n}\}_{n\in\mathbb{N}}$ as follows.
\begin{align}
\Lambda_{\mathcal{Z}}(t) & =\lim_{n\rightarrow\infty}\frac{1}{n}\log\mathbb{E}_{P_{n}}\left[2^{ntZ_{n}}\right]\nonumber \\
 & =\lim_{n\rightarrow\infty}\frac{1}{n}\log\mathbb{E}_{P_{n}}\left[\left(\frac{P_{n}}{Q_{n}}\right)^{t}(g(n))^{-t}2^{-ntr}\right]\nonumber \\
 & =\lim_{n\rightarrow\infty}\frac{1}{n}\log\sum_{x^{n},y^{n}}\left(P_{n}(x^{n},y^{n})\left(\frac{P_{n}(x^{n},y^{n})}{Q_{n}(x^{n},y^{n})}\right)^{t}(g(n))^{-t}2^{-ntr}\right)\nonumber \\
 & =\lim_{n\rightarrow\infty}\left(\frac{t}{n}D_{1+t}(P_{n}\|Q_{n})-tr-\frac{t\log g(n)}{n}\right)\nonumber \\
 & =t\left(I_{1+t}(X:Y)_{P_{XY}}-r\right).
\end{align}
Assume that $I(X:Y)_{P_{XY}}<r<I_{\infty}(X:Y)_{P_{XY}}$. By Lemma~\ref{lem:RenyiC-properties}~(\romannumeral4),
we see that $t\mapsto\Lambda_{\mathcal{Z}}(t)$ is differentiable
on $(0,+\infty)$, and
\begin{align}
\lim_{t\rightarrow0}\Lambda_{\mathcal{Z}}'(t)= & I(X:Y)_{P_{XY}}-r<0,\\
(\exists\,t_{0}>0)\lim_{t\rightarrow t_{0}}\Lambda_{\mathcal{Z}}'(t)\geq & I_{1+t_{0}}(X:Y)_{P_{XY}}-r>0.
\end{align}
Therefore, Lemma~\ref{lem:Gartner-Ellis} applies, giving that for
$r\in(I(X:Y)_{P_{XY}},I_{\infty}(X:Y)_{P_{XY}})$,
\begin{align}
 & \limsup_{n\rightarrow\infty}\frac{-1}{n}\log p_{n}\nonumber \\
= & \limsup_{n\rightarrow\infty}\frac{-1}{n}\log P_{n}(Z_{n}\geq0)\nonumber \\
\leq & \sup_{0<t<t_{0}}t\left(r-I_{1+t}(X:Y)_{P_{XY}}\right)\nonumber \\
\leq & \sup_{t\geq0}t\left(r-I_{1+t}(X:Y)_{P_{XY}}\right).\label{eq:pn-upper-bound}
\end{align}
Actually, Eq.~\eqref{eq:pn-upper-bound} can be extended to the whole
range $r\in\mathbb{R}$. When $r\leq I(X:Y)_{P_{XY}}$, the last line
of Eq.~\eqref{eq:pn-upper-bound} is $0$. Observing that $p_{n}$
is monotonically decreasing with $r$ and $0\leq p_{n}\leq1$, we
get that the first line is also $0$, by letting $r\searrow I(X:Y)_{P_{XY}}$.
When $r\geq I_{\infty}(X:Y)_{P_{XY}}$, the last line of Eq.~\eqref{eq:pn-upper-bound}
is $\infty$, which is of course an upper bound of the first line.
Thus, for $r\in\mathbb{R}$, we establish that
\begin{align}
\limsup_{n\rightarrow\infty}\frac{-1}{n}\log p_{n}\leq\sup_{t\geq0}t\left(r-I_{1+t}(X:Y)_{P_{XY}}\right).\label{eq:R-pn-upper-bound}
\end{align}

The combination of Eqs.~\eqref{eq:pn-lower-bound} and \eqref{eq:R-pn-upper-bound}
lets us finish the proof.
\end{IEEEproof}

\section{Alternative Simulation Scheme}

\label{app:altsimulation} \setcounter{equation}{0}
\global\long\def\theequation{C.\arabic{equation}}%
Combining the standard rejection sampling and the variant rejection
sampling proposed in~\cite{HJMR2007communication}, we design a two-phase
rejection sampling procedure. Using this we obtain an alternative
simulation scheme which can also let us derive the optimal exponents.

\subsubsection{Standard Rejection Sampling}

Let $P,Q$ be two distributions satisfying $\supp(P)\subseteq\supp(Q)$.
The rejection sampling works in the following way: at each iteration,
we fill the distribution $P$ with the best possible sub-distribution
$\lambda Q$ with a proper constant $\lambda\le1$, while maintaining
that our sum is always less than $P$ (note that since we are doing
rejection sampling, we can only create sub-distributions $\lambda Q$
at each iteration). After all iterations, eventually the output follows
$P$.

\begin{algorithm}[H]
\textbf{Computing Probabilities:} Given distributions $P,Q$ and a
positive integer $N$, the probabilities involved in the rejection
sampling are specified as follows.
\begin{enumerate}
\item Initially, we set $s_{0}=1$ and $p_{0}(x)=P(x)$ for all $x$.
\item For $j\in\{1,2,\ldots,N\}$, we let
\begin{align*}
a_{j}(x) & =\min\left\{ 1,\frac{p_{j-1}(x)}{s_{j-1}Q(x)}\right\} ,\\
p_{j}(x) & =p_{j-1}(x)-s_{j-1}Q(x)a_{j}(x),\\
s_{j} & =\sum_{x}p_{j}(x).
\end{align*}
\end{enumerate}
\textbf{Rejection Sampling Procedure:} Let $X_{j}\sim Q,0\le j\le N$
be sampled independently.
\begin{enumerate}
\item Compute the sequence of functions $a_{j},j\in\{1,2,\ldots,N\}$ as
given above.
\item For $j\leftarrow1$ to $N$ do

With probability $a_{j}(X_{j})$, output $j$ and halt.
\item Abort and output $j=0$.
\end{enumerate}
\caption{The standard rejection sampling with parameters $(P,Q,N)$.}
\label{alg:The-standard-rejection}
\end{algorithm}

We introduce the standard rejection sampling in detail; see Algorithm
\ref{alg:The-standard-rejection}. The performance of the standard
rejection sampling is analyzed now. Given $X_{j}=x$, the probability
that the procedure halts and yields an output at iteration $(j)$
is equal to $s_{j-1}a_{j}(x)$. The sub-distribution $p_{j}$ is the
remaining part of $P$ after the $j$-th iteration, and the sub-distribution
$p_{N}$ is the remaining part of $P$ after performing the procedure.
The value $s_{j}$ is the total probabilities of the remaining part
after the $j$-th iteration, and the procedure abort with probability
$s_{N}$. By definition, it holds that either $p_{j}(x)=p_{j-1}(x)-s_{j-1}Q(x)$
or $p_{j}(x)=0$. So, for every $j\in\{1,...,N\}$,
\begin{equation}
p_{j}(x)=\left|P(x)-\beta_{j}Q(x)\right|^{+},\,\forall x,
\end{equation}
where $\beta_{j}:=\sum_{i=0}^{j-1}s_{i}$.
\begin{claim}
For every $j\in\{0,...,N\}$,

$(a)$ the probability that the procedure halts within $j$ iterations
is exactly $1-s_{j}$;

$(b)$ for each $x$, the probability that the procedure halts within
$j$ iterations and outputs $j^{*}$ such that $X_{j^{*}}=x$ is exactly
$P(x)-p_{j}(x)$.
\end{claim}
In particular, for $j=N$, the procedure either aborts and outputs
$0$ with probability $s_{N}$, or halts and outputs $j^{*}$ such
that $X_{j^{*}}=x$ with probability
\begin{equation}
P(x)-p_{N}(x)=\min\{P(x),\beta_{N}Q(x)\}.
\end{equation}
So, the distribution $U$ of the final output of the procedure is
given by
\begin{equation}
U(x)=\min\{P(x),\beta_{N}Q(x)\}+s_{N}Q(x)\ge\min\{P(x),\beta_{N}Q(x)\}.\label{eq:-1}
\end{equation}
Note that $\beta_{N}$ could be significantly smaller than $N$. This
can be seen by considering the $n$-dimensional product distributions
$P^{\times n},Q^{\times n}$, for which case, $\beta_{N}$ is equal
to $2^{n(D(P\|Q)+o(1))}$ for any $N\ge2^{nD(P\|Q)}$. However, to
design a good simulation scheme which induces a smaller R{é}nyi
divergence $D_{1+s}(P\|U)$, we require $\beta_{N}$ to be of the
same order as $N$. To this end, we will utilize the variant of rejection
sampling proposed in \cite{HJMR2007communication}, which ensues a
large aborting probability $s_{N}$ and hence also a large $\beta_{N}$.

\subsubsection{Variant Rejection Sampling}

\begin{algorithm}[H]
\textbf{Computing Probabilities: }Given a distribution $Q$, a sub-distribution
$\hat{P}$ with total probability $\lambda=\sum_{x}\hat{P}(x)\in(0,1)$,
and a positive integer $N$, the probabilities involved in the rejection
sampling are specified as follows.
\begin{enumerate}
\item Initially, we set $\hat{s}_{0}=\lambda$ and $\hat{p}_{0}(x)=\hat{P}(x)$
for all $x$.
\item For $j\in\{1,2,\ldots,N\}$, we let
\begin{align*}
\hat{a}_{j}(x) & =\min\left\{ 1,\frac{\hat{p}_{j-1}(x)}{(1-\lambda+\hat{s}_{j-1})Q(x)}\right\} ,\\
\hat{p}_{j}(x) & =\hat{p}_{j-1}(x)-(1-\lambda+\hat{s}_{j-1})Q(x)\hat{a}_{j}(x),\\
\hat{s}_{j} & =\sum_{x}\hat{p}_{j}(x).
\end{align*}
\end{enumerate}
\textbf{Rejection Sampling Procedure:} Same as the one in the standard
rejection sampling but the probabilities involved replaced with the
above ones.

\caption{The variant rejection sampling with parameters $(\hat{P},Q,N)$ proposed
in \cite{HJMR2007communication}.}
\label{alg:The-variant-rejection}
\end{algorithm}

We now introduce the variant of rejection sampling proposed in \cite{HJMR2007communication};
see Algorithm \ref{alg:The-variant-rejection}. Similarly to the standard
rejection sampling, in the variant version, the sub-distribution $\hat{p}_{j}$
is the remaining part of $\hat{P}$ after the $j$-th iteration. The
value $\hat{s}_{j}$ is the total probabilities of the remaining part
of $\hat{P}$ after the $j$-th iteration. The procedure abort with
probability $1-\lambda+\hat{s}_{N}$. For every $j\in\{0,...,N\}$,
\begin{equation}
\hat{p}_{j}(x)=\left|\hat{P}(x)-\hat{\beta}_{j}Q(x)\right|^{+},\,\forall x,
\end{equation}
where $\hat{\beta}_{j}:=(1-\lambda)j+\sum_{i=0}^{j-1}\hat{s}_{i}$.
\begin{claim}
\cite[Claim 4.3]{HJMR2007communication} For every $j\in\{0,...,N\}$,

$(a)$ the probability that the procedure halts within $j$ iterations
is exactly $\lambda-\hat{s}_{j}$;

$(b)$ for each $x$, the probability that the procedure halts within
$j$ iterations and outputs $j^{*}$ such that $X_{j^{*}}=x$ is exactly
$\hat{P}(x)-\hat{p}_{j}(x)$.
\end{claim}
In particular, for $j=N$, the procedure either aborts and outputs
$0$ with probability $1-\lambda+\hat{s}_{N}$, or halts and outputs
$j^{*}$ such that $X_{j^{*}}=x$ with probability
\begin{equation}
\hat{P}(x)-\hat{p}_{N}(x)=\min\{\hat{P}(x),\hat{\beta}_{N}Q(x)\}.
\end{equation}
So, the distribution $\hat{U}$ of the final output of the procedure
is given by
\begin{align}
\hat{U}(x) & =\min\{\hat{P}(x),\hat{\beta}_{N}Q(x)\}+(1-\lambda+\hat{s}_{N})Q(x)\nonumber \\
 & \geq\min\{\hat{P}(x),\hat{\beta}_{N}Q(x)\}.
\end{align}

In the standard rejection sampling, the output is aimed to approximately
follow the target distribution $P$. However, in the variant version,
the output is aimed to approximately follow the target\emph{ sub-distribution}
$\hat{P}$. As mentioned above, the standard rejection sampling aborts
with probability $s_{N}$ which could be arbitrarily small when $N$
is large. In contrast, the variant rejection sampling aborts with
probability $1-\lambda+\hat{s}_{N}$ which is at least $1-\lambda$
for all $N$. This is a property we want.

\subsubsection{Two-phase Rejection Sampling}

We now introduce a two-phase rejection sampling procedure which are
based on the standard and variant versions given above; see Algorithm
\ref{alg:Two-phase-rejection-sampling}. Note that $\beta_{N}=\sum_{i=0}^{N-1}s_{i}\ge s_{N}N$
by the monotonicity of $s_{i}$ in $i$.

\begin{algorithm}[H]
\textbf{Computing Probabilities: }Given distributions $P,Q$ and positive
integers $N,K$, the probabilities involved in the rejection sampling
are specified as follows.
\begin{enumerate}
\item Compute probabilities $a_{j},p_{j},s_{j},0\le j\le N$ as in the standard
rejection sampling in Algorithm \ref{alg:The-standard-rejection}
with parameters $(P,Q,N)$.
\item Compute the sub-distribution $P_{0}(x)=\min\{P(x),\beta_{N}Q(x)\}$
with total probability $\lambda_{0}:=\sum_{x}P_{0}(x)=1-s_{N}\in(0,1)$.
\item Compute $\hat{P}(x)=P(x)-P_{0}(x)$ for all $x$ whose total probability
$\lambda=s_{N}$.
\item Compute probabilities $\hat{a}_{j},\hat{p}_{j},\hat{s}_{j},0\le j\le N$
as in the variant rejection sampling with parameters $(\hat{P},Q,N)$.
\end{enumerate}
\textbf{Rejection Sampling Procedure:} Perform the variant rejection
sampling procedure with parameters $(\hat{P},Q,N)$. However, differently,
when the variant procedure abort, it does not yield an output direction,
but instead goes into the standard rejection sampling procedure with
parameters $(P,Q,N)$. Similarly, when the standard procedure abort,
it does not yield an output direction, but instead repeats the standard
rejection sampling procedure with parameters $(P,Q,N)$. Repeat the
standard rejection sampling procedure up to $K$ times in total. If
finally the standard rejection sampling procedure abort, the procedure
outputs $j=0$. Note that all $X_{j}$'s in all the variant and standard
procedures are sampled independently, which means we have used $(K+1)N+1$
samples totally.

\caption{Two-phase rejection sampling with parameters $(P,Q,N,K)$.}
\label{alg:Two-phase-rejection-sampling}
\end{algorithm}

In the first phase, the variant procedure either aborts with probability
$1-\lambda+\hat{s}_{N}$, or halts and outputs $j^{*}$ such that
$X_{j^{*}}=x$ with probability
\begin{equation}
\hat{P}(x)-\hat{p}_{N}(x)=\min\{\hat{P}(x),\hat{\beta}_{N}Q(x)\},
\end{equation}
where $\hat{\beta}_{N}:=(1-\lambda)N+\sum_{i=0}^{N-1}\hat{s}_{i}\ge(1-\lambda)N$.

When the variant procedure aborts, we are going into the standard
procedures, i.e., the second phase. Conditionally on that we have
entered the second phase, the output of the $K$ standard procedures
follows a distribution $U$, which is given by
\begin{align}
U(x) & =(1+s_{N}+...+s_{N}^{K-1})P_{0}(x)+s_{N}^{K}Q(x)\nonumber \\
 & =\frac{1-s_{N}^{K}}{1-s_{N}}P_{0}(x)+s_{N}^{K}Q(x).
\end{align}

The final output of the two-phase procedure follows a distribution
$S$, which is given by
\begin{align}
S(x) & =\min\{\hat{P}(x),\hat{\beta}_{N}Q(x)\}+(1-\lambda+\hat{s}_{N})U(x)\nonumber \\
 & \ge\min\{\hat{P}(x),\hat{\beta}_{N}Q(x)\}+(1-s_{N}+\hat{s}_{N})\frac{1-s_{N}^{K}}{1-s_{N}}P_{0}(x)\nonumber \\
 & \geq\min\{\hat{P}(x),\hat{\beta}_{N}Q(x)\}+\left(1-s_{N}^{K}\right)P_{0}(x)\nonumber \\
 & \geq\left(1-s_{N}^{K}\right)\left(\min\{\hat{P}(x),\hat{\beta}_{N}Q(x)\}+P_{0}(x)\right).\label{eq:app-re-1}
\end{align}
Since $\hat{\beta}_{N}\ge(1-\lambda)N$ with $\lambda=s_{N}$, we
have

\begin{align}
 & \min\{\hat{P}(x),\hat{\beta}_{N}Q(x)\}+P_{0}(x)\nonumber \\
\geq & \min\{P(x),(1-s_{N})NQ(x)+P_{0}(x)\}\nonumber \\
= & \min\{P(x),(1-s_{N})NQ(x)+\min\{P(x),\beta_{N}Q(x)\}\}\nonumber \\
= & \min\{P(x),(1-s_{N})NQ(x)+\beta_{N}Q(x)\}\nonumber \\
\geq & \min\{P(x),(1-s_{N})NQ(x)+s_{N}NQ(x)\}\nonumber \\
= & \min\{P(x),NQ(x)\}.\label{eq:app-re-2}
\end{align}
Eqs.~\eqref{eq:app-re-1} and \eqref{eq:app-re-2} lead to
\begin{equation}
S(x)\geq\left(1-s_{N}^{K}\right)\min\{P(x),NQ(x)\}.\label{eq:for-next1}
\end{equation}
For the case $s=\infty$, we have
\begin{align}
D_{\infty}(P\|S)\leq & \log\max_{x}\frac{P(x)}{(1-s_{N}^{K})\min\{P(x),NQ(x)\}}\nonumber \\
\leq & \Big|\log\max_{x}\frac{P(x)}{NQ(x)}\Big|^{+}-\log(1-s_{N}^{K})\nonumber \\
= & |D_{\infty}(P\|Q)-\log N|^{+}-\log(1-s_{N}^{K}).
\end{align}
For the other case $s\in(0,\infty)$, the order-$(1+s)$ R{é}nyi divergence between $P$ and
the sampling distribution $S$ satisfies
\begin{align}
 & D_{1+s}(P\|S)\nonumber \\
\leq & \frac{1}{s}\log\left\{ \left(1-s_{N}^{K}\right)^{-s}P(\mathcal{O}^{c})+N^{-s}\sum_{x\in\mathcal{O}}P(x)\left(\frac{P(x)}{\left(1-s_{N}^{K}\right)Q(x)}\right)^{s}\right\} \nonumber \\
\leq & \frac{1}{s}\log\left\{ 1+N^{-s}\sum_{x}P(x)\left(\frac{P(x)}{Q(x)}\right)^{s}\right\} -\log\left(1-s_{N}^{K}\right)\nonumber \\
= & \frac{1}{s}\log\left\{ 1+\exp\{-s(\log N-D_{1+s}(P\|Q))\}\right\} -\log\left(1-s_{N}^{K}\right)\nonumber \\
\leq & \frac{1}{s\ln2}\exp\{-s(\log N-D_{1+s}(P\|Q))\}-\log\left(1-s_{N}^{K}\right).\label{eq:key-eq}
\end{align}

In summary, we have the following lemma. \begin{lemma}\label{lem:re-next}
The above two-phase rejection sampling procedure uses $\log((K+1)N+1)$
bits and induces an output distribution $S$ such that for $s\in(0,\infty)$,
\begin{align}
D_{1+s}(P\|S) & \le\frac{1}{s\ln2}\exp\{-s(\log N-D_{1+s}(P\|Q))\}-\log\left(1-s_{N}^{K}\right)
\end{align}
and
\begin{align}
D_{\infty}(P\|S)\leq|D_{\infty}(P\|Q)-\log N|^{+}-\log(1-s_{N}^{K}).
\end{align}
\end{lemma}

Therefore, to bound the sampling error, we need to estimate $s_{N}$.

\begin{lemma}\label{lem:} For any $t\geq0$, it holds that
\begin{equation}
s_{N}\leq\exp\left\{ \frac{-t}{1+t}(\log N-D_{1+t}(P\|Q))\right\} .
\end{equation}
\end{lemma}
\begin{IEEEproof}
We have
\begin{equation}
\sum_{x}\min\{P(x),\beta_{N}Q(x)\}=1-s_{N},
\end{equation}
where
\begin{equation}
\beta_{N}=\sum_{i=0}^{N-1}s_{i}\ge s_{N}N.
\end{equation}
So, $s_{N}$ satisfies
\begin{equation}
P(P<s_{N}NQ)\le\sum_{x}\min\{P(x),s_{N}NQ(x)\}\le1-s_{N}.
\end{equation}
Then, for any $t\geq0$, we have
\begin{align}
\log s_{N}\leq & \log P(P\geq s_{N}NQ)\nonumber \\
\leq & \log\sum_{x}P(x)\left(\frac{P(x)}{s_{N}NQ(x)}\right)^{t}\nonumber \\
= & -t\log N+tD_{1+t}(P\|Q)-t\log s_{N}.\label{eq:app-re-sn}
\end{align}
This directly implies the desired result.
\end{IEEEproof}
Let $\mathcal{W}:\mathcal{X}\to\mathcal{Y}$ be a discrete memoryless
channel. Now, we design our simulation scheme for $\mathcal{W}^{\times n}$
using the two-phase rejection sampling procedure. For given $s>0$
and $r\geq0$, let $K=\log n$, $N=2^{nr}$ and let $Q_{Y}$ be the minimizer of $\min_{Q_{Y}}\max_{x}D_{1+s}(\mathcal{W}_{Y|X}(\cdot|x)\|Q_{Y})$.
The simulation scheme is as follows.

\subsubsection{Simulation Scheme}
\begin{enumerate}
\item Alice and Bob share $Y_{j}^{(n)}\sim Q_{Y}^{\times n}$ for each $j\in\left\{ 0,1,\cdots,(K+1)N\right\} $.
\item Upon receiving the input $x^{n}$, Alice uses the two-phase rejection
sampling procedure with parameters $(\mathcal{W}_{Y|X}^{\times n}(\cdot|x^{n}),Q_{Y}^{\times n},N,K)$
to generate a random index $J\in\{0,1,...,(K+1)N\}$.
\item Alice sends $J$ to Bob using $\log((K+1)N+1)$ bits.
\item Bob generates $Y_{J}^{(n)}$ as the output if $r\geq I_{1+s}(\mathcal{W})$;
otherwise he uniformly picks a random element from $\mathcal{Y}^{\times n}$
as the output.
\end{enumerate}
We denote the above simulation by $\mathcal{N}^{(n)}$. The communication
rate is
\begin{equation}
\lim\limits_{n\to\infty}\frac{1}{n}\log\big\{\left(K+1\right)N+1\big\}=r.
\end{equation}

\textit{Case 1:} $s\in(0,\infty)$ and $r\in[I_{1+s}(\mathcal{W}),I_{\infty}(\mathcal{W}))$.
From Lemma~\ref{lem:re-next} and the definition of R{é}nyi capacity
(cf. Eq.~\eqref{eq:Ichannel}), we see that the simulation satisfies
\begin{align}
 & D_{1+s}\left(\mathcal{W}_{Y|X}^{\times n}(\cdot|x^{n})\big{\|}\mathcal{N}_{Y^{n}|X^{n}}^{(n)}(\cdot|x^{n})\right)\nonumber \\
\leq & \frac{1}{s\ln2}\exp\left\{ -ns\left(r-I_{1+s}(\mathcal{W})\right)\right\} -\log\left(1-s_{N}^{K}\right).\label{con-eq}
\end{align}
By Lemma \ref{lem:} and the definition of R{é}nyi capacity (cf.
Eq.~\eqref{eq:Ichannel}), we have
\begin{align}
s_{N}^{K} & \leq\exp\left\{ -\frac{nsK}{1+s}(r-I_{1+s}(\mathcal{W}))\right\} \nonumber \\
 & \dot{\leq}\exp\left\{ -ns(r-I_{1+s}(\mathcal{W}))\right\} ,\label{eq:eq-re-next}
\end{align}
The last line of Eq.~\eqref{eq:eq-re-next} is independent
of $x^{n}$. Hence, it holds that
\begin{align}
 & \max_{x^{n}}D_{1+s}\left(\mathcal{W}_{Y|X}^{\times n}(\cdot|x^{n})\big{\|}\mathcal{N}_{Y^{n}|X^{n}}^{(n)}(\cdot|x^{n})\right)\nonumber \\
\leq & \frac{1}{s\ln2}\exp\left\{ -ns\left(r-I_{1+s}(\mathcal{W})\right)\right\} -\log\big\{1-\exp\left\{ -ns(r-I_{1+s}(\mathcal{W}))\right\} \big\}\nonumber \\
\dot{=} & \frac{1}{s\ln2}\exp\left\{ -ns\left(r-I_{1+s}(\mathcal{W})\right)\right\} ,\label{eq:app-sc-for-next}
\end{align}
where the inequality follows from Eqs.~\eqref{con-eq} and \eqref{eq:eq-re-next},
and the last line comes from the fact that $-\ln(1-x)=x+o(x)$ as
$x\to0$. Thus, by the definition of $E_{{\rm rf}}^{(1+s)}(\mathcal{W},r)$
(cf. Eq.~\eqref{eq:def-rf}) and Eq.~\eqref{eq:app-sc-for-next},
we obtain
\begin{align}
E_{{\rm rf}}^{(1+s)}(\mathcal{W},r)\geq s\left(r-I_{1+s}(\mathcal{W})\right).\label{eq:key1}
\end{align}

\textit{Case 2:} $s=\infty$ and $r\geq I_{\infty}(\mathcal{W})$.
We have
\begin{equation}
	|D_{\infty}(\mathcal{W}_{Y|X}^{\times n}(\cdot|x^{n})\|Q_{Y}^{\times n})-r|^{+}=0.
\end{equation}
So, Lemma~\ref{lem:re-next} gives that
\begin{equation}\label{eq:dinf}
	D_{\infty}\left(\mathcal{W}_{Y|X}^{\times n}(\cdot|x^{n})\big{\|}\mathcal{N}_{Y^{n}|X^{n}}^{(n)}(\cdot|x^{n})\right)\leq-\log\left(1-s_{N}^{K}\right).
\end{equation}
By Lemma \ref{lem:} and the definition of R{é}nyi capacity (cf.
Eq.~\eqref{eq:Ichannel}),
\begin{equation}
	s_{N}^{K}\leq\exp\left\{ -\frac{nK}{2}(r-I_{2}(\mathcal{W}))\right\} .\label{eq:app-sn-upper-bound}
\end{equation}
Therefore, by the definition of $E_{{\rm rf}}^{(\infty)}(\mathcal{W},r)$
(cf. Eq.~\eqref{eq:def-rf}), Eqs.~\eqref{eq:dinf} and \eqref{eq:app-sn-upper-bound} lead to
\begin{align}
	E_{{\rm rf}}^{(\infty)}(\mathcal{W},r)\geq\liminf_{n\to\infty}\frac{K}{2}(r-I_{2}(\mathcal{W}))=\infty.\label{key0}
\end{align}

\textit{Case 3:} $s\in(0,\infty]$ and $r\in[0,I_{1+s}(\mathcal{W}))$.
$\mathcal{N}^{(n)}$ is given by
\begin{equation}
\mathcal{N}_{Y^{n}|X^{n}}^{(n)}(y^{n}|x^{n})=|\mathcal{Y}|^{-n}.
\end{equation}
Then
\begin{align}
 & D_{1+s}\left(\mathcal{W}^{\times n},\mathcal{N}^{(n)}\right)\nonumber \\
= & \max_{x^{n}}\frac{1}{s}\log\left(\sum_{y^{n}}\mathcal{W}_{Y|X}^{\times n}(y^{n}|x^{n})^{1+s}|\mathcal{Y}|^{ns}\right)\nonumber \\
\leq & \max_{x^{n}}\frac{1}{s}\log\left(|\mathcal{Y}|^{ns}\right)\nonumber \\
= & n\log|\mathcal{Y}|.
\end{align}
Hence, we obtain
\begin{equation}
E_{{\rm rf}}^{(1+s)}(\mathcal{W},r)\geq\liminf_{n\to\infty}-\frac{1}{n}\log(n\log|\mathcal{Y}|)=0.\label{eq:key}
\end{equation}

Combining Eqs.~\eqref{eq:key1} and \eqref{eq:key},
for any $s\in(0,\infty]$ and $r\in[0,I_\infty(\mathcal{W}))$, we obtain
\begin{align}\label{key-3}
E_{{\rm rf}}^{(1+s)}(\mathcal{W},r)\geq
\left|s\left(r-I_{1+s}(\mathcal{W})\right)\right|^{+}.
\end{align}
By the monotonicity of R{é}nyi divergence with respect to its order,
we have $E_{{\rm rf}}^{(1+s_{1})}(\mathcal{W},r)\geq E_{{\rm rf}}^{(1+s_{2})}(\mathcal{W},r)$
when $s_{1}\leq s_{2}$. Then, Eqs.~\eqref{key0} and \eqref{key-3} directly imply Theorem~\ref{thm:reliability-achi}.

\section{Miscellaneous Lemmas}

\label{app:Miscellaneous-Lemmas} \setcounter{equation}{0}
\global\long\def\theequation{D.\arabic{equation}}%
This appendix contains several technical lemmas that are used in the
proofs. At first, we collect some properties of the R{é}nyi divergence.

\begin{lemma}\label{lem:RenyiD-properties} Let $P,Q\in\mathcal{P}(\mathcal{X})$
and $P',Q'\in\mathcal{P}(\mathcal{X}')$. Then the R{é}nyi divergence
satisfies the following properties:

\begin{enumerate}[(i)]
\item Monotonicity w.r.t. the order~\cite{VanHarremos2014renyi}: If $0\leq\alpha\leq\beta$,
then $D_{\alpha}(P\|Q)\leq D_{\beta}(P\|Q)$;
\item Additivity~\cite{VanHarremos2014renyi}: For any $\alpha\in[0,\infty]$,
we have
\begin{equation}
D_{\alpha}(P\times P'\|Q\times Q')=D_{\alpha}(P\|Q)+D_{\alpha}(P'\|Q');
\end{equation}
\item Data processing inequality~\cite{AliSilvey1966general}: Let $\mathcal{W}:\mathcal{X}\rightarrow\mathcal{Y}$
be a channel. For any $\alpha\in[0,\infty]$, we have
\begin{equation}
D_{\alpha}(\mathcal{W}(P)\|\mathcal{W}(Q))\leq D_{\alpha}(P\|Q);
\end{equation}
\item Variational representation~\cite{CsiszarMatus2003information}: For
$\alpha\in(0,1)$, the R{é}nyi divergence can be written as
\begin{equation}
D_{\alpha}(P\|Q)=\min_{S\in\mathcal{P}(\mathcal{X})}\bigg\{\frac{\alpha}{1-\alpha}D(S\|P)+D(S\|Q)\bigg\}.
\end{equation}
\end{enumerate}
\end{lemma}

In the next lemma, we collect some properties of the R{é}nyi mutual
information and the R{é}nyi capacity. \begin{lemma}\label{lem:RenyiC-properties}
Let $P_{X}\in\mathcal{P}(\mathcal{X})$, and let $\mathcal{W}:\mathcal{X}\rightarrow\mathcal{Y}$
and $\mathcal{W}':\mathcal{X}'\rightarrow\mathcal{Y}'$ be discrete
memoryless channels. We have

\begin{enumerate}[(i)]
\item Additivity of R{é}nyi capacity~\cite{Arimoto1977information}:
For any $\alpha\in[0,\infty]$,
\begin{equation}
I_{\alpha}(\mathcal{W}\times\mathcal{W}')=I_{\alpha}(\mathcal{W})+I_{\alpha}(\mathcal{W}');
\end{equation}
\item Concavity and continuity in input distribution~\cite{HoVerdu2015convexity},
\cite{Nakibouglu2018renyi}: For $\alpha\in[1,\infty]$, $I_{\alpha}(X:Y)_{P_{X}\cdot\mathcal{W}_{Y|X}}$
is concave and continuous in $P_{X}$ for fixed $\mathcal{W}_{Y|X}$;
\item Convexity and continuity in channel~\cite{CoverThomas1991elements}:
$I(X:Y)_{P_{X}\cdot\mathcal{W}_{Y|X}}$ is convex and continuous in
$\mathcal{W}_{Y|X}$ for fixed $P_{X}$;
\item Properties of the R{é}nyi mutual information w.r.t. the order~\cite{Nakibouglu2018renyi}:
The function $\alpha\mapsto I_{\alpha}(X:Y)_{P_{X}\cdot\mathcal{W}_{Y|X}}$
is increasing and continuously differentiable on $(0,\infty)$, and
the function $t\mapsto tI_{1+t}(X:Y)_{P_{X}\cdot\mathcal{W}_{Y|X}}$
is convex on $(0,\infty)$;
\item Properties of the R{é}nyi capacity w.r.t. the order~\cite{Nakibouglu2018renyi}:
The function $\alpha\mapsto I_{\alpha}(\mathcal{W})$ is increasing
and continuous on $(0,\infty)$, and the function $t\mapsto tI_{1+t}(\mathcal{W})$
is convex on $(0,\infty)$.
\end{enumerate}
\end{lemma}

From Lemma \ref{lem:RenyiC-properties}, we derive the following lemma.

\begin{lemma}\label{lem:Renyi-maximization} Let $\mathcal{W}$ be
a discrete memoryless channel and $s>0$.

\begin{enumerate}[(i)]
\item If $r\geq\hat{R}(s)$, then
\begin{align}
\sup\limits_{t\geq0}t\left(r-I_{1+t}(\mathcal{W})\right)=\sup\limits_{t\geq s}t(r-I_{1+t}(\mathcal{W}));
\end{align}
\item If $0\leq r<\hat{R}(s)$, then
\begin{align}
\sup\limits_{t\geq s}t\left(r-I_{1+t}(\mathcal{W})\right)=s\left(r-I_{1+s}(\mathcal{W})\right);
\end{align}
\item We have
\begin{align}
\sup\limits_{t\geq0}t(\hat{R}(s)-I_{1+t}(\mathcal{W}))=s(\hat{R}(s)-I_{1+s}(\mathcal{W})).
\end{align}
\end{enumerate}
\end{lemma}
\begin{IEEEproof}
Consider the function $f(t):=t\left(r-I_{1+t}(\mathcal{W})\right)$.
The right derivative of $f(t)$ is
\begin{equation}
\partial_{t}^{+}f(t)=r-\hat{R}(t).
\end{equation}
Lemma~\ref{lem:RenyiC-properties}~(\romannumeral5) tells us that
$t\mapsto tI_{1+t}(\mathcal{W})$ is convex on $(0,\infty)$. So,
$t\mapsto\hat{R}(t)$ is increasing on $(0,\infty)$.

For the first statement, it is easy to see that $\partial_{t}^{+}f(t)\geq0$
for $t\leq s$. Therefore, the supremum can be restricted to $t\geq s$.
For the second statement, similarly we can see that $\partial_{t}^{+}f(t)<0$
for $t\geq s$. So, the supremum is attained at $t=s$. For the last
statement, we replace $r$ with $\hat{R}(s)$ and find that $\partial_{t}^{+}f(t)\geq0$
when $t<s$, and $\partial_{t}^{+}f(t)\leq0$ when $t>s$.
Therefore, the supremum is attained at $t=s$.
\end{IEEEproof}

The following lemma confirms the continuity property used in the proof of
Lemma~\ref{lem:infty-upper}.
\begin{lemma}\label{lem:continuity} Let $\{(a_{k},Q_{X}^{(k)},\widehat{\mathcal{W}}_{k})\}_{k=1}^{\infty}$
be a sequence in $\mathbb{R}\times\mathcal{P(X)}\times\mathcal{P(Y|X)}$
such that $a_{k}\searrow0$ and $(Q_{X}^{(k)},\widehat{\mathcal{W}}_{k})\to(Q_{X}^{*},\widehat{\mathcal{W}}^{*})$,
where $\mathcal{P(Y|X)}$ denotes the set of all discrete memoryless
channels from $\mathcal{X}$ to $\mathcal{Y}$. Define
\begin{equation}
f(a,Q_{X},\widehat{\mathcal{W}}):=\sum_{x\in\mathcal{X}}Q_{X}(x)D_{1+a}\left(\widehat{\mathcal{W}}_{Y|X}(\cdot|x)\big\|\widehat{\mathcal{W}}(Q_{X})\right).
\end{equation}
Then $\lim\limits_{k\to\infty}f(a_{k},Q_{X}^{(k)},\widehat{\mathcal{W}}_{k})=f(0,Q_{X}^{*},\widehat{\mathcal{W}}^{*})$.
\end{lemma}
\begin{IEEEproof}
To prove the statement, it suffices to prove that for any $\epsilon>0$,
there exists $N\in\mathbb{N}$ such that for any $k\geq N$, we have
\begin{equation}
|f(a_{k},Q_{X}^{(k)},\widehat{\mathcal{W}}_{k})-f(0,Q_{X}^{*},\widehat{\mathcal{W}}^{*})|\leq\epsilon.\label{eq:result}
\end{equation}
Since $\supp(\widehat{\mathcal{W}}^{*}(\cdot|x))\subseteq\supp(\widehat{\mathcal{W}}^{*}(Q_{X}^{*}))$
for any $x\in\supp(Q_{X}^{*})$, we have $f(a_{k},Q_{X}^{*},\widehat{\mathcal{W}}^{*})\to f(0,Q_{X}^{*},\widehat{\mathcal{W}}^{*})$.
Therefore, there exists $N_{1}\in\mathbb{N}$ such that for any $k\geq N_{1}$,
\begin{equation}
|f(a_{k},Q_{X}^{*},\widehat{\mathcal{W}}^{*})-f(0,Q_{X}^{*},\widehat{\mathcal{W}}^{*})|\leq\frac{\epsilon}{2}.\label{eq:continuity-4}
\end{equation}
Since $f(a_{N_{1}},Q_{X},\widehat{\mathcal{W}})$ is composed of continuous
functions in $(Q_{X},\widehat{\mathcal{W}})$, we have $f(a_{N_{1}},Q_{X}^{(k)},\widehat{\mathcal{W}}_{k})\to f(a_{N_{1}},Q_{X}^{*},\widehat{\mathcal{W}}^{*})$.
Therefore, there exists $N_{2}\in\mathbb{N}$ such that for any $k\geq N_{2}$,
\begin{equation}
|f(a_{N_{1}},Q_{X}^{(k)},\widehat{\mathcal{W}}_{k})-f(a_{N_{1}},Q_{X}^{*},\widehat{\mathcal{W}}^{*})|\leq\frac{\epsilon}{2}.\label{eq:continuity-3}
\end{equation}
Similarly, $f(0,Q_{X}^{(k)},\widehat{\mathcal{W}}_{k})\to f(0,Q_{X}^{*},\widehat{\mathcal{W}}^{*})$.
So, there exists $N_{3}\in\mathbb{N}$ such that for any $k\geq N_{3}$,
we have
\begin{align}
|f(0,Q_{X}^{(k)},\widehat{\mathcal{W}}_{k})-f(0,Q_{X}^{*},\widehat{\mathcal{W}}^{*})| & \leq\frac{\epsilon}{2}.\label{eq:continuity-5}
\end{align}
Let $N=\max\{N_{1},N_{2},N_{3}\}$. By the monotonicity of R{é}nyi
divergence with respect to its order, it follows that $f$ is non-decreasing
with respect to $a$. Hence, for any $k\geq N$, we have
\begin{align}
 & |f(a_{k},Q_{X}^{(k)},\widehat{\mathcal{W}}_{k})-f(0,Q_{X}^{*},\widehat{\mathcal{W}}^{*})|\nonumber \\
\leq & \max\{|f(a_{N_{1}},Q_{X}^{(k)},\widehat{\mathcal{W}}_{k})-f(0,Q_{X}^{*},\widehat{\mathcal{W}}^{*})|,|f(0,Q_{X}^{(k)},\widehat{\mathcal{W}}_{k})-f(0,Q_{X}^{*},\widehat{\mathcal{W}}^{*})|\}.\label{eq:continuity-1}
\end{align}
Using the triangle inequality, it holds that
\begin{align}
 & |f(a_{N_{1}},Q_{X}^{(k)},\widehat{\mathcal{W}}_{k})-f(0,Q_{X}^{*},\widehat{\mathcal{W}}^{*})|\nonumber \\
\leq & |f(a_{N_{1}},Q_{X}^{(k)},\widehat{\mathcal{W}}_{k})-f(a_{N_{1}},Q_{X}^{*},\widehat{\mathcal{W}}^{*})|+|f(a_{N_{1}},Q_{X}^{*},\widehat{\mathcal{W}}^{*})-f(0,Q_{X}^{*},\widehat{\mathcal{W}}^{*})|\leq\epsilon,\label{eq:continuity-2}
\end{align}
where the last inequality follows from Eqs.~\eqref{eq:continuity-4}
and \eqref{eq:continuity-3}. Inserting Eqs.~\eqref{eq:continuity-5}
and \eqref{eq:continuity-2} into Eq.~\eqref{eq:continuity-1} yields
Eq.~\eqref{eq:result}.
\end{IEEEproof}
Sion's minimax theorem has been used for several times in the proofs.

\begin{lemma}[\cite{Sion1958general}]\label{lem:minimax-Thm}
Let $\mathcal{A}$ be a compact convex set in a topological vector
space $\mathcal{V}$ and $\mathcal{B}$ be a convex subset of a vector
space $\mathcal{U}$. Let $f:\mathcal{A}\times\mathcal{B}\rightarrow\mathbb{R}$
be such that

\begin{enumerate}[(i)]
\item $f(a,\cdot)$ is quasi-concave and upper semi-continuous on $\mathcal{B}$
for each $a\in\mathcal{A}$, and
\item $f(\cdot,b)$ is quasi-convex and lower semi-continuous on $\mathcal{A}$
for each $b\in\mathcal{B}$.
\end{enumerate}
Then, we have
\begin{equation}
\inf_{a\in\mathcal{A}}\sup_{b\in\mathcal{B}}f(a,b)=\sup_{b\in\mathcal{B}}\inf_{a\in\mathcal{A}}f(a,b),\label{eq:min-max}
\end{equation}
and the infima in Eq.~\eqref{eq:min-max} can be replaced by minima.
\end{lemma}


\begin{thebibliography}{99}

\bibitem{BSST2002entanglement}
C.~H. Bennett, P.~W. Shor, J.~A. Smolin, and A.~V. Thapliyal,
  ``Entanglement-assisted capacity of a quantum channel and the reverse
  {Shannon} theorem,'' \emph{IEEE Trans. Inf. Theory}, vol.~48, no.~10, pp.
  2637--2655, 2002.

\bibitem{Winter2002compression}
A.~Winter, ``Compression of sources of probability distributions and density
  operators,'' \emph{arXiv:quant-ph/0208131}, 2002.

\bibitem{HJMR2007communication}
P.~Harsha, R.~Jain, D.~McAllester, and J.~Radhakrishnan, ``The communication
  complexity of correlation,'' in \emph{Twenty-Second Annual IEEE Conference on
  Computational Complexity (CCC'07)}.\hskip 1em plus 0.5em minus 0.4em\relax
  IEEE, 2007, pp. 10--23.

\bibitem{CLMW2011zero}
T.~S. Cubitt, D.~Leung, W.~Matthews, and A.~Winter, ``Zero-error channel
  capacity and simulation assisted by non-local correlations,'' \emph{IEEE
  Trans. Inf. Theory}, vol.~57, no.~8, pp. 5509--5523, 2011.

\bibitem{Cuff2013distributed}
P.~Cuff, ``Distributed channel synthesis,'' \emph{IEEE Trans. Inf. Theory},
  vol.~59, no.~11, pp. 7071--7096, 2013.

\bibitem{BDHSW2014quantum}
C.~H. Bennett, I.~Devetak, A.~W. Harrow, P.~W. Shor, and A.~Winter, ``The
  quantum reverse {Shannon} theorem and resource tradeoffs for simulating
  quantum channels,'' \emph{IEEE Trans. Inf. Theory}, vol.~60, no.~5, pp.
  2926--2959, 2014.

\bibitem{YGA2015channel}
M.~H. Yassaee, A.~Gohari, and M.~R. Aref, ``Channel simulation via interactive
  communications,'' \emph{IEEE Trans. Inf. Theory}, vol.~61, no.~6, pp.
  2964--2982, 2015.

\bibitem{HYBGA2016simulation}
F.~Haddadpour, M.~H. Yassaee, S.~Beigi, A.~Gohari, and M.~R. Aref, ``Simulation
  of a channel with another channel,'' \emph{IEEE Trans. Inf. Theory}, vol.~63,
  no.~5, pp. 2659--2677, 2016.

\bibitem{LiEi2018strong}
C.~T. Li and A.~El~Gamal, ``Strong functional representation lemma and
  applications to coding theorems,'' \emph{IEEE Trans. Inf. Theory}, vol.~64,
  no.~11, pp. 6967--6978, 2018.

\bibitem{YuTan2020on}
L.~Yu and V.~Y.~F. Tan, ``exact and {$\infty$}-{R\'enyi} common information,''
  \emph{IEEE Trans. Inf. Theory}, vol.~66, no.~6, pp. 3366--3406, 2020.

\bibitem{YuTan2020exact}
------, ``Exact channel synthesis,'' \emph{IEEE Trans. Inf. Theory}, vol.~66,
  no.~5, pp. 2799--2818, 2020.

\bibitem{CRBT2024channel}
M.~X. Cao, N.~Ramakrishnan, M.~Berta, and M.~Tomamichel, ``Channel simulation:
  Finite blocklengths and broadcast channels,'' \emph{IEEE Trans. Inf. Theory},
  vol.~70, no.~10, pp. 6780--6808, 2024.

\bibitem{Wyner1975common}
A.~Wyner, ``The common information of two dependent random variables,''
  \emph{IEEE Trans. Inf. Theory}, vol.~21, no.~2, pp. 163--179, 1975.

\bibitem{Fano1961transmission}
R.~M. Fano, \emph{Transmission of Information: A Statistical Theory of
  Communications}.\hskip 1em plus 0.5em minus 0.4em\relax MIT Press, Cambridge,
  1961.

\bibitem{Gallager1965simple}
R.~Gallager, ``A simple derivation of the coding theorem and some
  applications,'' \emph{IEEE Trans. Inf. Theory}, vol.~11, no.~1, pp. 3--18,
  1965.

\bibitem{SGB1967lower1}
C.~E. Shannon, R.~G. Gallager, and E.~R. Berlekamp, ``Lower bounds to error
  probability for coding on discrete memoryless channels. {I},'' \emph{Inf.
  Contr.}, vol.~10, no.~1, pp. 65--103, 1967.

\bibitem{Arimoto1973converse}
S.~Arimoto, ``On the converse to the coding theorem for discrete memoryless
  channels,'' \emph{IEEE Trans. Inf. Theory}, vol.~19, no.~3, pp. 357--359,
  1973.

\bibitem{DueckKorner1979reliability}
G.~Dueck and J.~K{\"o}rner, ``Reliability function of a discrete memoryless
  channel at rates above capacity,'' \emph{IEEE Trans. Inf. Theory}, vol.~25,
  no.~1, pp. 82--85, 1979.

\bibitem{Renyi1961measures}
A.~R{\'e}nyi, ``On measures of entropy and information,'' in \emph{Proc. 4th
  Berkeley Symp. Math. Statist. Probab.}, vol.~1.\hskip 1em plus 0.5em minus
  0.4em\relax University of California Press, Berkeley, 1961, pp. 547--561.

\bibitem{OCCB2024exponents}
A.~Oufkir, M.~X. Cao, H.-C. Cheng, and M.~Berta, ``Exponents for shared
  randomness-assisted channel simulation,'' \emph{arXiv:2410.07051}, 2024.

\bibitem{OYB2024exponents}
A.~Oufkir, Y.~Yao, and M.~Berta, ``Exponents for classical-quantum channel
  simulation in purified distance,'' \emph{arXiv:2410.10770}, 2024.

\bibitem{Csiszar1998method}
I.~Csisz{\'a}r, ``The method of types,'' \emph{IEEE Trans. Inf. Theory},
  vol.~44, no.~6, pp. 2505--2523, 1998.

\bibitem{Verdu2015alpha}
S.~Verd{\'u}, ``$\alpha$-mutual information,'' in \emph{Inf. Theory Appl.
  Workshop (ITA)}.\hskip 1em plus 0.5em minus 0.4em\relax IEEE, 2015, pp. 1--6.

\bibitem{HoVerdu2015convexity}
S.-W. Ho and S.~Verd{\'u}, ``Convexity/concavity of {R{\'e}nyi} entropy and
  $\alpha$-mutual information,'' in \emph{IEEE Int. Symp. Inf. Theory
  (ISIT)}.\hskip 1em plus 0.5em minus 0.4em\relax IEEE, 2015, pp. 745--749.

\bibitem{Arimoto1977information}
S.~Arimoto, ``Information measures and capacity of order $\alpha$ for discrete
  memoryless channels,'' in \emph{Topics in Information Theory}.\hskip 1em plus
  0.5em minus 0.4em\relax Amsterdam, The Netherlands: North Holland, 1977, pp.
  41--52.

\bibitem{Csiszar1995generalized}
I.~Csisz{\'a}r, ``Generalized cutoff rates and {R{\'e}nyi's} information
  measures,'' \emph{IEEE Trans. Inf. Theory}, vol.~41, no.~1, pp. 26--34, 1995.

\bibitem{Gallager1968information}
R.~Gallager, \emph{Information Theory and Reliable Communication}.\hskip 1em
  plus 0.5em minus 0.4em\relax John Wiley \& Sons, 1968.

\bibitem{bonnans2013perturbation}
J.~F. Bonnans and A.~Shapiro, \emph{Perturbation analysis of optimization
  problems}.\hskip 1em plus 0.5em minus 0.4em\relax Springer Science \&
  Business Media, 2013.

\bibitem{LiYao2024reliability}
K.~Li and Y.~Yao, ``Reliability function of quantum information decoupling via
  the sandwiched {R\'enyi} divergence,'' \emph{Commun. Math. Phys.}, vol. 405,
  no.~7, p. 160, 2024.

\bibitem{BCR2011the}
M.~Berta, M.~Christandl, and R.~Renner, ``The quantum reverse {Shannon} theorem
  based on one-shot information theory,'' \emph{Commun. Math. Phys.}, vol. 306,
  no.~3, pp. 579--615, 2011.

\bibitem{LiYao2025reliable}
K.~Li and Y.~Yao, ``Reliable simulation of quantum channels: the error
  exponent,'' \emph{IEEE Trans. Inf. Theory}, vol.~71, no.~1, pp. 518--529,
  2025.

\bibitem{Chen2000generalization}
P.-N. Chen, ``Generalization of {G\"{a}rtner--Ellis} theorem,'' \emph{IEEE
  Trans. Inf. Theory}, vol.~46, no.~7, pp. 2752--2760, 2000.

\bibitem{MosonyiOgawa2015two}
M.~Mosonyi and T.~Ogawa, ``Two approaches to obtain the strong converse
  exponent of quantum hypothesis testing for general sequences of quantum
  states,'' \emph{IEEE Trans. Inf. Theory}, vol.~61, no.~12, pp. 6975--6994,
  2015.

\bibitem{TomamichelHayashi2017operational}
M.~Tomamichel and M.~Hayashi, ``Operational interpretation of {R{\'e}nyi}
  information measures via composite hypothesis testing against product and
  {Markov} distributions,'' \emph{IEEE Trans. Inf. Theory}, vol.~64, no.~2, pp.
  1064--1082, 2017.

\bibitem{VanHarremos2014renyi}
T.~Van~Erven and P.~Harremos, ``R{\'e}nyi divergence and {Kullback--Leibler}
  divergence,'' \emph{IEEE Trans. Inf. Theory}, vol.~60, no.~7, pp. 3797--3820,
  2014.

\bibitem{AliSilvey1966general}
S.~M. Ali and S.~D. Silvey, ``A general class of coefficients of divergence of
  one distribution from another,'' \emph{J. Roy. Statist. Soc. Ser. B},
  vol.~28, no.~1, pp. 131--142, 1966.

\bibitem{CsiszarMatus2003information}
I.~Csisz{\'a}r and F.~Matus, ``Information projections revisited,'' \emph{IEEE
  Trans. Inf. Theory}, vol.~49, no.~6, pp. 1474--1490, 2003.

\bibitem{Nakibouglu2018renyi}
B.~Nakibo{\u{g}}lu, ``The {R{\'e}nyi} capacity and center,'' \emph{IEEE Trans.
  Inf. Theory}, vol.~65, no.~2, pp. 841--860, 2018.

\bibitem{CoverThomas1991elements}
T.~M. Cover and J.~A. Thomas, \emph{Elements of Information Theory}.\hskip 1em
  plus 0.5em minus 0.4em\relax John Wiley \& Sons, 1991.

\bibitem{Sion1958general}
M.~Sion, ``On general minimax theorems,'' \emph{Pac. J. Math.}, vol.~8, no.~1,
  pp. 171--176, 1958.

\end{thebibliography}

\end{document}